\documentclass[manuscript]{acmart}

\usepackage{enumitem}
\usepackage[justification=centering]{caption}
\usepackage{tcolorbox} 
\usepackage{multirow}
\usepackage{makecell}
\usepackage{url}
\usepackage{hyperref}

\usepackage{listings}
\usepackage{xcolor}

\tcbuselibrary{skins}
\newtcolorbox{rqbox}[2]{
  enhanced,
  colback=black!7,
  colframe=black!7,
  boxrule=0pt,
  borderline west={3pt}{0pt}{black!70},
  arc=2pt,
  left=10pt, right=10pt, top=6pt, bottom=6pt,
  before upper={%
    {\sffamily\bfseries\small #1\hspace{0.6em}\normalfont\sffamily\small #2}%
    \par\vspace{2pt}%
    {\color{black!60}\hrule height 0.8pt}\par\vspace{4pt}%
  }
}

\newtcolorbox{insightbox}{
  enhanced,
  colback=black!7,
  colframe=black!7,
  boxrule=0pt,
  borderline west={3pt}{0pt}{black!70},
  arc=2pt,
  left=10pt, right=10pt, top=6pt, bottom=6pt,
  before upper={%
    {\sffamily\bfseries\small Summary}%
    \par\vspace{2pt}%
    {\color{black!60}\hrule height 0.8pt}\par\vspace{4pt}%
  }
}

\AtBeginDocument{%
  }

\setcopyright{acmlicensed}
\copyrightyear{2025}
\acmYear{2025}
\acmDOI{XXXXXXX.XXXXXXX}

\newcommand{\firstrevision}[1]{\textcolor{black}{{#1}}}
\newcommand{\secondrevision}[1]{\textcolor{black}{{#1}}}

\begin{document}

\title{From Empirical Evaluation to Context-Aware Enhancement: Repairing Regression Errors with LLMs}

\author{Anh Ho}
\affiliation{
  \institution{The University of Melbourne}
  \country{Australia}
}
\email{anh.ho1@student.unimelb.edu.au}
\orcid{0000-0001-7483-7119}

\author{Thanh Le-Cong}
\affiliation{
  \institution{Singapore University of Technology and Design}
  \country{Singapore}
}
\email{congthanh_le@sutd.edu.sg}
\orcid{0000-0002-9566-324X}

\author{Bach Le}
\affiliation{
  \institution{The University of Melbourne}
  \country{Australia}
}
\email{bach.le@unimelb.edu.au}
\orcid{0000-0001-5044-1582}

\author{Christine Rizkallah}
\affiliation{
  \institution{The University of Melbourne}
  \country{Australia}
}
\email{christine.rizkallah@unimelb.edu.au}
\orcid{0000-0003-4785-2836}

\renewcommand{\shortauthors}{Ho et al.}

\begin{abstract}
Software systems constantly evolve to adapt to ever-changing customer demands and evolving markets. During the software evolution process, code changes across versions can unexpectedly break previously working functionalities, leading to so-called regression bugs.
Manually repairing regression bugs is both challenging and time-consuming, as the process often requires developers to traverse the history of the software in order to understand how the bugs were introduced and, consequently, how to repair them.
Despite the challenges of manual efforts, only a limited number of studies have investigated automated program repair (APR) for regression bugs, most of which date back several years and focus on traditional APR techniques.
Since then, various APR approaches, especially those leveraging the power of large language models (LLMs), have been rapidly developed to fix general software bugs. Unfortunately, the effectiveness of these advanced techniques in the context of regression bugs remains largely unexplored. This gap motivates the need for an empirical study evaluating the effectiveness of modern APR techniques in fixing real-world regression bugs.

In this work, we conduct an empirical study of APR techniques on regression bugs.
To facilitate our study, we introduce \mbox{\textsc{\firstrevision{RegressionBug4APR}}}, a high-quality benchmark of Java \firstrevision{and Python} regression bugs integrated into a framework designed to facilitate APR research. The current benchmark includes \firstrevision{200} regression bugs collected from widely used real-world GitHub repositories.
We begin by conducting an in-depth analysis of the benchmark, demonstrating its diversity and quality.
Building on this foundation, we empirically evaluate the capabilities of APR to regression bugs by assessing both traditional APR tools and advanced LLM-based APR approaches. Our experimental results show that classical APR tools fail to repair any bugs, while LLM-based APR approaches exhibit promising potential.
Motivated by these results, we investigate the impact of incorporating bug-inducing change information into LLM-based APR approaches for fixing regression bugs.
\firstrevision{We further conduct an ablation study to disaggregate the contribution of each contextual element within the bug-inducing change information.}
Our results highlight that this context-aware enhancement significantly improves the performance of LLM-based APR, \firstrevision{yielding 1.6×} more successful repairs compared to using LLM-based APR without such context.
\firstrevision{Moreover, our findings are consistent across both Java and Python benchmarks, providing preliminary evidence for the generalizability of our findings.}

\end{abstract}


\maketitle

\section{Introduction}
\label{sec:introduction}

Software systems are continuously evolving to meet user demands and adapt to rapidly changing market conditions, resulting in increasing system complexity and frequent source code modifications. Each modification usually requires interactions with an ever-growing number of components within the code base such as calling to existing functionality or integrating with other parts of the system. However, these code changes can unintentionally introduce regression bugs, disrupting previously working features of software systems~\cite{tan2015relifix}. 

Regression bugs often stem from the unintended consequences of modifications to existing code. 
This unique characteristic makes regression bugs particularly challenging, as developers must identify the specific changes responsible, i.e., bug-inducing changes, and trace the effects of bug-inducing changes on previously functioning features~\cite{zeller1999yesterday}.
In large-scale software projects, the challenges associated with regression bugs are intensive, as they often remain undetected for several years before being identified and resolved~\cite{yin2011fixes,bohme2014corebench}.
\firstrevision{Indeed, empirical studies highlight the significant prevalence of regression bugs in real-world systems.
Xiao et al.~\cite{xiao2019perspective} conducted a large-scale empirical study of 5,741 bug reports from the Linux kernel and found that approximately half of all bugs are regressions.
Similarly, Khattar et al.~\cite{khattar2015sarathi} reported that the proportion of regression bugs in the Google Chromium project reaches 51.09\%.
This underscores that regression bugs are not a niche concern but a dominant and persistent category of defects in large-scale real-world systems.
Furthermore, Xiao et al.~\cite{xiao2019empirical} studied 1,579 regression bugs across 57 Linux versions and found that bugs occurring in regression bug chains cost 2.4× more fixing time, involve 1.3× more developers, and generate 2.8× more discussion comments compared to isolated regression bugs, with 68\% of such chains propagating across software versions and compounding in cost over time.
Moreover, low-quality fixes might introduce new regressions, creating a recurring maintenance cycle that further amplifies the overall cost and effort~\cite{bohme2013regression,xiao2019empirical}.}
Consequently, regression bugs continue to be a persistent challenge in the software industry, affecting overall software quality and maintenance efforts~\cite{onoma1998regression, engstrom2010qualitative}.

Automated program repair (APR) techniques were developed to reduce the manual effort required to fix bugs and improve software quality by automating the repair process. Hence, applying APR to regression bugs would be particularly valuable, given their complexity and the effort required to address them. To date, there has been extensive research on APR for general software bugs~\cite{monperrus2018living, le2019automated, gao2022program}, including a growing interest in large language model (LLM)-based APR techniques in recent years~\cite{silva2023repairllama,jiang2023impact,ruan2024specrover, yin2024thinkrepair,bouzenia2025repairagent,li2024hybrid,xu2024aligning,luo2025federated,yang2025,huang2025comprehensive}.
However, the most recent work specifically targeting regression bug repair dates back several years. Notably, Relifix, developed for C/C++ programs, focuses solely on traditional pattern-based and search-based APR techniques~\cite{tan2015relifix}.
This naturally raises an important question:

\begin{center}
\textit{\textbf{How well do state-of-the-art APR techniques perform when applied \\ specifically to regression bugs?}}
\end{center}

\noindent
This motivates the need for an empirical study to revisit existing APR techniques to better understand their effectiveness on regression bugs, where they fall short and can be further improved.

One major challenge in conducting the empirical study lies in the lack of a high-quality and up-to-date benchmark that reflects both current software development practices and the ongoing progress in this research field.
Popular datasets commonly used for evaluating APR techniques, such as Defects4J~\cite{just2014defects4j}, Bugs.jar~\cite{saha2018bugs} and Bears~\cite{madeiral2019bears}, primarily target general software bugs and are associated with bug-triggering test cases that have been added/updated after the bug is reported~\cite{liu2021critical}. These datasets are not constructed with regression-specific characteristics in mind. The bugs are not revealed through regression testing, making it unclear whether they truly represent regressions. 
Hence, these datasets lack the temporal and behavioral properties that define regression bugs.
Existing datasets for regression bugs, such as CoREBench~\cite{bohme2014corebench} and CIBugs~\cite{kabadi2023future}, are outdated (bugs are before 2017) and difficult to reproduce, making them less suitable for studying regression repair.

To address these problems, we propose \mbox{\textsc{\firstrevision{RegressionBug4APR}}}, a high-quality Java \firstrevision{and Python} regression error benchmark integrated into a framework designed to facilitate APR research. 
\firstrevision{For Java, the benchmark is built upon an automated tool that mines regression bugs from software evolution history, namely \textsc{RegMiner}~\cite{song2022regminer}.
By leveraging \textsc{RegMiner}, we automatically collect potential regression errors from Java code repositories and then validate the regressions.
For Python, we developed a dedicated validation tool following the same high-level design as RegMiner, adapted for the Python ecosystem.}
Our benchmark collection framework ensures the reproducibility and quality of each mined bug instance, while also supporting future benchmark extensions.
Through this automated process, we have established \firstrevision{\textsc{RegressionBug4APR}}, which includes
\firstrevision{200 confirmed regression bugs collected from widely used real-world Java and Python GitHub repositories.}

We conduct an in-depth analysis of our benchmark \firstrevision{\textsc{RegressionBug4APR}} to demonstrate desired properties of the benchmark, including: \textit{reality}, where regression bugs are extracted from diverse real-world open-source software; \textit{currency}, with the bugs spanning from 2016 onward; \textit{diversity}, encompassing a broad range of projects and diverse repair operators; \textit{extensibility}, as our approach enables the dataset to expand over time with minimal human efforts by building upon \textsc{RegMiner}'s approach~\cite{song2022regminer}; and \textit{durable replicability}, with the framework offering easy access, extensibility, and snapshots of all projects and their dependencies, similar to that of Defects4J~\cite{just2014defects4j}. Prior study suggested that these properties are often required to enable fair evaluations~\cite{tomassi2019bugswarm}.

Using the diverse and high-quality benchmark, we conduct an empirical study to revisit APR techniques on regression errors. The evaluation is broadly categorized into two classes: \textit{traditional} and \textit{advanced LLM-based} techniques. For \textit{traditional} techniques, we examine template-based and search-based tools, including GenProg~\cite{le2011genprog}, Kali~\cite{qi2015analysis}, Cardumen~\cite{martinez2018ultra}, Arja~\cite{yuan2018arja}, jMutRepair~\cite{martinez2016astor}, and TBar~\cite{liu2019tbar}. 
For \textit{advanced} techniques, we focus on LLM-based approaches, further divided into two groups: \textit{fine-tuning-based} and \textit{prompt-based} techniques. In the \textit{fine-tuning} category, we evaluate full fine-tuning methods, including CodeGen models with 2B and 6B parameters and Incoder models with 1B and 6B parameters~\cite{xia2023automated}, as well as parameter-efficient fine-tuning methods, such as RepairLLaMA~\cite{silva2023repairllama}. 
For \textit{prompt-based} approaches, we explore two repair strategies, \textit{zero-shot prompting} and \textit{conversational interaction}, using advanced general-purpose LLMs: ChatGPT-3.5-Turbo and ChatGPT-4o. 
\firstrevision{Note that traditional and fine-tuning-based techniques are evaluated on Java regression bugs only, as they are inherently designed for Java and do not support Python. To ensure a fair comparison, prompt-based techniques are also evaluated on the Java regression bugs when compared against these approaches.}
Our experiments show that traditional pattern-based and search-based APR tools fail to repair any regression bugs in our dataset. In contrast, LLM-based approaches demonstrate promising performance: \firstrevision{RepairLLaMA, the top-performing fine-tuning-based technique, successfully repairs 15 bugs, while ChatGPT-4o with conversational prompting, the most effective prompt-based method, successfully repairs 18 bugs on the Java benchmark}. It is, however, noted that these LLM-based techniques are not inherently designed to tackle the nature of regression bugs, e.g., taking into account the bug-inducing changes to inform bug repair and thus leaving room for further improvements.

Finally, we further explore the impact of incorporating additional regression-specific context, e.g., bug-inducing change information, on improving the repair effectiveness of LLM-based APR.
We do so by leveraging prompt-based approaches, explicitly integrating bug-inducing changes into the prompt design. This choice is driven by the inherent flexibility of prompt-based methods in terms of model usage and prompt construction, as well as their data efficiency compared to fine-tuning-based techniques.
Our experiments show that incorporating this additional context significantly enhances the performance and precision of LLM-based techniques using prompting. \firstrevision{Specifically, the best-performing configuration, conversational ChatGPT-4o with full bug-inducing change (BIC) information, successfully repairs 39 out of 200 regression bugs in the \textsc{RegressionBug4APR} benchmark, representing a 1.6× improvement over the best-performing configuration without BIC context.} 
Interestingly, our qualitative analysis further reveals that bug-inducing change information provides crucial context that helps models better identify the root cause of the regression and determine whether to fully or partially revert to previous correct statements. In addition, it helps narrow the fix scope, thereby improving fault localization and reducing unnecessary code modifications.

\firstrevision{To further understand the contribution of each contextual element, we conduct an ablation study. Our results reveal that each component contributes positively and incrementally to repair performance, with code-level changes providing a larger gain than commit messages alone. Notably, commit messages and code changes capture complementary, non-redundant signals that together enhance repair effectiveness.}

In summary, the key contributions of this work are as follows:
\vspace{-0.3cm}
\begin{itemize}
    \item[1)] We introduce \mbox{\textsc{\firstrevision{RegressionBug4APR}}}, a high-quality Java \firstrevision{and Python} regression error benchmark integrated into a framework that supports reproducible studies in automated program repair and enables future extensions. We also provide a comprehensive analysis of the benchmark's characteristics.
    
    \item[2)] We perform an empirical study evaluating the effectiveness of several APR techniques, including recent state-of-the-art LLM-based approaches, on regression bugs.
    
    \item[3)] We investigate the impact of incorporating bug-inducing change information on the effectiveness of prompt-based repair techniques. Our study includes a qualitative analysis that reveals how this regression-specific context supports models in fixing regression errors.

    \item[4)] \firstrevision{We conduct an ablation study to investigate the individual contribution of each contextual element within the bug-inducing change information.}
\end{itemize}

\vspace{-0.3cm}
\textbf{Replication Packages:}
To support further research, we provide all artifacts associated with our dataset in a reproducible and easily accessible form. Moreover, the \mbox{\textsc{RegressionBug4APR}} benchmark is designed for extensibility, allowing the database to grow over time with minimal human effort. All artifacts related to this work are publicly available at the following links~\cite{regressionbug4apr-framework-java,regressionbug4apr-benchmark-java,regressionbug4apr-framework-python,regressionbug4apr-validation,regressionbug4apr-mining-python,regressionbug4apr-homepage}. In addition, the implementation of the prompt-based APR techniques for regression bugs used in our experiments is also publicly available~\cite{regressionbug4apr-repair}.

The remainder of this paper is organized as follows. Section~\ref{sec:background} provides background information and discusses related work on regression bugs and automated program repair.
Section~\ref{sec:studysetup} describes the study setup, including the research and experimental designs.
Section~\ref{sec:rq-1}, \ref{sec:rq-2}, \ref{sec:rq-3} \firstrevision{and \ref{sec:rq-4}} report the experimental results, followed by
Section~\ref{sec:discussion}, which explores the implications of our findings and the threats to validity.
Finally, Section~\ref{sec:conclusion} summarizes our conclusions and outlines directions for future work.

\vspace{-0.35cm}
\section{Background and Related Work}
\label{sec:background}

\firstrevision{In this section, we first introduce the concept of regression bugs, including their characteristics, categorization, and concrete examples for each type, along with existing benchmarks and tools for mining regression bugs.
We then discuss related work on program repair benchmarks and automated program repair (APR) techniques.
Finally, we provide an overview of regression bug repair, along with an illustrative example and formal problem formulation.}

\subsection{\firstrevision{Regression Bug}}

Regression bug is a common class of bugs that occur when a feature that previously worked correctly stops functioning after certain changes, such as bug fixes, new feature additions, code refactoring, or other software modifications. A regression test is designed to ensure that existing functionality or intended behavior remains intact during software evolution. When such a test fails, indicating that a recent change has unintentionally reverted the software to a “less developed state”, the associated bug is referred to as a \emph{regression bug}~\cite{nir2008locating}.

\vspace{-0.15cm}
\subsubsection{\firstrevision{Categorization of Regression Bug}}
\begin{figure}[!htbp]
\centerline{\includegraphics[width=0.55\columnwidth]{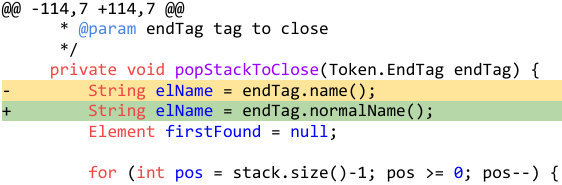}}
\vspace{-0.2cm}
\caption{Bug-inducing changes in \textit{RegressionBug-40} from our \textsc{RegressionBug4APR} benchmark.}
\label{fig:bic-rb40}
\end{figure}

To better understand the different causes of regression bugs, Tan et al.~\cite{tan2015relifix} categorizes regression bugs into three types based on how the regression was introduced: 

\textit{(i) Local.} A code modification directly breaks existing functionality within the modified program element. Intuitively, a local regression can often be fixed by reverting to the previous implementation. However, reverting may unintentionally undo newly intended updates associated with the modification. 
\firstrevision{\textit{RegressionBug-40}~\footnote{Available at: \url{https://brojackvn.github.io/RegressionBug4APR-Homepage/\#!/bug/RegressionBug-40/RegressionBug-40}} from \textit{jsoup}~\footnote{Available at: \href{https://github.com/jhy/jsoup}{\textit{https://github.com/jhy/jsoup}}} project, a Java HTML/XML parsing library, illustrates this type. As shown in Figure~\ref{fig:bic-rb40}, the bug-inducing commit modified one line in \texttt{popStackToClose()}, which matches a closing XML tag to its open element on the parse stack:}

\firstrevision{The origin of this bug stems from how tag names are handled during lookups. Matching a closing tag requires the name looked up on the parse stack to be normalized in the same way as the name stored when the element was pushed. Open elements are pushed under \texttt{settings.normalizeTag()}, which preserves tag-name case whenever the
parser is configured to do so, as \texttt{XmlTreeBuilder} is by default since XML is case-sensitive. The bug-inducing commit replaced the lookup on the closing side with \texttt{endTag.normalName()}, which lower-cases the name regardless of the parser settings, so the two sides no longer agree. A closing tag \texttt{</FOO>} now resolves to \texttt{foo}, the comparison against the stack element \texttt{FOO} fails, \texttt{firstFound} remains \texttt{null}, and the method returns without closing the element, corrupting the parse tree.}

\firstrevision{Resolving this issue requires careful consideration rather than a simple rollback. The bug-fixing commit restored correctness by replacing \texttt{endTag.normalName()} with \texttt{settings.normalizeTag(endTag.tagName)}. Notably, this is not a revert to the pre-commit \texttt{endTag.name()}, which returns the raw source name and would reintroduce a mismatch whenever the parser is configured to lower-case tag names. Instead, routing the lookup through \texttt{settings} instead makes the two sides agree under any configuration. This is the defining characteristic of a local regression: the failure is entirely confined to the modified method, with no other program element involved and no pre-existing latent fault triggered.}

\begin{figure}[!htbp]
\centerline{\includegraphics[width=0.56\columnwidth]{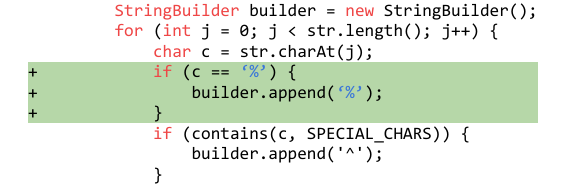}}
\vspace{-0.2cm}
\caption{Part of bug-inducing changes in \textit{RegressionBug-56} from our \textsc{RegressionBug4APR} benchmark.}
\label{fig:bic-rb56}
\end{figure}

\textit{(ii) Remote.} A code modification introduces a bug in an unchanged program element, leading to failures elsewhere in the codebase. 
\firstrevision{\textit{RegressionBug-56}~\footnote{Available at: \url{https://brojackvn.github.io/RegressionBug4APR-Homepage/\#!/bug/RegressionBug-56/RegressionBug-56}} from \textit{overthere}~\footnote{Available at: \href{https://github.com/xebialabs/overthere}{\textit{https://github.com/xebialabs/overthere}}} project, a Java library for remote file manipulation and command execution, illustrates this type.
As shown in Figure~\ref{fig:bic-rb56}, the bug-inducing commit modified the sanitize loop in \texttt{escapeSpecialCharacters()} method to escape the \texttt{\%} character by doubling it, a standard technique to prevent Windows shell variable expansion.}

\begin{figure}[!htbp]
\centerline{\includegraphics[width=0.62\columnwidth]{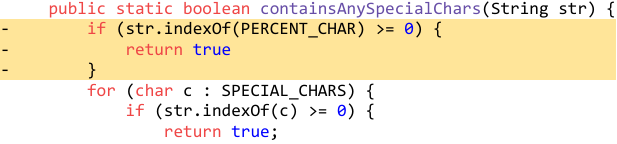}}
\vspace{-0.2cm}
\caption{Bug-fixing changes in \textit{RegressionBug-56} from our \textsc{RegressionBug4APR} benchmark.}
\label{fig:bfc-rb56}
\end{figure}

\firstrevision{This change is confined to the \texttt{escapeSpecialCharacters()} method, but it broke the entirely separate and unchanged method, \texttt{containsAnySpecialChars()}, which contains the check \texttt{if (str.indexOf(PERCENT\_CHAR) >= 0) \{return true;\}}.
The two methods divide responsibility for special characters: the escaper handles the characters it knows about, while \texttt{containsAnySpecialChars()} tells callers which remaining characters they must handle themselves, typically by quoting. Before the commit, \texttt{\%} belonged to the caller's side of that division, so flagging it was correct. The commit moved \texttt{\%} to the escaper's side but left the check in place, so both now claim it: callers apply a second round of escaping on top of the one the escaper already performed, producing double-escaped output. As shown in Figure~\ref{fig:bfc-rb56}, the bug-fixing commit restored correctness by removing the now-redundant \texttt{\%} check from \texttt{containsAnySpecialChars()}.}

\vspace{-0.2cm}
\firstrevision{This is a remote regression because the failure occurs in 
\texttt{containsAnySpecialChars()}, a program element that was not modified by the 
bug-inducing commit.  Unlike an unmask regression, there is no pre-existing latent bug: the method was correct before the commit but was not updated to account for the new handling of \texttt{\%} introduced by the sanitize modification, causing incorrect behavior elsewhere in the codebase. This illustrates a scenario where a developer modifies one part of the system without anticipating the assumptions that other unchanged parts depend upon.}

\begin{figure}[!htbp]
\centerline{\includegraphics[width=0.59\columnwidth]{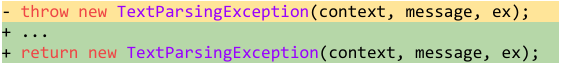}}
\vspace{-0.2cm}
\caption{Bug-inducing changes in \textit{RegressionBug-132} from our \textsc{RegressionBug4APR} benchmark: refactoring of \texttt{handleException()}.}
\label{fig:bic-rb132}
\end{figure}

\textit{(iii) Unmask.} A code modification exposes an existing bug that had previously not affected the behavior of some tests. For an unmask regression, where the earlier code effectively concealed an underlying bug, fixing the issue may involve restoring conditions that mask the problematic changes. In such cases, repair efforts focus on finding a scenario where the changes no longer impact test behavior, without simply reverting to the previous code, as doing so would re-hide the bug and fail to address the underlying problem. 
\firstrevision{\textit{RegressionBug-132}~\footnote{Available at: \url{https://brojackvn.github.io/RegressionBug4APR-Homepage/\#!/bug/RegressionBug-132/RegressionBug-132}} from \textit{univocity-parsers}~\footnote{Available at: \href{https://github.com/uniVocity/univocity-parsers}{\textit{https://github.com/uniVocity/univocity-parsers}}} project, a Java CSV/TSV parsing library, illustrates this type. As shown in
Figure~\ref{fig:bic-rb132}, the bug-inducing commit refactored \texttt{handleException()} in \texttt{AbstractParser} to enrich error diagnostics, changing it from a void method that throws to a method that returns the constructed exception, and updated its call site accordingly. Both the method and its call site were changed consistently, and the refactoring is correct and complete for its stated purpose.}

\begin{figure}[!htbp]
\centerline{\includegraphics[width=0.57\columnwidth]{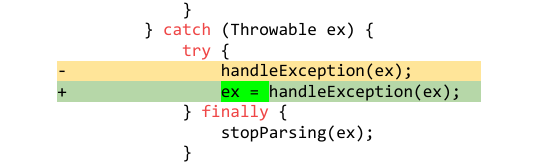}}
\vspace{-0.2cm}
\caption{Bug-inducing changes in \textit{RegressionBug-132} from our \textsc{RegressionBug4APR} benchmark: update to the \texttt{handleException()} call site.}
\label{fig:bic-rb132-callsite}
\end{figure}

\begin{figure}[!htbp]
\centerline{\includegraphics[width=0.52\columnwidth]{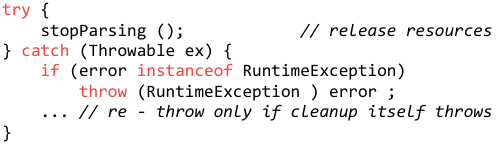}}
\vspace{-0.2cm}
\caption{Pre-existing latent fault in \texttt{stopParsing(Throwable error)} in \textit{RegressionBug-132} from our \textsc{RegressionBug4APR} benchmark.}
\label{fig:pre-existing-rb132}
\end{figure}

\firstrevision{As shown in Figure~\ref{fig:bic-rb132-callsite}, the call site sits inside a \texttt{try}/\texttt{finally} whose \texttt{finally} block invokes \texttt{stopParsing(Throwable error)} to release parser resources. That method, untouched by  the commit, carried a latent fault. It delegates resource release to the no-argument  \texttt{stopParsing()} and re-throws its \texttt{error} argument only from inside the  \texttt{catch} block, so the re-throw executes only when cleanup itself fails  (Figure~\ref{fig:pre-existing-rb132}).}

\begin{figure}[!htbp]
\centerline{\includegraphics[width=0.83\columnwidth]{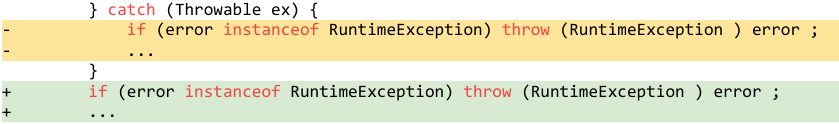}}
\vspace{-0.2cm}
\caption{Bug-fixing change in \textit{RegressionBug-132} from our \textsc{RegressionBug4APR} benchmark.}
\vspace{-0.4cm}
\label{fig:bfc-rb132}
\end{figure}

\firstrevision{Before the bug-inducing commit, this fault was harmless. Because \texttt{handleException()} threw, an exception was already propagating by the time the \texttt{finally} block ran, so whatever \texttt{stopParsing()} did or did not re-throw was irrelevant: the \texttt{TextParsingException} continued on to the caller under its own momentum. The commit removed exactly that guarantee. With \texttt{handleException()} returning rather than throwing, no exception is in flight when the \texttt{finally} block runs, and the \texttt{TextParsingException} reaches the caller only if \texttt{stopParsing(ex)} re-throws it. When cleanup succeeds, the \texttt{catch} block is never entered, the exception is silently discarded, and the caller receives no indication that parsing failed. As shown in Figure~\ref{fig:bfc-rb132}, the bug-fixing commit corrected \texttt{stopParsing(Throwable error)} by moving the re-throw outside the \texttt{catch} block so that it always executes after cleanup.}

\firstrevision{This is an unmask regression because the failure originates in  \texttt{stopParsing(Throwable error)}, a method not modified by the bug-inducing commit  whose fault pre-dated it entirely. Unlike a local regression, the bug-inducing change is itself incorrect: refactoring \texttt{handleException()} to return rather than throw is a valid and complete refactoring. Unlike a remote regression, the fault in  \texttt{stopParsing(Throwable error)} was not introduced in response to the new handling; it was present from the start, shielded from consequences by the exception propagation path that \texttt{handleException()}'s throw provided. The bug-inducing commit is innocent: it inadvertently removed that shield, exposing the dormant fault.}

\vspace{-0.1cm}
\subsubsection{Regression Bug Benchmark}
To facilitate regression error research, two notable datasets have been developed for studying regression bugs:
CoREBench~\cite{bohme2014corebench}, which was used by Relifix, and the more recent CIBugs~\cite{kabadi2023future}.
CoREBench, introduced in 2014, comprises a collection of 70 regression errors  extracted from four small to medium open-source C/C++ projects (less than 35KLOC): Make, Grep, Findutils, and Coreutils, which are publicly accessible through GNU homepage~\footnote{Available at: \href{http://savannah.gnu.org}{\textit{http://savannah.gnu.org}} and \href{http://debbugs.gnu.org}{\textit{http://debbugs.gnu.org}}}. The dataset has not been updated since its release. Moreover, due to quality concerns, such as dissatisfaction with how regression errors were defined and challenges with reproducibility, Relifix ultimately utilized only a subset of the dataset in its evaluation~\cite{tan2015relifix}.
The more recent benchmark, CIBugs, utilizes data from continuous integration (CI) systems. CIBug's strategy leverages the TravisTorrent database~\cite{beller2017travistorrent}, which contains metadata from 2,640,825 Travis builds synthesized from Travis CI's API and build logs, last updated on February 8, 2017. The regression errors captured in CIBugs are identified through continuous integration testing, where existing tests (i.e., regression tests) that previously passed begin to fail after code changes. However, CIBugs lacks information on bug-inducing commits necessary for forming regression bugs, and identifying these commits remains a challenging task~\cite{an2023fonte}. Furthermore, CIBugs requires strenuous human engineering efforts to manually curate the bugs, rendering it hard to scale.

\vspace{-0.1cm}
\subsubsection{RegMiner - A Regression Bug Mining Tool.}
While existing datasets rely heavily on manual curation, RegMiner offers an automated approach for identifying regression bugs from software evolution history~\cite{song2022regminer}. Specifically, RegMiner constructs a regression bug by searching in code repositories for a bug-fixing commit (\textit{bfc}), a bug-inducing commit (\textit{bic}), and a test case (\textit{t}) such that \textit{t} passes on \textit{bfc} and a version before \textit{bic}, but fails on \textit{bic} and a version before \textit{bfc}.

To begin, RegMiner collects bug-fixing commits that introduce new test cases from version histories. 
A commit is confirmed as a bug-fixing commit if the added test case passes on the current version and fails on the version before it, which serves as a prerequisite to search the regression bug.  
RegMiner then applies a measurement heuristic to prioritize bug-fixing commits that are more likely to be regressions.
Next, for each candidate \textit{bfc}, RegMiner searches for a corresponding bug-inducing commit \textit{bic}, where the test case fails on \textit{bic} but passes on the version before \textit{bic}. If such a commit is found, a regression instance is recorded as a tuple (\textit{bfc}, \textit{bic}, \textit{t}).

This process encounters several technical challenges:
\textit{(i) Futile search on non-regression bug-fixing commits:} Searching through commit history is time-consuming, so RegMiner introduces a heuristic to quantify the likelihood of a commit being a regression-fixing commit;
\textit{(ii) Test dependency migration:} Verifying whether a test fails or passes on prior versions is complicated as code and dependency changes over time, especially when regressions were introduced years earlier and the project has since undergone significant refactoring or library updates; and
\textit{(iii) High overhead for validation:} Confirming a regression requires checking out, compiling, and running tests across potentially thousands of commits, which incurs significant computational cost.

To overcome these challenges, RegMiner introduces three algorithms:
\textit{(i)} a heuristic for estimating potential bug-fixing commits,
\textit{(ii)} a test migration algorithm, and 
\textit{(iii)} a validation effort minimization algorithm.
These strategies enable scalable mining. The \textit{validation effort minimization} algorithm is a heuristic designed to reduce validation costs and to handle cases that suffer from incompatibility due to test migration. However, this approach can lead to incorrect identification of regression bugs. To address this problem, we develop a validation tool to ensure the reproducibility, correctness, and the overall quality of each mined regression instance.

\vspace{-0.1cm}
\subsection{Program Repair}
The following is an overview of existing program repair benchmarks and of automated program repair methodologies.

\vspace{-0.1cm}
\subsubsection{Program Repair Benchmarks}
High-quality datasets are critical for evaluating the effectiveness of APR techniques and facilitating rigorous comparisons across different approaches. Over time, researchers have introduced various benchmarks to meet these needs.

Defects4J~\cite{just2014defects4j}, one of the most widely used APR benchmarks, contains 835 real-world bugs from Java projects. 
Bugs.jar~\cite{saha2018bugs}, which includes 1,158 reproducible bugs collected from Apache projects, follows a similar construction process but expands the benchmark's size and covers different categories of Java applications.
Bears~\cite{madeiral2019bears} and BugSwarm~\cite{tomassi2019bugswarm} use automated processes to construct their datasets but differ in how bugs are reproduced.
Bears includes 251 reproducible bugs, while BugSwarm collects unprocessed bugs from failed CI builds, yielding 3,091 pairs of failing and passing builds.
However, a critical analysis~\cite{durieux2019critical} shows that only 50 Java and 62 Python bugs in BugSwarm are actually suitable for evaluating automated fault localization or program repair techniques.
A recent study~\cite{liu2021critical} indicated that approximately 90\% of the bugs in Defects4J, Bugs.jar, and Bears are associated with test cases that were added or modified after the bugs were reported. These datasets are not constructed with regression-specific characteristics in mind. The bugs are not revealed through regression testing, making it unclear whether they truly represent regressions. 
Therefore, these datasets lack the temporal and behavioral properties that define regression bugs.

In addition to Java benchmarks, several datasets have been developed for other programming languages.
For C/C++, ManyBugs~\cite{le2015manybugs} includes 185 bugs from nine large, popular open-source programs. 
IntroClass~\cite{le2015manybugs} contains 998 student-written programs with bugs, collected from small programming assignments in an introductory course. 
Codeflaws~\cite{tan2017codeflaws} includes 3,902 bugs extracted from Codeforces programming contests. 
Other notable benchmarks include BugsInPy~\cite{widyasari2020bugsinpy} for Python and BugsJS~\cite{gyimesi2019bugsjs} for JavaScript.

In contrast to these general software bug benchmarks, our proposed benchmark, \textsc{RegressionBug4APR}, is specifically focused on regression bugs. It includes regression-revealing test cases, along with the corresponding bug-inducing and bug-fixing commits. Unlike non-regression benchmarks, where each bug is typically identified by one or more failing test cases, \textsc{RegressionBug4APR} includes both a historical passing version and a subsequent failing version. This specificity distinguishes our benchmark and provides a valuable resource for understanding the characteristics of real-world regression bugs.

\vspace{-0.1cm}
\subsubsection{Automated Program Repair}
Traditional APR techniques can be broadly categorized into three main groups: \textit{search-based}, \textit{template-based}, and \textit{constraint-based} approaches.
\textit{Search-based} and \textit{Template-based approaches}
search for program variants that retain the required functionality while addressing the identified bug. Notable examples leveraging genetic programming include GenProg~\cite{le2011genprog}, HDRepair~\cite{le2016hisory}, Arja~\cite{yuan2018arja}, and jMutRepair~\cite{martinez2016astor}.
Another common strategy leverages predefined fix templates, either manually defined or mined from code repositories. Notable examples are TBar~\cite{liu2019tbar} and Cardumen~\cite{martinez2018ultra}. Relifix~\cite{tan2015relifix}, applied to C/C++ regression errors, reduces the search space by applying code transformations based on syntactic change patterns observed across multiple software versions.
\textit{Constraint-based approaches}
represent the space of patch candidates using symbolic constraints. By encoding the desired program behavior as a set of logical constraints, these approaches synthesize code that satisfies both the specification and the test suite. Notable techniques are SemFix~\cite{nguyen2013semfix}, Nopol~\cite{xuan2016nopol}, Angelix~\cite{mechtaev2016angelix}, and S3~\cite{le2017s3}. In our study, we do not evaluate semantic-based APR approaches as they often require the inclusion of manually-written models for system calls whose source codes are not available.

Shifting to \textit{deep learning-based} techniques, these methods are typically trained from scratch on large bug-fix corpora to automatically learn repair patterns from human-written patches. Once trained, the models can be applied to generate patches for buggy programs. Recent state-of-the-art techniques are CURE~\cite{jiang2021cure}, RewardRepair~\cite{ye2022neural}, Recoder~\cite{zhu2021syntax}, and KNOD~\cite{jiang2023knod}, NeverMore~\cite{alhefdhi2025}, among others.
More recently, the rapid advancement of large language models (LLMs) has given rise to a new line of \textit{LLM-based APR techniques}, which have demonstrated substantial improvements over prior deep learning-based methods~\cite{jiang2023impact}. Due to their superior capabilities, our study places significant emphasis on LLM-based APR. These LLM-based methods typically utilize large pretrained models for code generation and can be grouped into two main categories: \textit{fine-tuning-based} and \textit{prompt-based} techniques.
\textit{Fine-tuning-based approaches} focus on refining the weights of code-specific LLMs, such as \secondrevision{CodeLLaMA}~\cite{roziere2023code}, Incoder~\cite{fried2022incoder} and CodeGen~\cite{nijkamp2022codegen}, to improve performance on bug-fixing tasks.
Notably, Jiang et al.~\cite{jiang2023impact} demonstrated that full-parameter fine-tuning significantly enhances APR performance, achieving remarkable improvements over the base models.
\secondrevision{Silva et al.}~\cite{silva2023repairllama} investigated the effectiveness of parameter-efficient fine-tuning using QLoRA~\cite{dettmers2023qlora}. Their work evaluated various input-output formatting strategies and introduced \secondrevision{RepairLLaMA}, a fine-tuned version of CodeLLaMA tailored specifically for program repair.
\textit{Prompt-based approaches} aim to directly leverage LLMs without fine-tuning, primarily via prompting. \secondrevision{This direction was pioneered by AlphaRepair~\cite{xia2022less}, which formulated repair as a zero-shot cloze-style infilling task, and has since been extended with richer prompting strategies to provide repair information to the model}. Xia et al.~\cite{xia2024automated} proposed ChatRepair, a conversation-driven APR approach that interacts with ChatGPT using contextual feedback such as test names, relevant test code, and error messages to iteratively guide the repair process. Yin et al.~\cite{yin2024thinkrepair} applied chain-of-thought prompting and few-shot learning techniques to further enhance the performance of LLMs on APR tasks.
\secondrevision{Beyond single-shot and conversational prompting, a growing line of works casts repair as an \textit{agentic} process in which an LLM autonomously invokes tools to localize faults, inspect the codebase, and validate candidate patches, as in RepairAgent~\cite{bouzenia2025repairagent} and SpecRover~\cite{ruan2024specrover}.
Others supply structural information beyond the isolated buggy function, whether through program analysis~\cite{li2024hybrid}, repository-level context~\cite{yang2025}, fix templates~\cite{huang2025template}, or high-level bug specifications~\cite{ye2025adverintent}.}
Recent surveys have systematically reviewed LLM-based APR techniques, providing comprehensive taxonomies of existing approaches, discussing their applications, and identifying current challenges and future research directions~\cite{yang2025survey,zhang2026systematic}.

\subsection{\firstrevision{Regression Bug Repair}}

\subsubsection{\firstrevision{Overview of Regression Bug Repair}}
\firstrevision{Regression bug repair is a specialized form of test-based repair that targets regression bugs specifically. Similar to general-purpose test-based repair, it takes a buggy program and a test suite as input and aims to generate a patch that satisfies all tests. What distinguishes regression repair from general-purpose repair is that it is permitted to leverage additional regression-specific context, most notably, bug-inducing change information that general-purpose approaches do not consider.}

\firstrevision{This additional context is valuable because regression bugs do not introduce entirely new faults; rather, they violate previously correct functionality by modifying code that was once working correctly. Unlike general bugs, each regression bug is linked to a bug-inducing commit that records the code changes introduced at the point where the regression was first observed. Identifying this commit provides context about why the failure occurred, enabling repair techniques to distinguish unintended behavioral deviations from intended modifications and to reason about how the regression was introduced. Prior work has shown that many bug-fixing changes correspond to reverting or modifying the changes introduced in the bug-inducing commit~\cite{wen2019exploring, tan2015relifix}, suggesting that such information provides valuable guidance for automated repair.}
\firstrevision{In practice, developers routinely use tools such as \texttt{git bisect}~\footnote{Available at: \url{https://git-scm.com/docs/git-bisect}} to identify bug-inducing commits. For example, in \textit{PyRegression-2} in our benchmark, \textit{Issue~\#62520}~\footnote{Available at: \url{https://github.com/pandas-dev/pandas/issues/62520}} of the pandas project, a regression is reported where \texttt{Series.pow} works correctly in pandas~2.x but fails in pandas~3.x. A core contributor applies \texttt{git bisect} and identifies commit \texttt{e4ca40511c} (\textit{``API: mode.nan\_is\_na to consistently distinguish NaN-vs-NA''}) as the bug-inducing commit. Based on this information, contributors narrow the investigation to the specific changes in that commit and involve the relevant code owner. The code owner identifies the faulty line in \texttt{ArrowEA.\_arith\_method} and implements a fix in \textit{PR~\#62572}~\footnote{Available at: \url{https://github.com/pandas-dev/pandas/pull/62572}}. This illustrates a clear workflow: (1) identifying the bug-inducing commit, (2) narrowing down the faulty changes, and (3) generating a fix guided by that context.}

\firstrevision{Compared to general-purpose bug repair, regression repair introduces both opportunities and challenges. On the opportunity side, the availability of a bug-inducing commit provides richer contextual information to guide the repair process. On the challenge side, regression bug repair introduces difficulties that do not arise in general settings. First, repair techniques must reason across multiple program versions to understand how recent modifications introduced the regression. Second, not all changes in a regression-inducing commit are incorrect; some are intended improvements. Repair approaches must carefully identify which changes are responsible for the regression while preserving valid modifications. For example, blindly reverting all changes in a bug-inducing commit may undo intended improvements alongside the faulty ones, leading to semantically incorrect patches.}

\subsubsection{An Illustrative Example}
We present an example of a regression bug, illustrating how code changes unexpectedly create errors while introducing new features or bug fixes. This example, corresponding to \textit{RegressionBug-42} in our benchmark, comes from the \textit{jsoup} project, a Java library designed to simplify working with real-world HTML and XML. 

\begin{figure}[!htbp]
\centerline{\includegraphics[width=0.93\columnwidth]{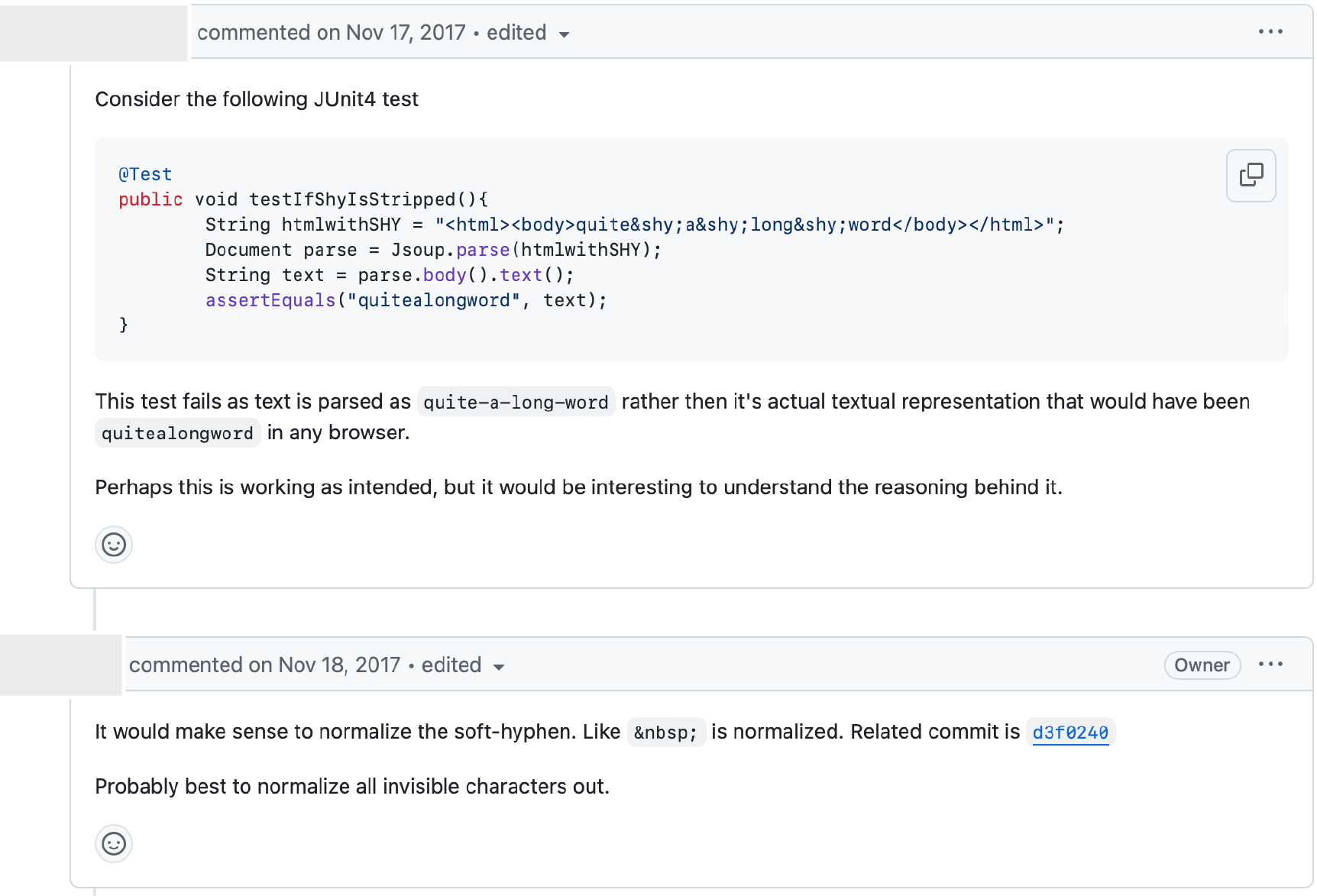}}
\vspace{-0.1cm}
\caption{An issue reported requiring developers to resolve; however, after fixing it, another bug is inadvertently introduced. See the original issue at: \href{https://github.com/jhy/jsoup/issues/978}{\textit{https://github.com/jhy/jsoup/issues/978}}}
\label{issue-working}
\end{figure}

\begin{figure}[!htbp]
\centerline{\includegraphics[width=0.67\columnwidth]{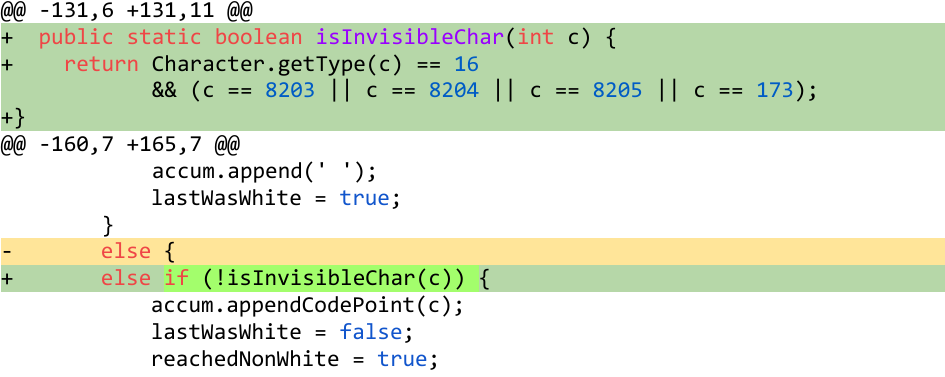}}
\vspace{-0.2cm}
\caption{Regression-inducing changes where the intention was to fix an issue, but an unintended regression was introduced}
\label{bic}
\end{figure}

\begin{figure}[!htbp]
\centerline{\includegraphics[width=0.93\columnwidth]{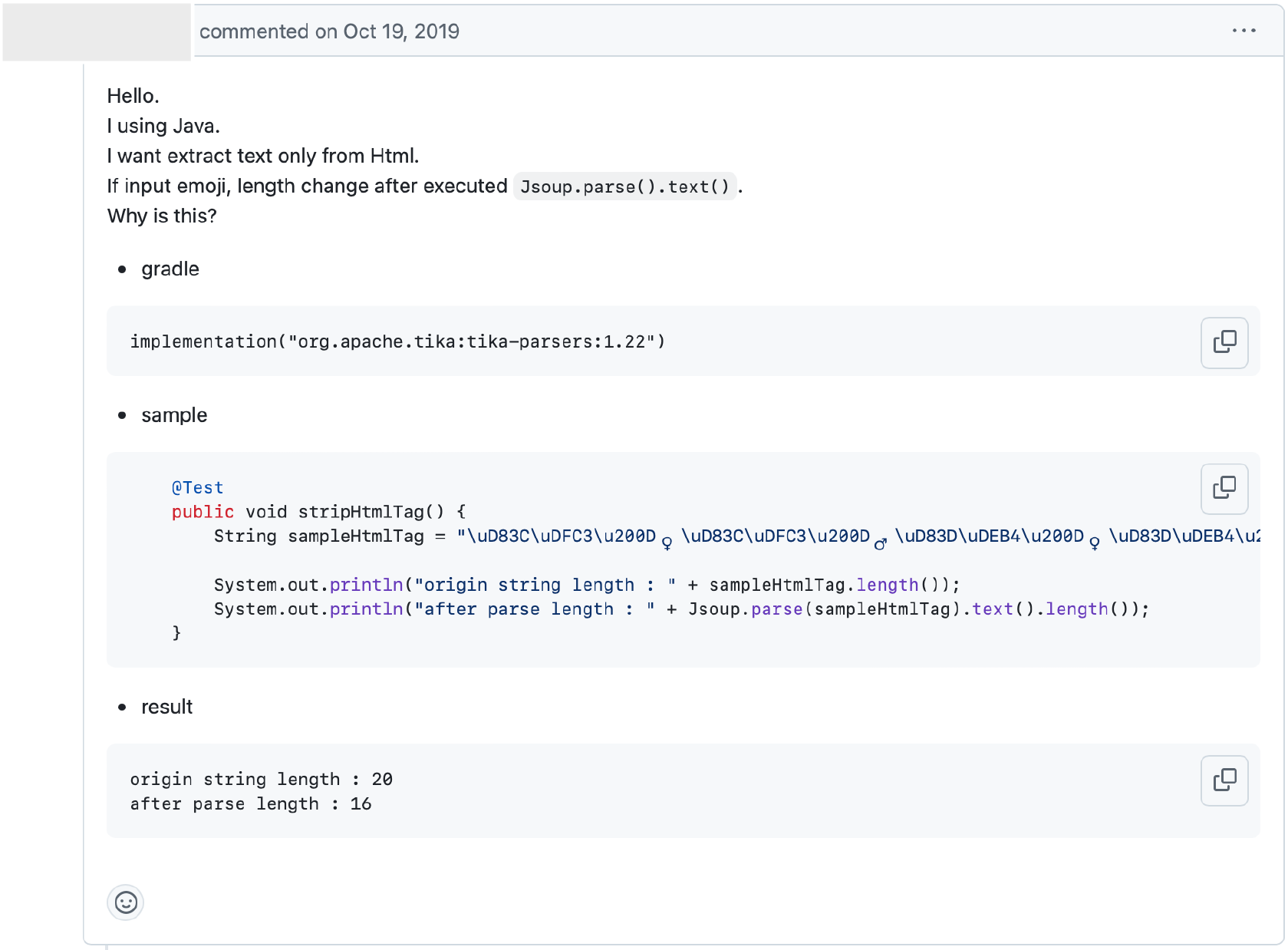}}
\vspace{-0.1cm}
\caption{An issue reported requiring developers to resolve. This issue is caused by the fix of a problem introduced two years prior (see Figure~\ref{issue-working}). See the original issue at: \href{https://github.com/jhy/jsoup/issues/1269}{\textit{https://github.com/jhy/jsoup/issues/1269}}.}
\label{issue-at-buggy}
\end{figure}

\textit{Bug-inducing changes} – This regression bug occurred when developers attempted to fix an issue raised by the user as shown in Figure~\ref{issue-working}. In this issue, users pointed out that the current version requires an enhancement of text rendering and formatting features as the \textit{\&shy} character was displaying as a hyphen (-). To resolve this, developers constructed the \texttt{isInvisibleChar} method to normalize text by omitting invisible characters, including \textit{zero-width space (ZWSP)}, \textit{zero-width non-joiner (ZWNJ)}, \textit{zero-width joiner (ZWJ)}, and \textit{soft hyphen (SHY)}, as illustrated in Figure~\ref{bic}. Unfortunately, removing the \textit{ZWNJ} and \textit{ZWJ} caused a change in the emoji's length after applying this function; an unintended behavior that was only identified in other issues after several years (see Figure~\ref{issue-at-buggy}).

\begin{figure}[!htbp]
\centerline{\includegraphics[width=0.68\columnwidth]{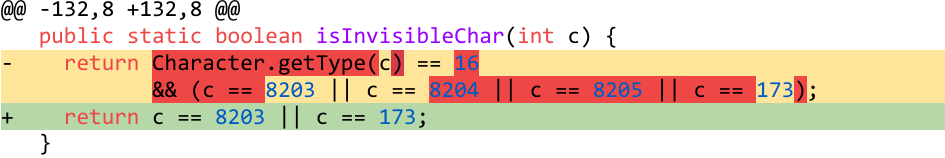}}
\vspace{-0.1cm}
\caption{Bug-fixing changes made by developers to resolve the regression bug.}
\label{bfc}
\end{figure}

\textit{Bug-fixing changes} - As illustrated in Figure~\ref{bfc}, to fix the regression bug, developers modified the \texttt{isInvisibleChar} method to selectively only drop the characters requiring removal, specifically \textit{SHY} and \textit{ZWSP}, while preserving functional invisible characters such as \textit{ZWJ} and \textit{ZWNJ}. Although this fix is relatively straightforward, it necessitates a deeper understanding of the underlying bugs and the initial purpose of bug-inducing changes. The root cause of the issue lies in the overly broad handling of invisible characters in the \texttt{isInvisibleChar} method. The original intent of the text normalizer was to remove unnecessary formatting characters (e.g., \textit{ZWSP} and \textit{SHY}) to facilitate cleaner text processing. However, the approach unexpectedly removed \textit{ZWJ} and \textit{ZWNJ} characters, which are essential for joining characters into meaningful sequences. In particular, in contexts such as emojis or in scripts such as Arabic. The bug fix, therefore, needs to ensure that \textit{ZWJ} and \textit{ZWNJ} are retained, while still allowing for the removal of \textit{ZWSP} and \textit{SHY}.

We can see that the generation of this fix required more than simply adjusting the buggy version to pass the test cases. For instance, reverting the code to the working commit would indeed pass the test case, but it would also undo the feature improvements made by the developers. Thus, careful consideration was needed to ensure the changes did not negatively impact other functions. This was accomplished by examining the bug-inducing changes and comparing the working version before the modifications with the failures introduced afterwards. 

\subsubsection{\firstrevision{Problem Formulation}}

\firstrevision{Formally, each regression bug instance in our benchmark consists of the following components: \textit{(i)} a buggy program version $P_b$; \textit{(ii)} a test suite $T$, including at least one failing regression-witnessing test case; and \textit{(iii)} a bug-inducing commit, from which the bug-inducing changes are extracted and represented as a diff file. The goal of the repair task is to generate a patch $p$ such that the patched program $P_b + p$ passes all tests in $T$. The bug-inducing commit is the key distinguishing element of our benchmark compared to general-purpose APR benchmarks, as it provides historical context about how the regression was introduced, which is not available in traditional bug repair settings.}

\firstrevision{In our setting, the bug-inducing commit is assumed to be already identified, for example via tools such as \textit{git bisect} or \textit{SZZ}~\cite{sliwerski2005changes}. This assumption allows us to isolate and evaluate the impact of bug-inducing change information on repair effectiveness, without conflating it with the orthogonal challenge of commit identification. A similar assumption is commonly made in APR research with respect to fault localization. Specifically, by assuming perfect fault localization, existing techniques isolate repair capability from localization capability, enabling a more focused evaluation of patch generation~\cite{silva2023repairllama,xia2024automated,jiang2023impact,liu2019tbar}. In the same spirit, our assumption of a known bug-inducing commit allows us to study how regression-specific context influences the repair process in isolation.}

\section{Study Setup}
\label{sec:studysetup}
In this section, we describe the setup of our empirical study. We begin by outlining the research questions and presenting the overall workflow of the study. We then provide details of the benchmark construction process. Following that, we describe the selection criteria and experimental settings of the automated program repair (APR) techniques under our evaluation. Finally, we present the evaluation metrics used to assess the effectiveness of the repair methods.

\subsection{Research Design}
Our study is guided by the following research questions:

\begin{itemize}
    \item[] \textbf{RQ1.} \textit{What is the degree of quality and diversity of the \mbox{\textsc{RegressionBug4APR}} benchmark for APR studies?}
    This research question aims to examine the overall quality and characteristics of the \textsc{RegressionBug4APR} benchmark, with the goal of assessing its suitability for supporting research in APR.
    
    \item[] \textbf{RQ2.} \textit{How effective are existing APR techniques at repairing regression bugs?} 
    This research question aims to investigate the effectiveness of traditional and advanced APR techniques in repairing regression bugs.
    
    \item[] \textbf{RQ3.} \textit{What is the impact of providing additional context, particularly bug-inducing change information, on APR techniques?} 
    This research question aims to explore whether providing additional context, specifically bug-inducing change information, can enhance the repair performance.

    \item[] \firstrevision{\textbf{RQ4.} \textit{What is the contribution of each contextual element in the bug-inducing change information to repair effectiveness?}}
    \firstrevision{This research question investigates the individual contribution of different components of the bug-inducing change information to repair effectiveness through an ablation study.}
\end{itemize}

Our study begins with the construction of the \firstrevision{\textsc{RegressionBug4APR}} benchmark, which includes \firstrevision{200} regression bugs collected from widely used real-world Java \firstrevision{and Python} GitHub repositories. Since data availability has long posed a challenge for APR research~\cite{just2014defects4j, madeiral2019bears, tomassi2019bugswarm, kabadi2023future}, existing regression bug datasets are outdated and difficult to reproduce. In addressing \textbf{RQ1 (Section~\ref{sec:rq-1})}, we analyze our benchmark construction process and conduct an in-depth analysis of human-written patches to demonstrate the quality and diversity of the dataset.
Building on this high-quality benchmark, we address \textbf{RQ2 (Section~\ref{sec:rq-2})} by empirically evaluating several APR techniques, including both traditional and advanced LLM-based approaches, on their ability to repair regression bugs. These evaluations are performed without considering the regression-inducing context, aligning with the conventional APR workflow, which typically take as input the buggy program and its associated test suite.
In addressing \textbf{RQ3 (Section~\ref{sec:rq-3})}, we focus on prompt-based APR techniques, motivated by their demonstrated effectiveness, inherent flexibility in model usage and prompt construction, and greater data efficiency compared to fine-tuning-based approaches. We investigate whether incorporating regression-specific context, specifically bug-inducing change information, into the prompt design can enhance repair effectiveness. Furthermore, we conduct a qualitative analysis of plausible patches to examine how models leverage this contextual information to repair regression bugs.
\firstrevision{Finally, in addressing \textbf{RQ4 (Section~\ref{sec:rq-4})}, we build on the findings of RQ3 and conduct an ablation study on the best-performing configuration, conversational ChatGPT-4o, to isolate and evaluate the individual contribution of each contextual element within the bug-inducing change information. This allows us to determine which components are most informative for guiding LLM-based regression repair and to better understand the sources of improvement observed in RQ3.}

\subsection{Benchmark Construction}
\label{sec:benchmark-construction}
\begin{figure}[htbp]
    \centerline{\includegraphics[width=\columnwidth]{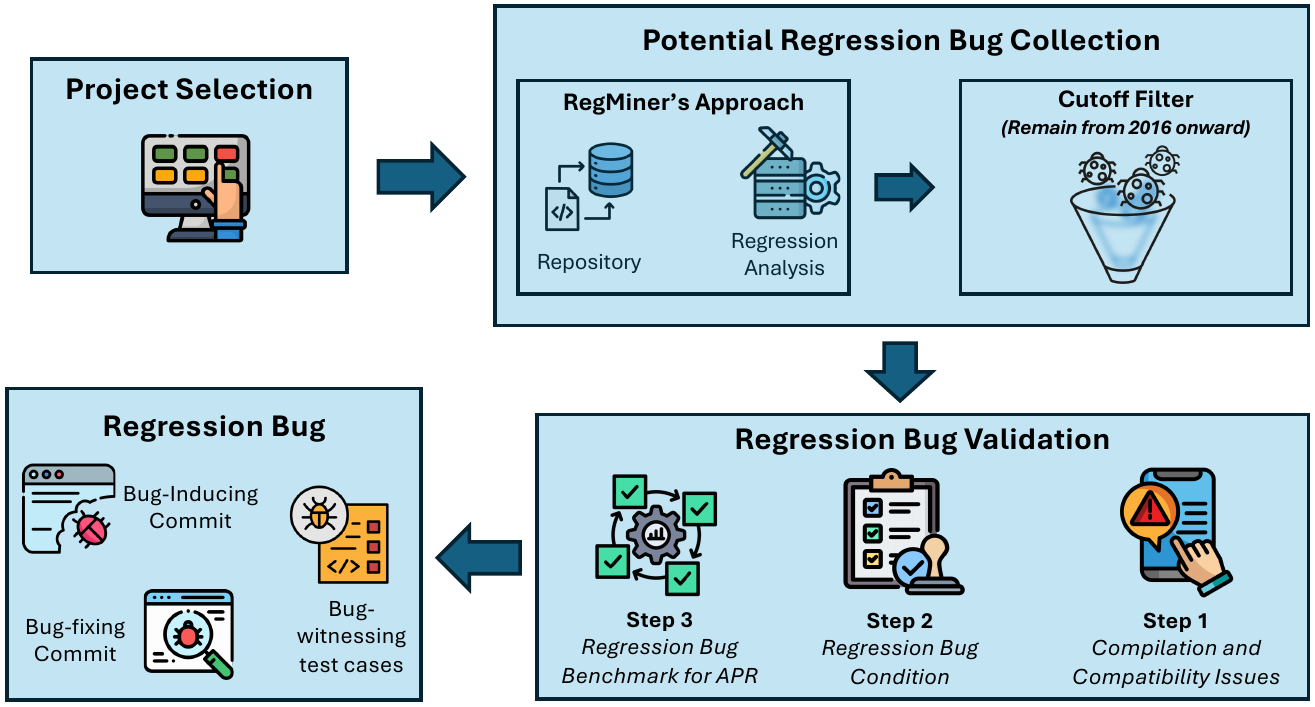}}
    \caption{Overview of the \firstrevision{\textsc{RegressionBug4APR}} benchmark construction.}
    \label{fig:overviewRegMiner4APR}
\end{figure}

In this section, we outline the process for constructing a benchmark of regression bugs.
Figure~\ref{fig:overviewRegMiner4APR} illustrates an overview of the benchmark construction.

\vspace{0.1cm}

\vspace{-0.2cm}

\subsubsection{Project Selection}
In this study, we focus on Java and Python regression bugs.
\firstrevision{Both languages are among the most widely used programming languages today, as consistently ranked in the TIOBE Index~\footnote{Available at: \url{https://www.tiobe.com/tiobe-index/}}, with Python ranked first and Java ranked fourth at the time of this study.}
While there has been a substantial body of research on regression errors in C/C++ projects~\cite{zeller1999yesterday, nir2008locating, yin2011fixes, bohme2014corebench, tan2015relifix}, most of it dates back several years, leaving a gap in up-to-date studies on Java and Python regression errors.

\firstrevision{\textbf{Java.}} We selected well-known libraries and tools hosted on GitHub, implemented in Java, that meet the following criteria:
\textit{(i)} The project has more than 500 commits, used as a proxy to ensure sufficient development activity and code evolution;
\textit{(ii)} The project uses Maven or Gradle as the build system and JUnit test cases, ensuring compatibility with the RegMiner tool and test execution environment;
\textit{(iii)} The project is not related to mobile or distributed platforms (e.g., Android), which often require specialized environments (e.g., emulators or network configurations) that complicate reproducibility; and
\textit{(iv)}  The project is not structured as a multi-module build, as multi-module builds introduce complexity in dependency resolution and test isolation, which can hinder accurate mining, compiling, and test execution during regression analysis.
Following these criteria, we sampled and selected \firstrevision{79} Java repositories~\footnote{Available at: \url{https://github.com/brojackvn/RegressionBug4APR-Framework/blob/main/GITHUB_REPO_MINING.MD}} hosted on GitHub. This sample size was chosen due to the time-intensive nature of the bug mining tool, \textsc{RegMiner}, as well as the manual effort required for subsequent steps such as patch minimization and bug analysis.

\firstrevision{\textbf{Python.} We selected two widely used open-source projects: \textit{pandas}~\footnote{Available at: \url{https://pandas.pydata.org}}, a widely adopted data analysis and manipulation library, and \textit{Django}~\footnote{Available at: \url{https://github.com/django/django}}, a high-level web framework. These projects were selected because they represent diverse application domains and, among the Python projects we evaluated, they provided the highest number of successfully reproducible commit histories. Due to environment reproduction challenges specific to Python projects, we explored only the feasible portion of each project's commit history to identify regression bugs. Specifically, for \textit{pandas}, we
examined commits from \texttt{b89f1d0d05} (March 6, 2024) to \texttt{b015a3bfb4} (October 21, 2025) on the main branch; for \textit{Django}, we examined commits from \texttt{306607d5b9} (September 16, 2021) to \texttt{8a0315fab7} (January 10, 2026) on the main branch.}

\subsubsection{Potential Regression Bug Collection}
\label{subsec:potential-bug-collection}
For Java, we leverage RegMiner~\cite{song2022regminer}, a well-known regression mining tool, to identify potential regression bugs from the selected GitHub repositories.
In total, we identified \firstrevision{1,631} potential regression bug candidates from these repositories.
To ensure that our dataset is both timely and up-to-date, we filtered this set to retain only those whose bug-fixing commits were introduced from \firstrevision{2016} onward.
We chose this cutoff because the most recent existing dataset on Java regression errors, CIBugs~\cite{kabadi2023future}, primarily includes bugs from before 2017, and we aim to provide a more up-to-date benchmark. After applying this filter, we retained \firstrevision{758} potential regression bugs for further validation.

\firstrevision{For Python, we developed a validation tool~\footnote{Available at: \url{https://github.com/brojackvn/PyRegression-Mining-Tool}} that follows the same spirit as RegMiner. For simplicity, certain optimizations for minimizing search overhead, such as the heuristic algorithm used in RegMiner, are not implemented due to the complexity of porting those components from Java to Python. Specifically, the tool identifies regression bugs by migrating test cases across commit histories to detect regressions within a reproducible execution environment. Due to the complexity of Python environment reproduction, we focus on a curated list of reproducible commits rather than exhaustively searching the full commit history. A validation step is applied at the end of the process to confirm that each identified instance satisfies the regression bug conditions described in Section~\ref{subsec:regression-bug-validation}. Through this process, we identified 50 confirmed Python regression bugs.}

\subsubsection{Regression Bug Validation}
\label{subsec:regression-bug-validation}
However, it is possible that not all extracted candidates represent true regressions since RegMiner employs heuristics in its mining process. Therefore, after collecting the set of potential regression bugs, we developed an automated validation tool to identify true regression bugs. This tool applies a series of filtering criteria to evaluate the relevance and accuracy of each case as follows:
\begin{itemize}
    \item \textit{Step 1: Compilation and Compatibility Issues (Executability).}
    This step is necessary because RegMiner uses test cases extracted from the bug-fixing commit, which need to be migrated to earlier versions using rule-based heuristics. Our validation tool takes as input the potential bug-inducing commit, bug-fixing commit, and the corresponding test case that exposes the regression. It then attempts to reconstruct the four relevant commit versions with migrated test cases. However, \firstrevision{236} of the \firstrevision{758} potential regression bugs encountered failures during this step, falling into two main categories: 
    \begin{itemize}
        \item \firstrevision{\textit{Compilation failures}, where the migrated test case cannot be compiled against earlier program versions due to structural incompatibilities introduced between commits. For example, in the \textit{w3c/epubcheck} project (bug-inducing commit \texttt{1268a8fb}, bug-fixing commit \texttt{356fac0f}, test case \texttt{com.adobe.epubcheck.ops.OPSCheckerTest} \texttt{\#testCustomElements}), the bug-fixing commit refactored \texttt{OPSCheckerTest} by introducing new method signatures, renaming symbols, and adding the \texttt{Messages} class, none of which exist in the earlier bug-inducing commit. As a result, the migrated test file references symbols and method overloads that cannot be resolved against the older codebase, causing \texttt{maven-compiler-plugin} to fail with \texttt{cannot find symbol} and \texttt{ambiguous method reference} errors.}
        
        \item \firstrevision{\textit{Dependency compatibility failures}, where required dependencies are no longer available in remote repositories (e.g., Maven Central), preventing the project from being built. In some cases, dependencies are removed within days of the bug being mined. Even if the project compiled successfully at the time of mining, subsequent reconstruction attempts may fail with errors such as: \textit{``[ERROR] Caused by: The following artifacts could not be resolved: ...``}. This highlights the inherent instability of relying on remote dependency repositories for historical version reconstruction, and directly motivates our design decision in Section~\ref{subsec:data-storage} to clone each required commit version along with all necessary dependencies, rather than relying on remote repositories at runtime.}
    \end{itemize}
    After filtering out these cases, \firstrevision{522} potential regression bugs remained.

    \item \textit{Step 2: Regression Bug Conditions (Validity).}
    This step ensures the behavioral conditions of a true regression bug are satisfied, as RegMiner may yield false positives. Specifically, our validation tool verifies that the test case passes on both the bug-fixing commit and the version prior to the bug-inducing commit, and fails on both the bug-inducing commit and the version prior to the bug-fixing commit. Bugs that do not satisfy this condition are discarded. After this step, \firstrevision{264} potential regression bugs remained.
    
    \item \textit{Step 3: Regression Bug Benchmark for APR (Utility).}
    This step ensures that the regression bugs included in the benchmark are suitable for automated program repair (APR). We integrated an additional condition into our validation tool to check that the buggy version fails only the test cases that exposes the regression, and that the bug-fixing commit passes all test cases. Although this may exclude some complex but valid regressions (e.g., those affecting multiple test cases beyond the ones exposing the regression), it helps construct a clean and regression-focused benchmark tailored for APR research.
\end{itemize}

Ultimately, through this process, we curated a dataset of \firstrevision{150 confirmed Java regression bugs, drawn from \firstrevision{79} distinct GitHub repositories. Combined with \firstrevision{50 confirmed Python regression bugs}, this yields the \textsc{RegressionBug4APR} benchmark comprising a total of 200 regression bugs}.

\subsubsection{Data Storage}
\label{subsec:data-storage}
To ensure extensibility and durable reproducibility, we store the validated regression bugs in a structured database hosted on GitHub. This centralized storage facilitates easy access and management of the regression bug dataset. In addition, we have developed an interface on top of this database to streamline usage for researchers, support reproducible studies in program repair, and enable future extensions without requiring environment-specific setup for each individual bug. The key characteristics of our storage design include:

\begin{itemize}
    \item \textit{Durable reproducibility:}
    To guard against future dependency issues (e.g., broken builds due to removed or updated dependencies), we clone each required commit version along with all necessary resources and dependencies. For Maven-based projects, we use the built-in command \texttt{mvn dependency:copy-dependencies}; for Gradle-based projects, we implemented a custom function to extract and preserve all required dependencies.
    
    \item \textit{Extensibility:}
    The database is structured to support future expansions of the benchmark. Each regression bug is represented by four key snapshots: the \textit{bug-inducing} commit and its immediate predecessor, and the \textit{bug-fixing} commit and its immediate predecessor. Each regression instance is stored within a branch in our GitHub repository, allowing new cases to be added easily.
    
    \item \textit{Easy of Use:}
    To support ongoing research, we integrated the benchmark into a framework that provides essential commands to interact with the benchmark without extensive configuration. This framework offers uniform access to build and execution tasks by abstracting the underlying build systems. Supported commands include \texttt{info}, \texttt{env}, \texttt{checkout}, \texttt{compile}, and \texttt{test}.
\end{itemize}

At this stage, we finalize the construction and storage of the \textsc{RegressionBug4APR} benchmark. To enable further analysis and better support APR studies, we extract both the bug-fixing changes (i.e., human-written patches) and the bug-inducing changes.

\subsubsection{Bug-inducing and Bug-fixing Change Extraction and Minimization.}
Since developers may introduce unrelated edits alongside bug fixes, isolating the critical changes associated with each regression error is essential for accurate benchmark analysis and for supporting APR studies (e.g., patch validation). To achieve this, we extract both the bug-inducing and bug-fixing changes by performing the following two steps, which aim to best capture the modifications relevant to each regression error.

\textit{Step 1: Test Execution and Coverage Collection.}
We begin by executing the bug-witnessing test cases on both the bug-inducing and bug-fixing commits to collect line-level code coverage. By comparing the coverage data with the corresponding code changes, we discard any lines not covered by the test cases and retain only those directly involved in triggering or fixing the bug.

Specifically, to isolate the bug-inducing and bug-fixing changes, we first collect the code coverage data from the relevant commits by executing the bug-witnessing test cases using \textit{JaCoCo}\footnote{Available at: \url{https://www.jacoco.org/jacoco/}}, an industry-grade Java code coverage library, \firstrevision{for Java bugs, and \textit{Coverage.py}\footnote{Available at: \url{https://coverage.readthedocs.io}} for Python bugs. The coverage data is collected by running the targeted test cases with the respective coverage tool, which produces a coverage report containing the raw line-level coverage information for the executed classes and methods.}

From the resulting coverage report, we extract the set of lines covered by test cases. The overall coverage information is represented as:
$
\mathcal{G} = \{ \text{file}_i \mapsto \{ \text{cov}_{i1}, \text{cov}_{i2}, \dots \} \mid i = 1, \dots, n \},
$
where $n$ is the number of files covered by the test cases, and $\text{cov}_{ij}$ denotes a line number covered in $\text{file}_i$.
We then use a diff tool to obtain the code changes in both the bug-inducing and bug-fixing commits. These changes are computed as diffs against each commit’s immediate predecessor. The code changes are represented as:
$
\mathcal{H} = \{ \text{file}_i \mapsto \{ \text{chg}_{i1}, \text{chg}_{i2}, \dots \} \mid i = 1, \dots, m \},
$
where $m$ is the number of files modified, and $\text{chg}_{ij}$ denotes a line number modified in $\text{file}_i$.
Finally, for each $\text{file}_i$, the set of relevant changes covered by the test cases is computed as the intersection:
$
\mathcal{R}_i = \mathcal{G}(\text{file}_i) \cap \mathcal{H}(\text{file}_i),
$
where $\mathcal{R}_i$ denotes the set of changed lines in $\text{file}_i$ that are covered by the targeted test cases.
Each commit, bug-inducing or bug-fixing, has its own computed set $\mathcal{R}_i$, representing the code changes directly exercised by the bug-witnessing test cases. These refined sets are crucial for pinpointing the specific code modifications responsible for the observed regression behavior. For specific types of changes, however, we apply the following additional considerations:

\firstrevision{When computing $\mathcal{R}_i$, we apply the following additional considerations based on the type of change:}

\begin{itemize}
    \item \firstrevision{\textit{Bug-inducing changes.} We focus on identifying lines that may cause the bug. If a changed branch of an if/else statement is not covered by any test case, we remove it, as it cannot be responsible for the observed failure.}
    \item \firstrevision{\textit{Bug-fixing changes.} Uncovered lines require more careful consideration. Even if a line is not directly exercised by the test cases, it may still constitute a semantically related component of the fix. For example, when test cases cover only one branch of an if/else statement, but the other branch is still part of the fix. To ensure correctness, we validate uncertain removals by reinserting them into the buggy version and running all unit tests. If reinserting a removed change fixes the bug, we conclude that the change is bug-fixing and retain it. Otherwise, it can be safely discarded. This validation step relies on test execution rather than human judgment.}
\end{itemize}

\textit{Step 2: Patch Minimization through Rules.}
To further enhance the clarity and effectiveness of our patch analysis, we manually identify and remove code changes that do not affect the bug-inducing or bug-fixing behavior. Following Defects4J~\footnote{Available at: \url{https://github.com/rjust/defects4j/blob/master/framework/bug-mining/Patch-Minimization-Guide.md}}, we define the following rules to guide our patch minimization efforts:

\begin{itemize}
    \item \firstrevision{\textit{Code Formatting:} Code changes that alter only the appearance of the code without affecting its behavior may be excluded. This includes comments, unnecessary new lines, and whitespace or indentation changes.}
    
    \item \firstrevision{\textit{Simple Code Refactoring:} Code changes that modify the code structure syntactically without introducing any behavioral difference are removed. For example, in \textit{RegressionBug-106}, we removed a change from \texttt{if (!excludePrivate || !foundTld.isPrivate)} to \texttt{if (!(excludePrivate \&\& foundTld.isPrivate))}. Although these two expressions are logically equivalent, the change affects only the syntactic structure of the condition and is therefore excluded as a refactoring change unrelated to the bug.}
\end{itemize}

\firstrevision{While this step involves human intervention and may introduce subjective bias, we took two measures to mitigate this risk. First, we applied pre-defined rules to guide and constrain minimization decisions, reducing ambiguity and ensuring consistency across all bug instances. Second, all manual minimization decisions were independently reviewed by a second author, and any disagreements were resolved through discussion.}

\firstrevision{After applying both steps, the overall patch size decreased from 7,368 to 3,385 lines after Step~1, and further to 2,853 lines after Step~2 for human patches, representing an overall reduction of approximately 61.28\%. For bug-inducing changes, the size decreased from 45,995 to 14,045 lines after Step~1, and further to 13,334 lines after Step~2, representing an overall reduction of approximately 71.01\%.}
\subsection{APR Technique Selection and Experimental Settings}
\label{sec:APR-technique-selection}

In this section, we present the APR techniques and their experimental settings evaluated in our empirical study. We revisit both traditional and LLM-based approaches to assess their applicability to regression bugs.

\subsubsection{Traditional APR Techniques}
To examine whether traditional search-based and pattern-based approaches, utilizing mutation operators and fix patterns, can effectively repair regression errors, we selected a set of widely studied heuristic-based APR tools. Specifically, we include five search-based techniques: GenProg~\cite{le2011genprog}, Kali~\cite{qi2015analysis}, Cardumen~\cite{martinez2018ultra}, Arja~\cite{yuan2018arja}, and jMutRepair~\cite{martinez2016astor}, following the selection made by Kabadi et al.~\cite{kabadi2023future}.
For template-based APR, we include TBar~\cite{liu2019tbar}, which systematically aggregates fixing patterns derived from prior template-based APR techniques and is widely regarded as a representative approach in the program repair literature. We do not evaluate semantic-based APR approaches in our study as they often require the inclusion of manually-written models for system calls whose source codes are not available.

To conduct our experiments, we utilize Cerberus, a unified framework for running and evaluating traditional APR tools, \firstrevision{which has been shown to reproduce identical results as originally reported by the respective tool authors on Defects4J, confirming the reliability of our experimental setup~\cite{shariffdeen2023cerberus}}. 
All experiments for these tools are performed on the server equipped with an AMD CPU (16 cores) and 64GB of RAM. We retain the default configuration settings provided by Cerberus for each tool to ensure \firstrevision{consistency, reproducibility, and tool-agnostic comparisons~\cite{martinez2016astor,liu2019tbar}}.
\firstrevision{While some parameters could be tuned, such as the maximum number of generations and population size for genetic-programming-based approaches (e.g., jGenProg, jKali, jMutRepair, Cardumen, and Arja), as well as the set of predefined repair patterns for TBar, aggressive per-tool tuning may introduce bias and reduce comparability across tools.
We also note that the default configuration for genetic-programming-based tools already uses a maximum of 1,000,000 generations and an overall timeout of 30 minutes, a budget that aligns with developers' expectations for APR tools in practice~\cite{eladawy2024automated,noller2022trust}.}

\vspace{-0.1cm}
\subsubsection{Fine-tuning-based APR Techniques}
To investigate how well fine-tuning-based approaches perform when applied specifically to regression errors, we revisit several recently proposed fine-tuning-based APR methods. These techniques have demonstrated promising results compared to earlier deep learning-based methods~\cite{jiang2023impact}.
Following the work of Jiang et al.~\cite{jiang2023impact}, we include several full-parameter fine-tuned models that exhibit significant performance improvements over their original pretrained counterparts. Specifically, we evaluate Incoder-1B and Incoder-6B~\cite{fried2022incoder}, as well as CodeGen-2B and CodeGen-6B~\cite{nijkamp2022codegen}.
In addition, we evaluate RepairLLaMA~\cite{silva2023repairllama}, a fine-tuned version of CodeLLaMA~\cite{roziere2023code} trained using QLoRA~\cite{dettmers2023qlora}, a parameter-efficient fine-tuning method. This technique is particularly interesting due to its reduced memory and computation requirements during training while still achieving strong repair performance.

\vspace{-0.1cm}
\begin{figure}[htbp]
    \centering
    \captionsetup{justification=centering}
    \makebox[\textwidth][c]{%
        \begin{minipage}[b]{0.58\textwidth}
            \centering
            \includegraphics[width=\linewidth]{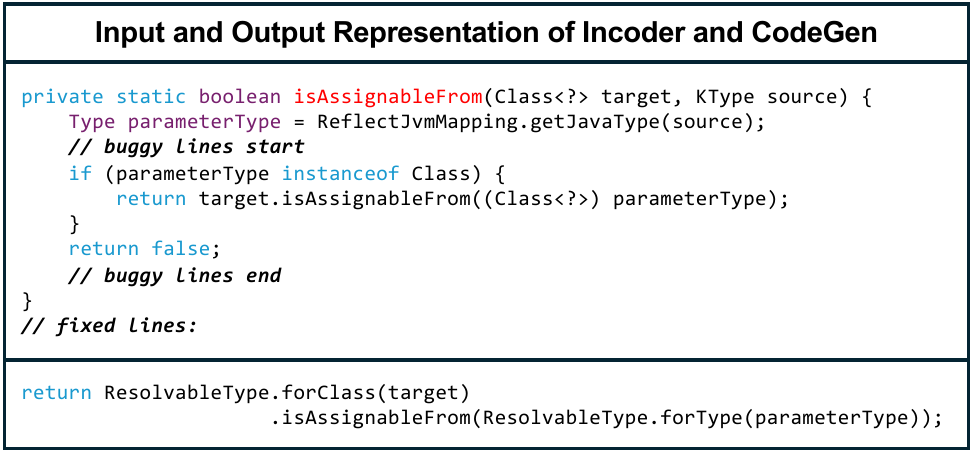}
            \caption{The input and output representation\\ of Incoder and CodeGen.}
            \label{fig:ir-or-incoder-codegen}
        \end{minipage}%
        \begin{minipage}[b]{0.58\textwidth}
            \centering
            \includegraphics[width=\linewidth]{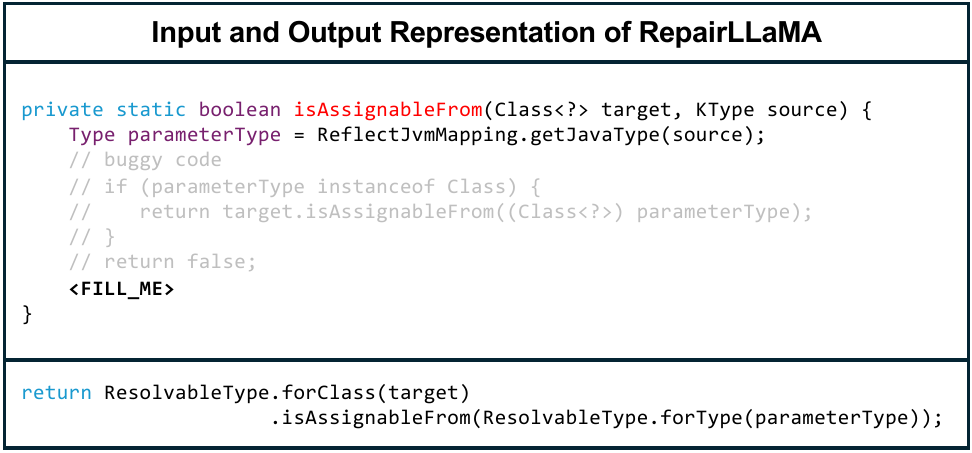}
            \caption{The input and output representation\\ of RepairLLaMA.}
            \label{fig:ir-or-repairllama}
        \end{minipage}%
    }
\end{figure}
\vspace{-0.2cm}

For input and output representation, we adopt model-specific configurations aligned with the respective fine-tuning procedures, as illustrated in Figure~\ref{fig:ir-or-incoder-codegen} and Figure~\ref{fig:ir-or-repairllama}.
In particular, \textit{Incoder} and \textit{CodeGen}~\cite{jiang2023impact} require the entire buggy function along with annotations marking the buggy location. The buggy segment is highlighted using the markers \textit{"// buggy lines start}" and \textit{"// buggy lines end"}, followed by a placeholder line starting with \textit{"// fixed lines:"}, which designates where the fix should be generated. The model is expected to output the corresponding fixed code chunk that replaces the buggy code.
In contrast, \textit{RepairLLaMA}~\cite{silva2023repairllama} adopts a different input format tailored to its infilling-based approach. The input consists of a code prefix followed by a comment beginning with the marker \textit{"// buggy code"}, which introduces the buggy lines retained in commented-out form. This is followed by the special token \textit{"<FILL\_ME>"}, which indicates the location where the model should generate the fix. The model is expected to output the corresponding fixed code chunk that replaces the buggy segment while preserving the original context.

All experiments are conducted using FLAMES~\cite{lecong2024flames}, a unified framework for executing and configuring APR evaluations with large language models. We run all fine-tuning-based techniques on our highest-performance machine, which is equipped with a single NVIDIA A100 GPU (80GB VRAM), 250GB of system RAM, and a 32-core Intel Xeon CPU running at 2.90GHz. To maintain consistent generation quality across models, we set the temperature hyperparameter to 1.0 to control output diversity. Additionally, to prevent out-of-memory crashes on our setup, we limit the number of generated candidates using a beam size of 10.

\subsubsection{Prompt-based APR Techniques}
To investigate how well prompt-based approaches perform when applied specifically to regression errors, we explore techniques that leverage general-purpose large language models (LLMs), such as ChatGPT, without requiring any fine-tuning. These approaches rely solely on prompt engineering to guide the model in generating plausible patches.
In this category, two common strategies have emerged: \textit{zero-shot prompting APR} and \textit{conversational APR}. Zero-shot prompting APR involves issuing a single prompt to the LLM to generate a patch, without relying on any interactive feedback or prior conversational context~\cite{xiang2024far}. In contrast, conversational APR techniques incorporate multiple dimensions of feedback to iteratively query the model. These methods maintain a conversation history and leverage prior failed patching attempts to improve future generations through prompting~\cite{xia2024automated,yin2024thinkrepair}.

In our experiments, we select advanced models from the ChatGPT family to implement prompt-based repair. Following prior work by Xu et al.~\cite{xu2024aligning}, we note that the next-token prediction objective used by decoder-only LLMs (e.g., GPT-4) is misaligned with the masked span prediction objective employed in current infilling-style methods. Since these models are used without any fine-tuning, the effectiveness of LLMs can be improved by aligning the repair task with their training objective; that is, allowing them to refine the entire program without explicitly providing buggy code hunks.

\begin{figure}[htbp]
    \centerline{\includegraphics[width=0.6\columnwidth]{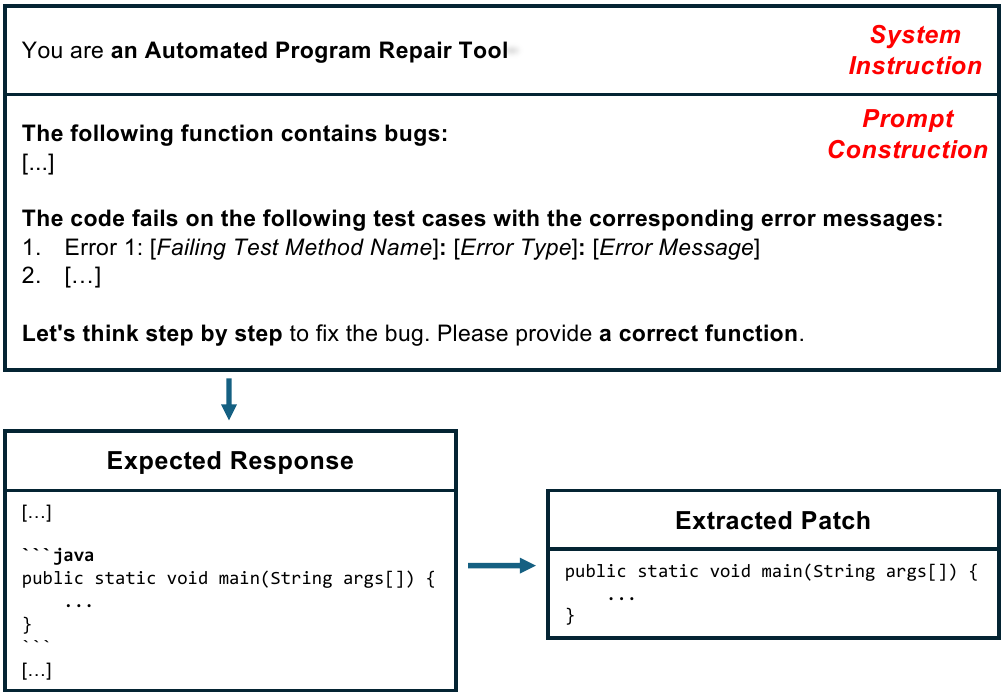}}
    \caption{Structure of our prompt-based APR query. The code snippet and error messages are omitted for clarity. The expected response is formatted to allow reliable extraction of the repaired function from the LLM’s output.}
    \label{fig:prompt-construction}
\end{figure}

To this end, we input the entire buggy method along with relevant debugging information as part of the prompt.
Following prior works~\cite{xiang2024far,xia2024automated,yin2024thinkrepair}, we manually examined a few alternative approaches using selected bugs through the web-based interface of ChatGPT~\footnote{Available at: \url{https://chatgpt.com}}. Based on these insights, we construct our prompt as illustrated in Figure~\ref{fig:prompt-construction}.
More specifically, we initialize ChatGPT with the system instruction \textit{"You are an Automated Program Repair Tool"} to explicitly define its role. We then construct the user prompt by providing various types of contextual information related to the buggy function. This includes the full body of the buggy function, followed by a list of failing test cases along with their associated error messages. Initially, only functional errors are present in the benchmark. In this case, our framework \textsc{RegressionBug4APR} supports the extraction of key diagnostic information, including the name of the failing test case (which often serves as a concise summary of the function under test), the error type (e.g., \textit{org.junit.ComparisonFailure}), and the detailed error message (e.g., \textit{expected:<[6]> but was:<[1]>}), which provides concrete evidence of the failure.
Finally, the prompt concludes with a chain-of-thought indicator, \textit{"Let's think step by step to fix the bug"}, followed by the task description, \textit{"Please provide a correct function"}. This prompt structure encourages the model to reason about the bug and generate a complete and syntactically correct replacement for the entire function, rather than returning an unformatted or incomplete patch.

\begin{table}[htbp]
\centering
\small
\caption{Prompt Templates for Feedback Cases}
\vspace{-0.2cm}
\begin{tabular}{|l|p{10cm}|}
\hline
\textbf{Case} & \textbf{Prompt Template} \\
\hline
\textbf{Compilation Error} & 
The fixed version is not compilable. The code has the following compilation error:
\par \textbf{\{error\_message\}} \par
Please provide the correct function along with any required imports. \\
\hline
\textbf{Functional Error} & 
The fixed version is still not correct. The code fails on the following test cases with the following error messages:
\par \textbf{\{error\_message\}} \par
Please provide the correct function again. \\
\hline
\textbf{No Response Code} & 
The response does not provide the code function. \par
Please provide the correct function again. \\
\hline
\textbf{Timeout} & 
The fixed version is still not correct and ran out of time. \par
Please provide the correct function again. \\
\hline
\end{tabular}
\label{tab:feedback-prompts}
\end{table}

In the case of zero-shot prompting APR, we use only the initial prompt to query ChatGPT and extract a candidate patch from the model’s output. In contrast, conversational APR techniques begin with the same initial prompt but continue with an interleaved process of patch generation and validation-driven feedback. Each generated patch is immediately validated by compiling and running the program against the test suite. If the patch fails, we construct a feedback prompt that includes both the previously generated incorrect patch and the associated failure information, which is then used to prompt the next generation.
We observe that there are four main cases in which feedback is provided to the model: \textit{(i) \textbf{Compilation Error}}, when the patch cannot be compiled;  \textit{ (ii) \textbf{Functional Error}}, when the patch compiles but fails one or more test cases; \textit{(iii) \textbf{No Response Code}}, when the model does not return a valid code function; and \textit{(iv) \textbf{Timeout}}, when the patch execution exceeds a predefined time limit, often indicating an infinite loop or inefficient implementation. For each case, we formulate a corresponding natural language feedback prompt that communicates the type of failure and requests a corrected function. These feedback templates are summarized in Table~\ref{tab:feedback-prompts}.

The repair tool is developed in Python and communicates with the ChatGPT API endpoint~\footnote{Available at: \url{https://platform.openai.com/docs/overview}}. We use \textit{gpt-3.5-turbo-0125} and \textit{gpt-4o-2024-08-06}, which were the most up-to-date and cost-effective models at the time of implementation. Inference and evaluation scripts are executed on a server equipped with a 16-core AMD CPU and 64GB of RAM.
To control a moderated randomness of text generation, we follow prior work and set the \textit{temperature} hyperparameter to 1.0, which is also the default value provided by OpenAI. Additionally, we adopt a relatively small sampling size of 10 candidates per bug, given the significantly higher inference cost (in terms of both time and financial budget) of LLMs compared to other APR tools. This setting also ensures a fair comparison with the fine-tuning-based approaches evaluated in this study.

For the conversational strategy, since ChatGPT has a limited context window~\footnote{Available at: \url{https://platform.openai.com/docs/guides/chat}}, meaning it cannot process arbitrarily long input sequences, we introduce a threshold on conversation length (i.e., the number of prompt-response exchanges in a continuous repair session). Once this maximum conversation length is reached, the repair process is restarted from the initial prompt. A maximum conversation length of 1 corresponds to zero-shot prompting APR. As the maximum conversation length increases, the model can incorporate more repair history (e.g., previous patches and corresponding feedback) into its prompts. In our experiments, we retain the sampling size of 10 and set the maximum conversation length to 5. This setting still ensures a fair comparison, as it assumes that each new prompt in the conversation builds upon the model’s previous patch rather than generating a completely new patch.

\subsubsection{Prompt-based APR Techniques Augmented with Bug-Inducing Change Information}
\label{sec:apr-with-bic}

To explore the potential of leveraging regression bug information in automated program repair, we focus on prompt-based techniques, which are motivated by their effectiveness, inherent flexibility in model usage and prompt construction, and greater data efficiency compared to fine-tuning-based approaches. In particular, large language models, such as ChatGPT, have demonstrated remarkable generalization capabilities and an ability to understand and generate code from natural language instructions~\cite{liu2024refining}. These strengths make LLMs especially well-suited for our investigation. We hypothesize that providing LLMs with additional contextual information, specifically bug-inducing change information, can significantly enhance their ability to fix regression bugs. Identifying where and how a regression was introduced offers valuable insight that can support the patch generation process.

In our experiments, we extend existing prompt-based APR frameworks by adapting the prompt design to explicitly incorporate regression-specific context. While our implementation maintains the two repair strategies, \textit{zero-shot prompting} and \textit{conversational APR}, we design the prompt to align with our research objective of investigating how augmenting prompts with bug-inducing change information affects repair effectiveness. The overall interaction structure, inference process, and configuration settings remain consistent with the prompt-based APR techniques described earlier in this section.

\begin{figure}[htbp]
    \hspace*{-0.01\columnwidth}
    \centering
    \makebox[\textwidth][c]{%
        \begin{minipage}[t]{0.53\textwidth}
            \centering
            \includegraphics[width=\linewidth]{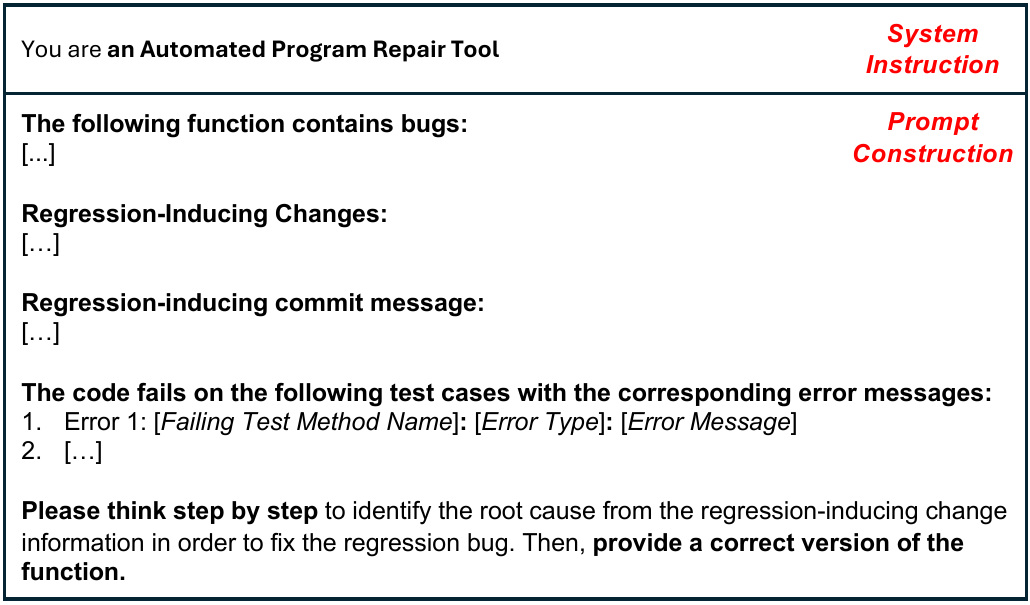}
            \textit{\footnotesize(a) Regression-inducing commit includes changes \\ to the buggy function.}
        \end{minipage}%
        \hspace{4pt}
        \begin{minipage}[t]{0.53\textwidth}
            \centering
            \includegraphics[width=\linewidth]{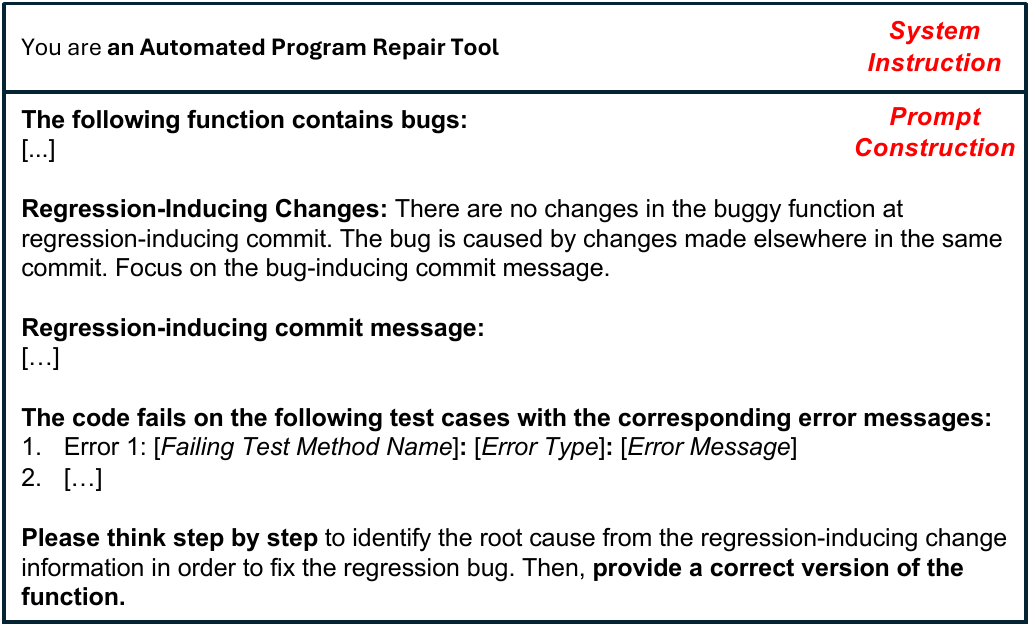}
            \textit{\footnotesize(b) Buggy function is unchanged, and the fault is introduced by other changes in the regression-inducing commit.}
        \end{minipage}%
    }
    \vspace{-0.2cm}
    \caption{Prompt templates designed for prompt-based APR with regression-inducing change information.}
    \label{fig:bic-prompt-pair}
\end{figure}

The primary distinction in this approach lies in the design of the prompt. Figure~\ref{fig:bic-prompt-pair} illustrates two prompt templates used in prompt-based APR for regression bugs, each corresponding to a different scenario based on how the regression was introduced. Specifically, we initialize ChatGPT with the system instruction \textit{"You are an Automated Program Repair Tool"} to clearly define its role. The user prompt is then constructed by providing contextual information related to the buggy function, including its full body and a list of failing test cases along with their corresponding error messages. At the initial stage, our benchmark includes only functional errors. To support this process, we use \textsc{RegressionBug4APR}, which extracts diagnostic information from the failing test cases, similar to the prompt-based APR setup described earlier.
In the case where the regression-inducing commit includes changes to the buggy function, we extract the corresponding code diff and commit message to provide additional context, as shown in Figure~(\ref{fig:bic-prompt-pair}a). This additional information helps guide the model toward understanding how the changes may have introduced the regression.
In contrast, some regression bugs are caused by changes made elsewhere in the same regression-inducing commit, leaving the buggy function itself unchanged. In such cases, as illustrated in Figure~(\ref{fig:bic-prompt-pair}b), we explicitly note that the buggy function was not modified and highlight the commit message as a key signal for localizing the root cause of the errors.
Finally, each prompt concludes with a specialized chain-of-thought instruction tailored for regression scenarios: \textit{"Let's think step by step to identify the root cause from the regression-inducing change information in order to fix the regression bug"}, followed by the task instruction: \textit{"Then, provide a correct version of the function"}. This structure encourages the model to reason about the root cause of the regression and generate a complete and syntactically correct replacement for the function, rather than producing an unstructured or partial patch.

\subsection{Evaluation Metrics}
Following prior work, we adopt two widely used metrics for evaluating APR tools: \textit{(i) the number of correct patches} and \textit{(ii) the number of plausible patches}. A \textit{plausible patch} is one that passes all test cases, while a \textit{correct patch} is semantically or syntactically equivalent to the developer’s reference patch. We follow common practice in APR and manually assess semantic equivalence to determine correct patches.

Specifically, we compute three derived metrics: plausible rate (PR), correct rate (CR), and precision (P). The correct rate and plausible rate indicate the overall effectiveness of APR tools, while precision reflects the reliability of the generated patches with respect to the overfitting problem~\cite{le2018overfitting, smith2015cure}.

\begin{equation*}
\renewcommand{\arraystretch}{1.5}
\begin{array}{ccc}
PR = \dfrac{\# \text{Plausible Patches}}{\# \text{Bugs in Dataset}} & \quad
CR = \dfrac{\# \text{Correct Patches}}{\# \text{Bugs in Dataset}} & \quad
P  = \dfrac{\# \text{Correct Patches}}{\# \text{Plausible Patches}}
\end{array}
\end{equation*}

\vspace{-0.2cm}
\section{RQ1: Quality and Diversity of the \textsc{\firstrevision{RegressionBug4APR}} Benchmark}
\label{sec:rq-1}

To validate the quality and diversity of the \textsc{\firstrevision{RegressionBug4APR}} benchmark, we conduct an analysis of its construction process and perform a detailed study of its overall characteristics. Our methodology consists of two parts: assessing the quality of the benchmark construction and characterizing the features of the benchmark.

\vspace{-0.1cm}
\subsection{Benchmark Quality}
\vspace{-0.1cm}

\begin{figure}[htbp]
    \centerline{\includegraphics[width=\columnwidth]{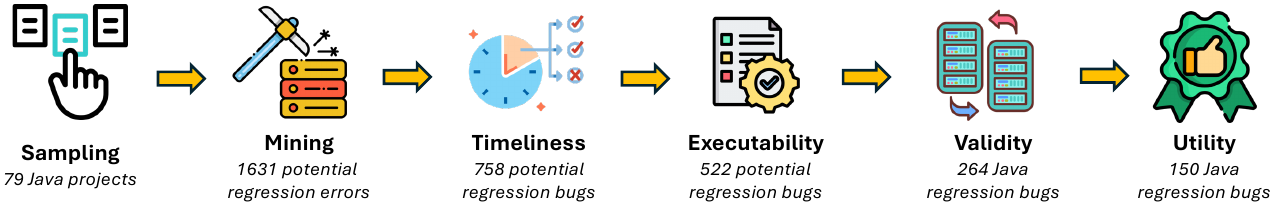}}
    \caption{\firstrevision{Number of bugs filtered out during the Java regression bug validation process.}}
    \label{fig:bug-filter-out}
\end{figure}

\vspace{-0.2cm}
We systematically analyze each filtering and validation step applied during the benchmark construction process, as detailed in Section~\ref{sec:benchmark-construction}.
\firstrevision{Note that this analysis covers the Java regression bug collection process only. The Java regression bugs are initially mined by RegMiner and subsequently re-validated using our own quality criteria to ensure they satisfy the conditions required for APR research. 
For Python, the validation process is integrated directly into the tool, such that any identified bug is already a confirmed regression bug suitable for APR research, without requiring additional validation steps, as described in Section~\ref{subsec:potential-bug-collection}.}
At each step, we record the number and proportion of potential regression bugs filtered out, as presented in Figure~\ref{fig:bug-filter-out}. This allows us to quantitatively assess the effectiveness of each criterion and demonstrate the overall reliability and soundness of the resulting \textsc{RegressionBug4APR} benchmark.

Overall, our validation process filters out \firstrevision{90.8\%} of the initially mined Java candidates, retaining only those that meet the criteria for true regression bugs. The resulting dataset not only preserves the defining behavioral characteristics of regression errors but is also specifically designed to support research on automated regression bug repair.

\subsection{Benchmark Characterization}
We characterize the features of \firstrevision{\textsc{RegressionBug4APR}} benchmark by conducting an in-depth analysis of its key properties. Specifically, we examine overall benchmark statistics and provide a detailed study of human-written patches, focusing on their associated repair operators.

\begin{figure}[!htbp]
\includegraphics[width=1\columnwidth]{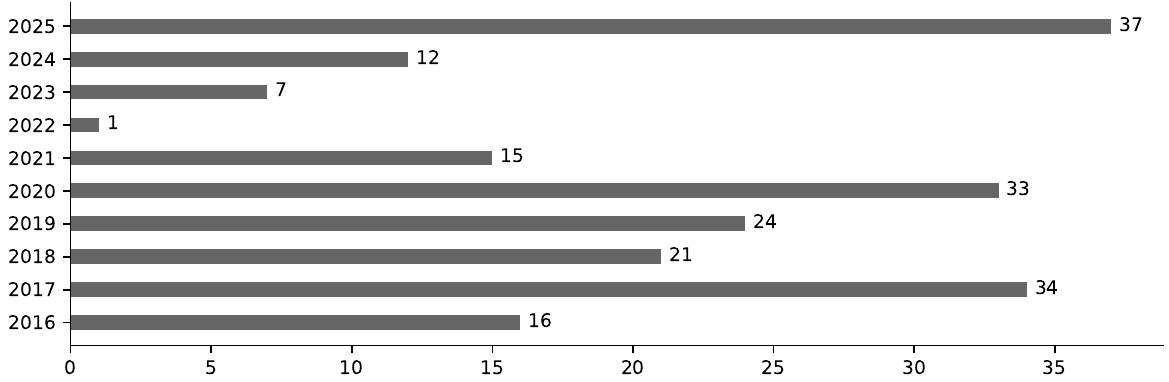}
\caption{Time distribution of the regression bugs in our benchmark}
\label{fig:dateOfBug}
\end{figure}

\subsubsection{Subject Dataset: \firstrevision{\textsc{RegressionBug4APR}}}
Our benchmark consists of 200 regression bugs extracted from widely used real-world Java and Python GitHub repositories. For each bug, \textsc{RegressionBug4APR} provides four snapshots of the project versions: the bug-fixing commit, the bug-inducing commit, and their immediate predecessors. Additionally, \textsc{RegressionBug4APR} bugs are characterized as follows: 
\textit{(i)} they have been selected after 2016, as shown in Figure~\ref{fig:dateOfBug};
\textit{(ii)} they are related to source code fixes, with exclusions for changes within the build system, configuration files, documentation, or tests; 
\textit{(iii)} they are reproducible, as each bug contains at least one test case that satisfies the conditions for regression errors mentioned in Section~\ref{sec:benchmark-construction}; 
and \textit{(iv)} they are isolated, ensuring that bug-inducing changes and bug-fixing patches do not include unrelated modifications, such as new features or refactoring.

The detailed information about each bug are included in our artifacts, as provided in Section~\ref{sec:introduction}. \firstrevision{Each bug in \textsc{RegressionBug4APR} is
referenced using a simple notation: \textit{``RegressionBug-''} followed by a bug
ID ranging from 1 to 150 for Java regression bugs, and \textit{``PyRegression-''}
followed by a bug ID ranging from 1 to 50 for Python regression bugs.}

\subsubsection{Data Statistics}
We begin by analyzing basic properties of patches in our benchmark. In particular, we focus on patch size and spreading scope, which help us understand the complexity and challenges of bug fixes.

\begin{table}[!htbp]
\small
\caption{Descriptive statistics for patch size and scope}
\vspace{-0.2cm}
\begin{center}
\begin{tabular}{lccccccc}
\hline
& \textbf{Min} & \textbf{25\%} & \textbf{50\%} & \textbf{75\%} & \textbf{90\%} & \textbf{95\%} & \textbf{Max} \\
\hline
\textbf{\# Added lines} & 0 & 1 & 3 & 7 & 15 & 23 & 159 \\ 
\textbf{\# Removed lines} & 0 & 0 & 1 & 2 & 8 & 16 & 111 \\
\textbf{\# Patch size} & 0 & 2 & 4 & 10 & 22 & 42 & 226 \\ 
\hline
\textbf{\# Chunks} & 1 & 1 & 1 & 3 & 6 & 12 & 93 \\ 
\textbf{\# Modified files} & 1 & 1 & 1 & 1 & 2 & 3 & 24 \\
\textbf{\# Modified classes} & 1 & 1 & 1 & 1 & 3 & 7 & 24 \\
\textbf{\# Modified methods} & 0 & 1 & 1 & 1 & 3 & 5 & 66 \\
\hline
\end{tabular}
\label{table:descriptiveStatistic}
\end{center}
\end{table}

\textbf{Size of the RegressionBug4APR patches.} This is presented by the number of lines added, removed and patch size, as shown in Table~\ref{table:descriptiveStatistic}.
\begin{itemize}
    \item \textit{Added lines:} 25\% of patches contain no added lines, while half of the patches add at most three lines. Patches with more than \firstrevision{15} added lines are less frequent.
    \item \textit{Removed lines:} Half of the patches contain no removed lines, and \firstrevision{95\%} of the patches have no more than 16 removed lines.
    \item \textit{Patch size:} The total size of a patch, calculated as the sum of added and removed lines, involves at most 4 lines for half of the patches. For 90\% of patches, no more than \firstrevision{22} lines are affected.
\end{itemize}

\textbf{Scope of the RegressionBug4APR patches.} This is analyzed by the number of chunks, modified methods, modified classes, and modified files, as demonstrated in Table~\ref{table:descriptiveStatistic}.
\begin{itemize}
    \item \textit{Chunks:} Defined as continuous lines of code changes, 75\% of patches contain three or fewer chunks.
    \item \textit{Modified methods:} \firstrevision{155 patches (77.5\%) change one method.}
    \item \textit{Modified classes:} We observe that the number of modified classes is closely related to the number of modified files in each patch.
    \item \textit{Modified files:} \firstrevision{26} patches require changes in at least two files to address the bug. Of these, \firstrevision{9} follow a unique regression-fixing pattern, namely \textit{Revert to previous statement}.
\end{itemize}

Overall, the median patch size in \textsc{RegressionBug4APR} is four lines, with approximately 10\% of patches affecting more than 22 lines. Among the \firstrevision{200} patches, \firstrevision{100} contain a single chunk, \firstrevision{155} modify only one method, and \firstrevision{174} are confined to a single file. The occurrence of larger patches is attributed to a specific regression bug-fixing pattern, namely \textit{Revert to previous statement}, as highlighted in our bug analysis. 
This patch size distribution implies that the benchmark is suitable for evaluating a wide range of APR techniques. For instance, traditional APR tools typically produce small patches, while LLM-based approaches can handle larger modifications. Moreover, the presence of a few much larger patches highlights the need for future research into techniques capable of addressing more substantial repairs.

\subsubsection{Bug Categorization}
We further justify the diversity of the \textsc{RegressionBug4APR} benchmark by characterizing the constituent bugs. We dissect bugs in the benchmark into several different repair operators, of which some are obtained from Java defect classes proposed by Pan et al.~\cite{pan2009toward} and Sobreira et al.~\cite{sobreira2018dissection}. Each operator is characterized by the type of code change performed to fix the defect. \firstrevision{These operators correspond to foundational constructs that are common across imperative programming languages. Since Java and Python share the same core imperative constructs, such as assignments, conditionals, loops, method calls, and return statements and etc, we apply the same set of operators to annotate patches from both languages in our benchmark.} Table~\ref{table:repairActions} summarizes the groups of repair operators observed in the benchmark. We describe the details of each operator group below.

\begin{table}[!htbp]
\small
\caption{Repair Operators}
\vspace{-0.3cm}
\begin{center}
\begin{tabular}{lll}
\hline
Acronym & Group & Action \\ 
\hline
asgn & Assignment  & A/R/M \\
cnd  & Condition   & A/R/M \\
lp   & Loop        & A/R/M \\ 
mc   & Method call & A/R/M \\ 
md   & Method definition & A/R/M \\
obj  & Object Instantiation & A/R/M \\
exp  & Exception & A/R \\
ret  & Return & A/R/M \\
var  & Variable & A/R/M \\
rev  & Revert to previous statement & N.A \\
\hline
\multicolumn{3}{l}{$^{\mathrm{abbreviation}}$A: Add; R: Remove; M: Modify.}
\end{tabular}
\label{table:repairActions}
\end{center}
\end{table}

\begin{itemize}
    \item \textit{Assignment:} Changes involving a simple assignment operator (=), unary increment/decrement operators, or compound assignment operators with arithmetic operations. Actions applicable to this group include \textit{addition} and \textit{removal} at the statement level, and \textit{modification} at the expression level (e.g., modifying the assigned value).
    \item \textit{Condition:} Changes related to conditional branches, such as \texttt{if-else}, \texttt{switch-case}, and the ternary operator. Actions applicable to this group include \textit{addition} and \textit{removal} at the statement level, and \textit{modification} at the expression level (e.g., modifying, expanding, or reducing the conditional expression). 
    \item \textit{Loop:} Changes related to loop constructions, such as \texttt{for} and \texttt{while}. Actions applicable to this group include \textit{addition}, \textit{removal} at statement level, and \textit{modification} at both statement and expression levels.
    \item \textit{Method Call:} Changes related to method calls. Actions applicable to this group include \textit{addition} and \textit{removal} at the statement level, and \textit{modification} at the expression level (e.g., adding, removing, swapping, and modifying parameters).
    \item \textit{Method definition:} Changes related to method definitions and signatures. Actions applicable to this group include \textit{addition} and \textit{removal} at the statement level, and \textit{modification} at the expression level (e.g., adding or removing parameters, modifying parameter types, changing the return type, and altering the method's scope).
    \item \textit{Object Instantiation:} Changes related to the instantiation of objects are observed through the keyword \texttt{new}. Actions applicable to this group include \textit{addition} and \textit{removal} at the statement level, and \textit{modification} at the expression level (e.g., modifying the parameters passed in objects).
    \item \textit{Exception:} A statement related to exception handling and throwing. Actions applicable to this group include \textit{addition} and \textit{removal} at the statement level (e.g., adding or removing \texttt{try-catch} or \texttt{throw} blocks).
    \item \textit{Return:} Changes related to \texttt{return} statements. Actions applicable to this group include \textit{addition} and \textit{removal} at the statement level (e.g., wrapping with a conditional \texttt{if-else}), and \textit{modification} at the expression level (e.g., modifying the logic expression of the return).
    \item \textit{Variable:} Changes related to variable declaration and usage. Actions applicable to this group include \textit{addition} and \textit{removal} at the statement level, and \textit{modification} at the expression level (e.g., modifying the variable type, replacing it with another variable or method call, and altering the assigned value).
    \item \textit{Revert to previous statement:} Changes related to rolling back at the statement level. The rationale behind this action is that developers may have made mistakes that should have been left intact, resulting in the regression bug at hand. This is particularly unique for regression bugs, as it involves observing multiple versions and reasoning through the changes.
\end{itemize}

\begin{figure}[!htbp]
\includegraphics[width=1\columnwidth]{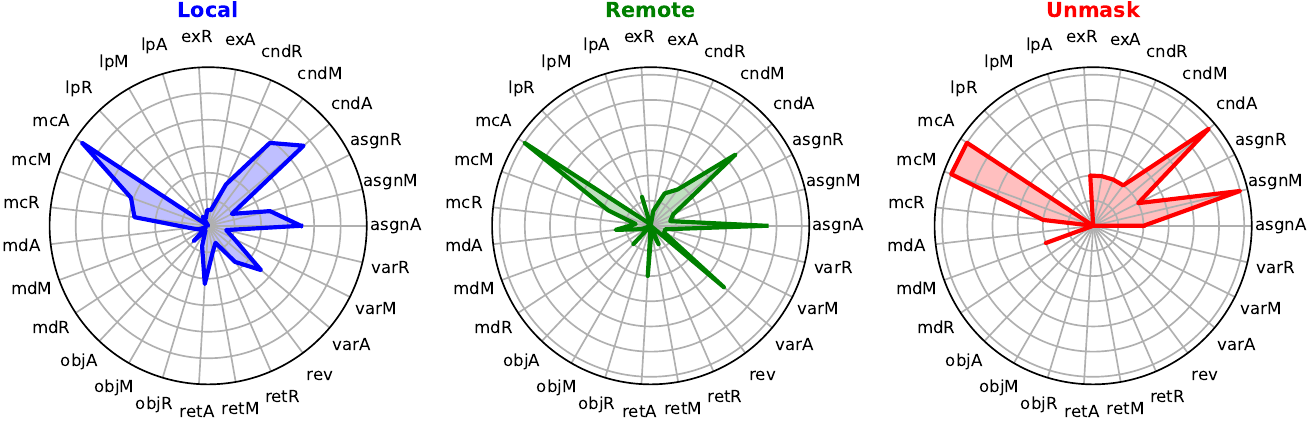}
\caption{Distribution of repair operators on different types of regression bug. Note that, we abbreviated the names of the repair action groups following Table~\ref{table:repairActions}. 
For example, ``mcA" denotes \textit{Method Call Addition}.}
\label{fig:distributionOfEachType}
\end{figure}
\vspace{-0.5cm}
Following the regression categorization by Tan et al.~\cite{tan2015relifix}, based on how the regression was introduced (as described in Section~\ref{sec:background}), our dataset contains \firstrevision{139 \textit{local} bugs, 50 \textit{remote} bugs, and 7 \textit{unmask} bugs, with 4 bugs left uncategorized due to the complexity of their bug-inducing and bug-fixing changes}.
Figure~\ref{fig:distributionOfEachType} illustrates the overall distribution of repair operators for each type of regression bug. The radial axis corresponds to the repair operators discussed in the previous section, indicating their occurrence across the different bug types. Common patterns, such as \textit{Conditional Addition}, \textit{Method Call Addition}, \textit{Method Call Modification}, \textit{Assignment Modification} and \textit{Assignment Addition}, are prevalent in all three types of regression bugs. Notably, the \textit{Revert to Previous Statement} operator is more frequently observed in the \textit{Local} bug category.

\begin{figure}[!htbp]
\includegraphics[width=0.91\columnwidth]{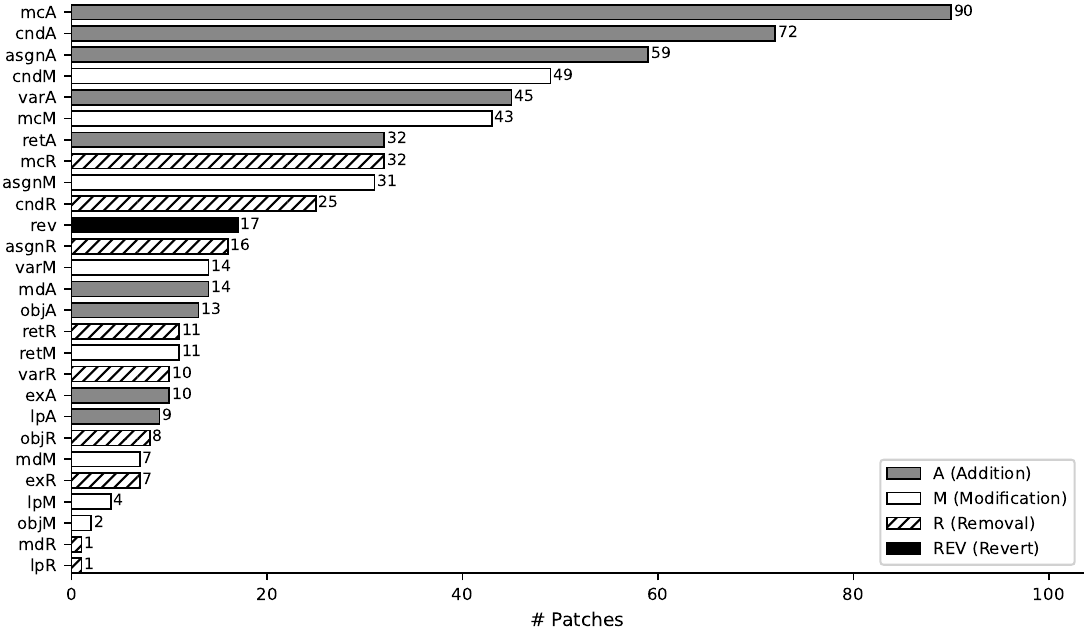}
\vspace{-0.2cm}
\caption{Occurrence of repair operators in \textsc{RegressionBug4APR}. Note that, we abbreviated the names of the repair action groups following Table~\ref{table:repairActions}. 
For example, ``mcA" denotes \textit{Method Call Addition}.}
\label{fig:distributionOfRepairOperators}
\end{figure}

Figure~\ref{fig:distributionOfRepairOperators} demonstrates the ranking of repair operators based on human patches, with the vertical axis representing the operators and the horizontal axis indicating the number of patches in which they are found. To enhance clarity, grey bars represent addition actions, diagonal striped bars indicate removal actions, white bars signify modification actions, and black bars denote revert actions. \firstrevision{\textit{Method Call Addition}} is the most prevalent repair operator, appearing in 90 patches, followed by \firstrevision{\textit{Conditional Addition} with 72 patches and \textit{Assignment Addition} with 59 patches}. The \textit{Revert to Previous Statement} operator is exclusive to \textit{local} regression bugs, appearing in \firstrevision{17} patches. In all cases, addition (\firstrevision{54.1\%}) exceeds removal and modification by \firstrevision{36.4\%} and \firstrevision{28.5\%}, respectively.

\begin{figure}[!htbp]
\includegraphics[width=0.95\columnwidth]{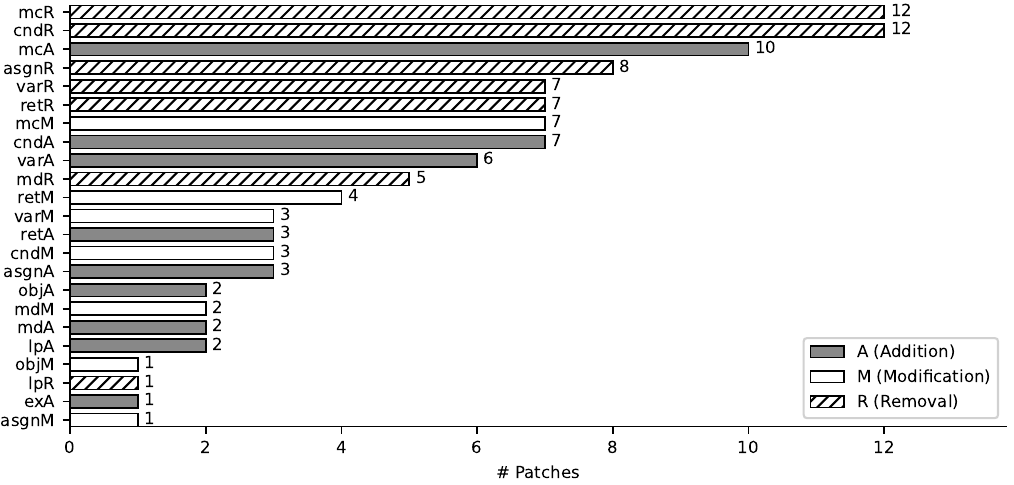}
\vspace{-0.2cm}
\caption{Detailed examination of a repair operator, namely \textit{revert to previous statement}. Note that, we abbreviated the names of the repair action groups following Table~\ref{table:repairActions}. 
For example, ``mcA" denotes \textit{Method Call Addition}.}
\label{fig:examineRevert}
\end{figure}

\vspace{-0.1cm}
We further analyze the patches that utilize the \textit{revert to previous statement} operator. In our dataset, there are \firstrevision{17} regression bugs that were patched by humans using only this repair operator. Additionally, we analyze these patches to identify the repair operators that humans would have employed if the revert operator had not been applied. Figure~\ref{fig:examineRevert} illustrates all the repair operators used by humans to address these bugs. In total, \firstrevision{114} repair operators were employed to fix these \firstrevision{17} bugs, accounting for \firstrevision{17.81\%} of the total operators used across all bugs. This underscores the importance of recognizing this pattern in addressing regression bugs effectively.

\begin{figure}[htbp]
\includegraphics[width=0.8\columnwidth]{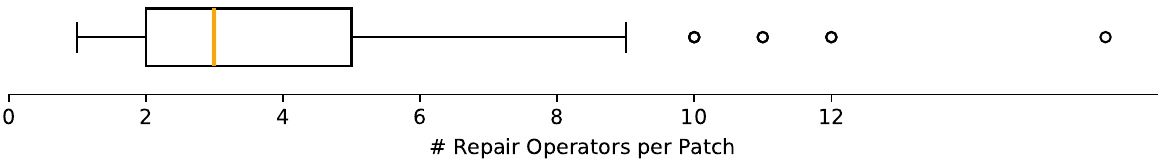}
\vspace{-0.3cm}
\caption{Distribution of the number of repair operators per patch.}
\label{fig:operatorPerPatch}
\end{figure}
\vspace{-0.1cm}

Figure~\ref{fig:operatorPerPatch} presents the distribution of the number of repair operators per patch. The whiskers and outliers indicate that the data spans from 1 (lower whisker) to \firstrevision{9} (upper whisker), with several outliers reaching up to \firstrevision{15}. The median (highlighted in \firstrevision{orange}) is \firstrevision{3}, while the box extends from \firstrevision{2} to \firstrevision{5}, indicating that the middle 50\% of the data lies within this range. The distribution is skewed to the right, with a few patches exhibiting a much higher number of operators than the majority.

Overall, the results highlight the diversity of repair operators required to fix bugs in our dataset. Compared with Defects4J~\cite{sobreira2018dissection}, \firstrevision{our benchmark covers all repair operators observed in Defects4J}. Notably, for \textit{Local} regression bugs, the \textit{Revert to Previous Statement} operator proves effective. Furthermore, a substantial portion of the patches involve the application of multiple repair operators, indicating the non-trivial nature of many fixes.

These results demonstrate that \textsc{RegressionBug4APR} captures a wide range of patch sizes and scopes. With several patches exhibiting distinctive regression-fixing patterns, the dataset includes both representative and challenging edge cases that are rarely addressed by existing tools. This diversity makes \textsc{RegressionBug4APR} a valuable benchmark for evaluating and advancing automated program repair techniques.

\vspace{-0.2cm}
\begin{rqbox}{\textbf{RQ1}}{\textbf{Quality and Diversity of the RegressionBug4APR Benchmark}}
\textit{The analysis demonstrates the high quality of the \textsc{RegressionBug4APR} benchmark for APR research, highlighting key characteristics such as reality (sourced from real-world Java \firstrevision{and Python} repositories), timeliness (extending the timeframe of the most recent and relevant benchmarks), and diversity (in terms of patch size, scope, and repair operators). In addition, our structured data storage process ensures durable reproducibility of all bugs.}
\end{rqbox}
\section{RQ2: Effectiveness of APR Techniques on Regression Bugs}
\label{sec:rq-2}
To investigate the effectiveness of existing APR techniques in the context of regression errors, we evaluate both traditional and LLM-based approaches in repairing buggy programs, without providing additional regression-specific context. This setup aligns with the conventional APR workflow, which relies solely on the buggy program and its associated test suite for validation.
\firstrevision{Note that this research question focuses exclusively on Java bugs, as the traditional APR tools and fine-tuning-based LLM approaches evaluated here are inherently designed for Java and do not support Python.}
The traditional APR tools include jGenProg~\cite{le2011genprog}, jKali~\cite{qi2015analysis}, Cardumen~\cite{martinez2018ultra}, Arja~\cite{yuan2018arja}, jMutRepair~\cite{martinez2016astor} and TBar~\cite{liu2019tbar}. The LLM-based approaches include Incoder-1B and Incoder-6B~\cite{fried2022incoder}, CodeGen-2B and CodeGen-6B~\cite{nijkamp2022codegen}, and RepairLLaMA~\cite{silva2023repairllama}. Additionally, we evaluate prompt-based techniques, including zero-shot prompting and conversational APR, using two advanced models: ChatGPT-3.5-Turbo and ChatGPT-4o. The selection, configuration, and implementation details of these tools are described in Section~\ref{sec:APR-technique-selection}. Table~\ref{table:empirical-results} summarizes the overall performance of APR techniques on the \textsc{\firstrevision{RegressionBug4APR}} benchmark, reporting the number of plausible patches, the number of correct patches among them, and the corresponding precision scores.

\subsection{Experimental Results}
\begin{figure}[htbp]
    \centering
    \begin{minipage}[b]{0.49\linewidth}
        \centering
        \includegraphics[width=\linewidth]{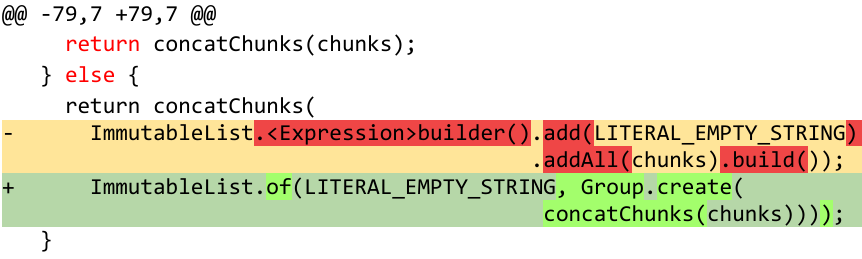}
        \vspace{2pt}
        \textit{\footnotesize(a) Regression-inducing changes in RegressionBug-84 from our \textsc{RegressionBug4APR} benchmark.}
        \label{fig:bic84}
    \end{minipage}
    \hfill
    \begin{minipage}[b]{0.49\linewidth}
        \centering
        \includegraphics[width=\linewidth]{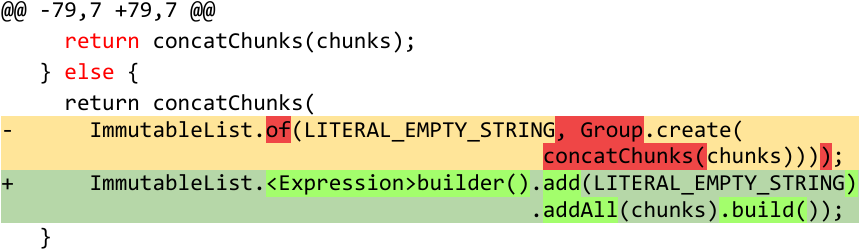}
        \vspace{2pt}
        \textit{\footnotesize(b) Regression-fixing changes in RegressionBug-84 from our \textsc{RegressionBug4APR} benchmark.}
        \label{fig:bfc84}
    \end{minipage}
    \caption{Illustration of the \textit{Revert to Previous Statement} repair operator in \textit{RegressionBug-84} from the \textsc{RegressionBug4APR} benchmark.}
    \label{fig:regressionbug-84}
\end{figure}

Our experiments indicate that \textit{traditional APR techniques} were unable to repair any bugs within our dataset. A deeper analysis suggests the limitation stems from two main factors: \textit{(i)} the unavailability of fixes within the current version of the program, as many fixes may only exist in earlier versions~\cite{barr2014plastic}, and \textit{(ii)} the inadequacy of existing bug-fixing patterns, which do not encompass many of the bugs present in our dataset.
\firstrevision{For instance, Figures~(\ref{fig:regressionbug-84}a) and~(\ref{fig:regressionbug-84}b) present the bug-inducing and bug-fixing changes for \textit{RegressionBug-84} from the \textsc{RegressionBug4APR} benchmark. In this case, the error emerged when developers sought to enhance system performance but inadvertently introduced implementation errors, resulting in subtle semantic bugs. The most straightforward resolution involved reverting the previous changes to eliminate the fault. However, if this bug is examined in isolation, without considering the associated regression-inducing commit, as is often the case with many APR tools, the bug-fixing pattern appears more complex, requiring numerous modifications that fall outside the scope of existing fix templates. This complexity likely explains the limited applicability of such patterns in state-of-the-art APR techniques. Conversely, if the regression-inducing commit is included in the input, a simple \textit{“Revert to Previous Statement”} pattern could effectively resolve the issue.}

\begin{table*}[!htbp]
\centering
\fontsize{7.5pt}{9.5pt}\selectfont
\caption{Performance Evaluation of APR Methods on the \firstrevision{Java Regression Bugs of the \textsc{RegressionBug4APR} benchmark.}}
\label{table:empirical-results}
\begin{tabular}{>{\raggedright\arraybackslash}p{5.2cm} c c c}
\Xhline{0.7pt}
\multirow{2}{*}{\textbf{Method}} 
& \textbf{\#Plausible Patches} & \textbf{\#Correct Patches} & \multirow{2}{*}{\textbf{Precision (\%)}} \\
& \textbf{(Plausible Rate \%)} & \textbf{(Correct Rate \%)} \\
\Xhline{1pt}

\multirow{2}{*}{\makecell[l]{\textbf{ChatGPT-4o + Conversation}}} & 60 & 18 & \multirow{2}{*}{30.00\%}      \\
 & \textit{(40.00\%)} & \textit{(12.00\%)} & \\ \cline{1-4}
  
\multirow{2}{*}{\makecell[l]{\textbf{ChatGPT-4o + Zero-shot Prompting}}} & 35 & 13 & \multirow{2}{*}{37.14\%} \\
& \textit{(23.33\%)} & \textit{(8.67\%)} & \\ \cline{1-4}

\multirow{2}{*}{\makecell[l]{\textbf{ChatGPT-Turbo-3.5 + Conversation}}} & 41 & 9 & \multirow{2}{*}{21.95\%} \\
& \textit{(27.33\%)} & \textit{(6.00\%)} & \\ \cline{1-4}

\multirow{2}{*}{\makecell[l]{\textbf{ChatGPT-Turbo-3.5 + Zero-shot Prompting}}} & 28 & 8 & \multirow{2}{*}{28.57\%} \\
& \textit{(18.67\%)} & \textit{(5.33\%)} & \\ \cline{1-4}

\Xhline{1pt}
\multirow{2}{*}{\textbf{RepairLLaMA}} & 31 & 15 & \multirow{2}{*}{48.39\%} \\
& \textit{(20.67\%)} & \textit{(10.00\%)} & \\ \cline{1-4}

\multirow{2}{*}{\textbf{Incoder-6B}} & 14 & 5 & \multirow{2}{*}{35.71\%} \\
& \textit{(9.33\%)} & \textit{(3.33\%)} & \\ \cline{1-4}

\multirow{2}{*}{\textbf{Incoder-1B}} & 13 & 5 & \multirow{2}{*}{38.46\%} \\
& \textit{(8.67\%)} & \textit{(3.33\%)} & \\ \cline{1-4}

\multirow{2}{*}{\textbf{CodeGen-6B}} & 12 & 3 & \multirow{2}{*}{25.00\%} \\
& \textit{(8.00\%)} & \textit{(2.00\%)} & \\ \cline{1-4}

\multirow{2}{*}{\textbf{CodeGen-2B}} & 11 & 4 & \multirow{2}{*}{36.36\%} \\
& \textit{(7.33\%)} & \textit{(2.67\%)} & \\

\Xhline{1pt}

\multirow{2}{*}{\textbf{TBar}} & 0 & 0 & \multirow{2}{*}{N/A} \\
& \textit{(0\%)} & \textit{(0\%)} & \\ \cline{1-4}

\multirow{2}{*}{\textbf{jGenProg}} & 0 & 0 & \multirow{2}{*}{N/A} \\
& \textit{(0\%)} & \textit{(0\%)} & \\ \cline{1-4}

\multirow{2}{*}{\textbf{Cardumen}} & 0 & 0 & \multirow{2}{*}{N/A} \\
& \textit{(0\%)} & \textit{(0\%)} & \\ \cline{1-4}

\multirow{2}{*}{\textbf{jKali}} & 0 & 0 & \multirow{2}{*}{N/A} \\
& \textit{(0\%)} & \textit{(0\%)} & \\ \cline{1-4}

\multirow{2}{*}{\textbf{Arja}} & 0 & 0 & \multirow{2}{*}{N/A} \\
& \textit{(0\%)} & \textit{(0\%)} & \\ \cline{1-4}

\multirow{2}{*}{\textbf{jMutRepair}} & 0 & 0 & \multirow{2}{*}{N/A} \\
& \textit{(0\%)} & \textit{(0\%)} & \\

\Xhline{1pt}
\end{tabular}
\end{table*}

Moving to more advanced approaches, APR techniques based on large language models demonstrate significantly improved performance over traditional tools. We categorize these techniques into two main families: \textit{fine-tuning-based} and \textit{prompt-based} methods. Our results show promising outcomes, some bugs can be fixed with minimal context, requiring only the buggy function and debugging information. However, other bugs necessitate richer contextual information to understand the cause of the regression and to reuse relevant variables or logic from earlier versions to construct the correct fix (e.g., reverting a previous change as in \textit{RegressionBug-84}).

\textit{Fine-tuning-based APR tools}, which employ LLMs further fine-tuned on large-scale bug-fix corpora, generate between \firstrevision{11 and 31 plausible patches}, with \firstrevision{3 to 15 correct patches} and precision rates ranging from \firstrevision{25.00\% to 48.39\%}. Among them, RepairLLaMA achieves the highest performance, producing \firstrevision{15 correct patches} and attaining the highest precision rate of \firstrevision{48.39\%}.
\firstrevision{Incoder outperforms CodeGen in terms of correct patches, with both Incoder-6B and Incoder-1B generating 5 correct patches each; however, Incoder-1B demonstrates slightly better precision (38.46\% vs. 35.71\%). For CodeGen, CodeGen-2B achieves 4 correct patches compared to 3 for CodeGen-6B, with CodeGen-2B also demonstrating higher precision (36.36\% vs. 25.00\%).
Notably, RepairLLaMA is able to fix every bug resolved by Incoder and CodeGen, with the exception of \textit{RegressionBug-48} and \textit{RegressionBug-139}, for which it generates a plausible but incorrect patch.}

On the other hand, \textit{prompt-based APR techniques}, including both \textit{zero-shot prompting} and \textit{conversational} strategies, also demonstrate strong performance. These methods yield between \firstrevision{28 and 60 plausible patches}, with \firstrevision{8 to 18 correct patches}, and precision rates ranging from \firstrevision{21.95\% to 37.14\%}. 
Among them, ChatGPT-4o with conversational interaction achieves the highest number of correct patches (18 in total).
Overall, ChatGPT-4o outperforms ChatGPT-Turbo-3.5 in terms of correct patches: ChatGPT-4o achieves \firstrevision{18} and \firstrevision{13} correct patches under the conversational and zero-shot configurations, respectively, compared to \firstrevision{9} and \firstrevision{8} for ChatGPT-Turbo-3.5.
For both models, the conversational configuration generally generates more plausible patches than zero-shot prompting. Interestingly, the zero-shot prompting configuration results in higher precision for both models (\firstrevision{37.14\%} vs. \firstrevision{30.00\%} for ChatGPT-4o, and \firstrevision{28.57\%} vs. \firstrevision{21.95\%} for ChatGPT-Turbo-3.5), compared to their conversational counterparts.
This discrepancy may be due to the tendency of conversational strategies to overfit to the given test cases, as they rely on iterative feedback from test executions to guide patch generation, resulting in a higher number of plausible patches but relatively few correct ones among them.

\begin{insightbox}
\textit{In sum, ChatGPT-4o (Conversation) achieves the highest number of correct repairs, successfully fixing \firstrevision{18} bugs, while RepairLLaMA achieves \firstrevision{15} correct repairs with the highest precision of \firstrevision{48.39\%}, making it the most reliable tool among all evaluated techniques. The union of both tools results in \firstrevision{27} unique bugs being fixed in the Java portion of the RegressionBug4APR benchmark. Notably, RepairLLaMA is able to repair \firstrevision{9} bugs that ChatGPT-4o (Conversation) fails to fix, while ChatGPT-4o (Conversation) is able to repair \firstrevision{12} bugs that RepairLLaMA fails to fix, highlighting the complementary nature of the two approaches.}
\end{insightbox}

\subsection{\firstrevision{Performance Breakdown by Regression Bug Type}}
\begin{figure}[!htbp]
\centerline{\includegraphics[width=1\columnwidth]{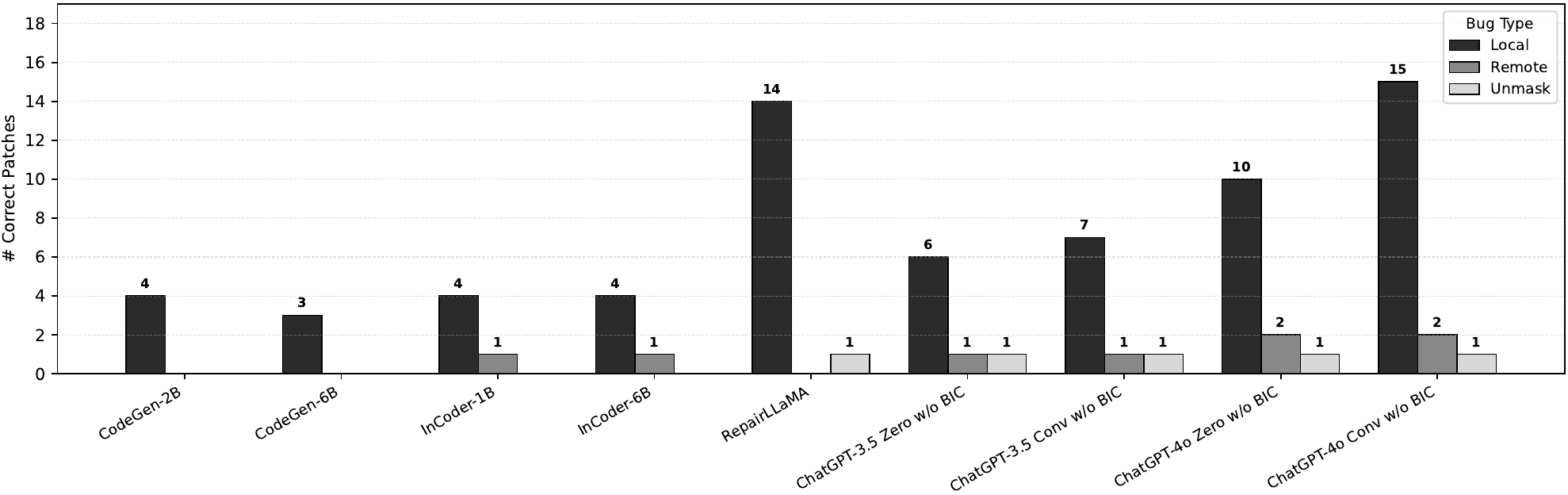}}
\vspace{-0.2cm}
\caption{\firstrevision{Number of correct patches per regression bug type (Local, Remote, and Unmask) for each evaluated APR technique in RQ2 on the Java regression bugs of the \textsc{RegressionBug4APR} benchmark.}}
\label{fig:rq2-breakdown-bug-type}
\end{figure}

\begin{figure}[!htbp]
\centerline{\includegraphics[width=1\columnwidth]{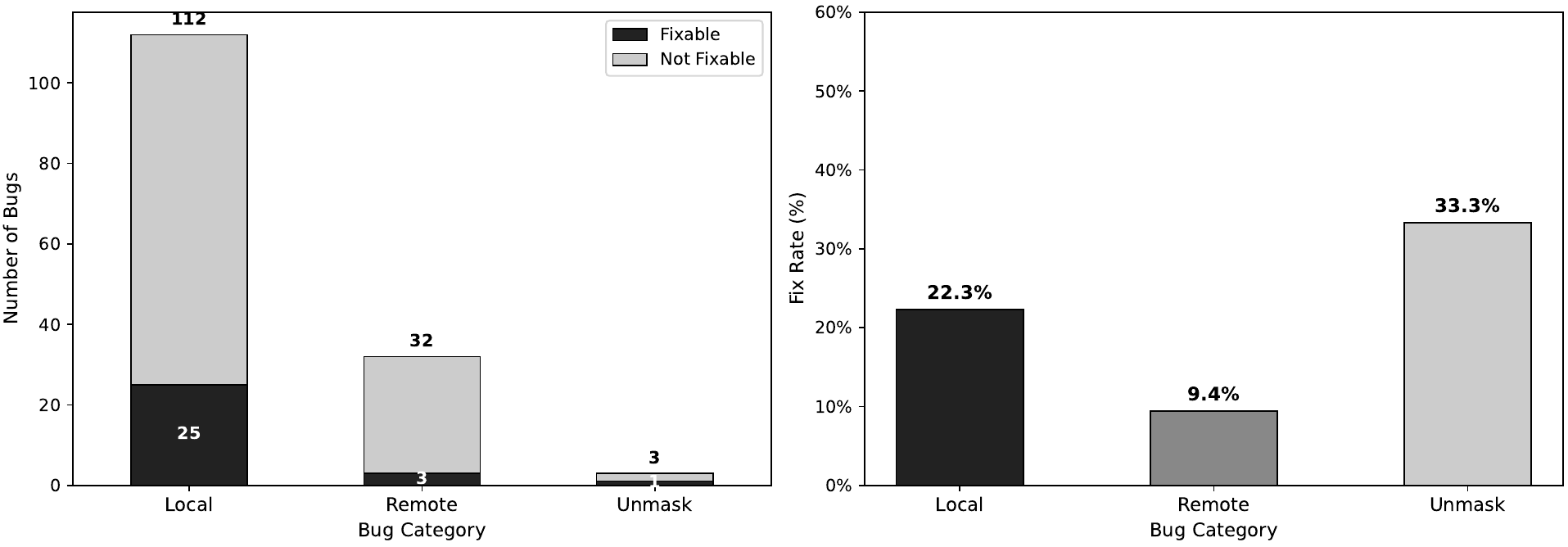}}
\vspace{-0.2cm}
\caption{\firstrevision{Number of fixable bugs (left) and fix rate (right) per regression bug type (Local, Remote, and Unmask) across all evaluated APR techniques in RQ2 on the Java regression bugs of the \textsc{RegressionBug4APR} benchmark.}}
\label{fig:rq2-fix-rate-bug-type}
\end{figure}

\firstrevision{To further understand which regression bug types are most challenging for existing APR techniques, we provide a breakdown of repair results across Local, Remote, and Unmask categories on the Java of the \textsc{RegressionBug4APR} benchmark, as illustrated in Figures~\ref{fig:rq2-breakdown-bug-type} and~\ref{fig:rq2-fix-rate-bug-type}. The Java portion of the \textsc{Regression4APR} benchmark comprises 112 Local, 32 Remote, and 3 Unmask Java regression bugs, with 3 additional bugs left uncategorized. No evaluated technique successfully repairs the uncategorized bugs, as they require modifications across multiple files. Across all evaluated techniques, Local bugs consistently account for the majority of correct repairs, as shown in Figure~\ref{fig:rq2-breakdown-bug-type}. ChatGPT-4o (Conversation) achieves the highest number of Local bug repairs (15), followed by RepairLLaMA with 14. Remote bugs prove particularly challenging, with most techniques producing at most 1--2 correct repairs for this category. Unmask bugs show moderate repairability despite their small sample size, with RepairLLaMA, ChatGPT-Turbo-3.5, and ChatGPT-4o each correctly repairing 1 Unmask bug. As shown in Figure~\ref{fig:rq2-fix-rate-bug-type}, Remote bugs achieve the lowest fix rate (9.4\%) among all three bug types, while Local bugs achieve 22.3\% and Unmask bugs achieve 33.3\%.}

\begin{insightbox}
\firstrevision{\textit{On the Java regression bugs of the \textsc{RegressionBug4APR} benchmark, considering the union of all evaluated approaches without BIC context, Local bugs achieve a fix rate of 22.3\% (25 out of 112), Remote bugs achieve 9.4\% (3 out of 32), and Unmask bugs achieve 33.3\% (1 out of 3). Remote bugs achieve the lowest fix rate among all three bug types.}}   
\end{insightbox}

\subsection{\firstrevision{Analysis of Representative Repair Instances}}
\noindent\firstrevision{\textbf{\textit{Bugs fixed by RepairLLaMA but not ChatGPT-4o (Conversation).}}
RepairLLaMA is able to repair bugs that ChatGPT-4o (Conversation) fails to fix, as illustrated by \textit{RegressionBug-95} and \textit{RegressionBug-30}.}
In \textit{RegressionBug-95}, the correct fix involves adjusting a numeric constant: \texttt{"IOUtils.getChars(nanos, 30 - (9 - nanoSize), buf);"} to \texttt{"IOUtils.getChars(nanos, 29 - (9 - nanoSize), buf);"}. This change precisely aligns the buffer offset with the expected location of the \textit{nanos} value in timestamp formatting. Although ChatGPT-4o (Conversation) effectively utilizes the failure message to exactly pinpoint the faulty location, it generates an overfitting patch: \texttt{"IOUtils.getChars(nanos, buf.length, buf);"}. This version only passes the test case when \textit{"nanoSize"} is \textit{9}, making it a coincidental success rather than a semantically correct fix.
In \textit{RegressionBug-30}, ChatGPT-4o (Conversation) produces a plausible patch after five attempts and three rounds of interaction. The final patch is unnecessarily complex and lacks the core corrective logic expected in the human-written fix. As a result, it introduces semantic divergence from the intended behavior, despite passing the test suite.

\noindent\firstrevision{\textbf{\textit{Bugs fixed by ChatGPT-4o (Conversation) but not RepairLLaMA.}}
ChatGPT-4o (Conversation) is able to repair bugs that RepairLLaMA fails to fix, as illustrated by \textit{RegressionBug-85}.}
ChatGPT-4o successfully repaired this bug after four attempts and two rounds of interaction. In the first round, it produced a partial fix and responded with the following insight: 
\begin{itemize}
    \item[] \small \textit{"... The error message indicates that the expected output should have the `metadata` argument only in the `if` condition and not in the `else` condition. The current implementation mistakenly includes `metadata` in both branches of the conditional expression..."}
\end{itemize}
Although this patch did not yet pass the test case, the response demonstrated its ability to reason about the structure of the bug.
In the second round, ChatGPT-4o refined its fix with the following response:
\begin{itemize}
    \item[] \small \textit{"... To address the error you mentioned, it seems the issue lies in how the conditional expression is being constructed, particularly regarding the parameters passed to the `construct` method for the `SOY\_VISUAL\_ELEMENT`. The test case indicates that when `GOOG\_DEBUG` is `true`, the `SOY\_VISUAL\_ELEMENT` should be constructed with both the ID and the name, while when `GOOG\_DEBUG` is `false`, it should be constructed with only the ID..."}
\end{itemize}
This revised fix passed the test case and demonstrates how ChatGPT-4o with conversation effectively leverages failing test case information and interaction history to guide its repair process. The example highlights the benefit of interactive, test-driven repair in addressing complex semantic errors.

\noindent\firstrevision{\textbf{\textit{A difficult instance fixed by no technique.}}
We present \textit{RegressionBug-76} as a representative difficult instance that no evaluated technique could fix.
This bug is from the \textit{cron-utils} project, which supports Quartz cron expressions including "overflow ranges", i.e., ranges where the start value exceeds the end value, representing a wrap-around (e.g., 20--10 for seconds means seconds 20 to 59 and 0 to 10). A bug-inducing commit (\textit{``Issue \#305: Fix bug regarding bad year handling for next execution''}) added \texttt{.withStrictRange()} to the seconds, minutes, hours, day-of-month, and year field definitions in the Quartz definition builder, inadvertently rejecting overflow ranges as invalid and breaking backward compatibility. The failures are observed below:}

\firstrevision{\textit{Prompt-based approaches} consistently conflate the overflow range rejection with an off-by-one error in the valid range bounds, leading them to reduce \texttt{(1,32)} to \texttt{(1,31)} and \texttt{(1,13)} to \texttt{(1,12)}. These transformations appear semantically natural from general domain knowledge but are factually incorrect in the Quartz specification, where these bounds are intentional to permit step values equal to the full field span. The few patches that correctly removed \texttt{.withStrictRange()} always compounded the error by simultaneously reducing these bounds, introducing new failures in period-step validation.}

\firstrevision{\textit{Fine-tuning-based approaches} reveal a complementary failure mode. CodeGen and InCoder models produce severely truncated or structurally malformed patches, with over 80\% of outputs timing out due to incomplete method bodies, and others fail to compile due to field duplication, wrong field ordering, or hallucinated API signatures (e.g., \texttt{.withStrictRange(1,32)}). RepairLLaMA, while generating syntactically complete and compilable code (9/10 patches compile successfully), fails to produce any semantically meaningful change: all nine compilable patches unconditionally retain \texttt{.withStrictRange()} on seconds, minutes, and hours, with the only modifications being irrelevant reordering of method chains or incorrect additions of .optional() to mandatory fields.}

\firstrevision{The root cause is that no model produced the correct fix required by the ground truth, selectively removing \texttt{.withStrictRange()} from exactly four fields while preserving it on the year field. This distinction is derivable only from the BIC context, which is unavailable in the current repair setting. This case illustrates a fundamental limitation shared by both prompt-based and fine-tuning-based APR: when the observable test failure does not directly reflect the underlying root cause of the regression, and when the correct repair requires both library-specific semantic knowledge and an understanding of the intent behind the bug-inducing context, existing techniques, whether based on fine-tuning or conversational prompting, are insufficient to generate the correct patch.}

\firstrevision{This example motivates future work: \textit{(i)} developing repair approaches that incorporate bug-inducing context beyond the currently available information (e.g., buggy method and test failure); and \textit{(ii)} incorporating domain-specific knowledge to support cases where the correct fix requires library-level understanding that cannot be inferred from the available context or the trained knowledge of LLMs.}

\begin{rqbox}{\textbf{RQ2}}{\textbf{Effectiveness of APR Techniques on Regression Bugs}}
\firstrevision{\textit{Our experimental results of 15 APR methods on 150 Java regression bugs demonstrate that traditional techniques fail to repair any bugs in the dataset, largely due to the limitations of their underlying bug-fixing patterns. In contrast, advanced APR tools leveraging large language models show promising results. Notably, ChatGPT-4o (Conversation) achieves the highest number of correct repairs, successfully fixing 18 bugs, while RepairLLaMA attains the highest precision of 48.39\%, making it the most reliable tool in terms of generating correct patches.}}
\end{rqbox}
\vspace{-0.5cm}
\section{RQ3: Impact of Bug-Inducing Change Information on APR Performance}
\label{sec:rq-3}

To explore whether providing additional context, specifically bug-inducing change information, can enhance regression error repair performance, we focus on prompt-based techniques and adapt the prompt design to explicitly incorporate regression-specific context.
We choose this approach based on its demonstrated effectiveness (as shown in RQ2), its inherent flexibility in both model usage and prompt construction, and its greater data efficiency compared to fine-tuning-based methods.
The selection, configuration, and implementation details of these techniques are provided in Section~\ref{sec:APR-technique-selection}.

\vspace{-0.4cm}
\subsection{Experimental Results}
\label{subsec:results-rq3}
Our experimental results show that for both base models, including ChatGPT-4o and ChatGPT-Turbo-3.5, incorporating regression-specific context, specifically bug-inducing change (BIC) information, improves the plausible rate, correct rate and precision compared to their counterparts configured without this information.
Additionally, when BIC context is included, the conversational configuration consistently outperforms the zero-shot prompting configuration in both base models.

\firstrevision{Table~\ref{table:bic-results-java} summarizes the overall performance of prompt-based APR techniques on the Java benchmark, reporting the number of plausible patches, the number of correct patches and the corresponding precision scores.}
For ChatGPT-4o, the conversational strategy with BIC successfully repairs \firstrevision{29 bugs (19.33\%)}, compared to only \firstrevision{18 bugs (12.00\%)} without BIC, an increase of approximately \firstrevision{7.33\%}. The precision also increases by \firstrevision{10.28\% (from 30.00\% to 40.28\%)}. Similarly, zero-shot prompting with BIC achieves \firstrevision{21 correct repairs}, outperforming the version without BIC, which fixes \firstrevision{13 bugs, with a precision gain of 4.04\% (from 37.14\% to 41.18\%).}
A similar trend is observed for ChatGPT-Turbo-3.5. The conversational strategy with BIC fixes \firstrevision{14 bugs (9.33\%), compared to 9 bugs (6.00\%) without BIC, accompanied by a 4.47\% increase in precision (from 21.95\% to 26.42\%)}. Likewise, zero-shot prompting with BIC results in \firstrevision{12 correct repairs, outperforming its counterpart without BIC, which fixes 8 bugs, with a corresponding precision gain of 6.72\% (from 28.57\% to 35.29\%).}
When comparing configurations within each base model, zero-shot prompting with BIC consistently outperforms both the conversational and zero-shot configurations without BIC, further underscoring the value of incorporating regression-specific context.
Lastly, it is important to note that ChatGPT-4o under conversation configuration augmented with BIC achieves the best overall performance across all evaluated configurations, in terms of both correct repairs and precision.

\begin{table*}[!htbp]
\centering
\fontsize{6.4pt}{7.2pt}\selectfont
\caption{Performance Evaluation of Prompt-based APR Approaches With and Without \firstrevision{Full} Bug-Inducing Change (BIC) Information \firstrevision{on Java regression bugs of the \textsc{RegressionBug4APR} benchmark}.}
\label{table:bic-results-java}
\begin{tabular}{llccc}
\Xhline{0.5pt}
\multirow{2}{*}{\textbf{Base Model}}& \multirow{2}{*}{\textbf{Method}} 
& \textbf{\#Plausible Patches} & \textbf{\#Correct Patches} & \multirow{2}{*}{\textbf{Precision (\%)}} \\
& & \textbf{(Plausible Rate \%)} & \textbf{(Correct Rate \%)} \\
\Xhline{0.5pt}
\multirow{8}{*}{\textbf{ChatGPT-4o}}
& \multirow{2}{*}{\textbf{\textit{Zero-shot Prompting w/o BIC}}} & 35 & 13 & \multirow{2}{*}{\textit{37.14\%}} \\
& & \textit{(23.33\%)} & \textit{(8.67\%)} & \\ \cline{2-5}
& \multirow{2}{*}{\textbf{\textit{Conversation w/o BIC}}} & 60 & 18 & \multirow{2}{*}{\textit{30.00\%}} \\
& & \textit{(40.00\%)} & \textit{(12.00\%)} & \\ \cline{2-5}
& \multirow{2}{*}{\textbf{\textit{Zero-shot Prompting with BIC}}} & 51 & 21 & \multirow{2}{*}{\textit{41.18\%}} \\
& & \textit{(34.00\%)} & \textit{(14.00\%)} & \\ \cline{2-5}
& \multirow{2}{*}{\textbf{\textit{Conversation with BIC}}} & 72 & 29 & \multirow{2}{*}{\textit{40.28\%}} \\
& & \textit{(48.00\%)} & \textit{(19.33\%)} & \\ 
\Xhline{0.5pt}
\multirow{8}{*}{\textbf{ChatGPT-Turbo-3.5}}
& \multirow{2}{*}{\textbf{\textit{Zero-shot Prompting w/o BIC}}} & 28 & 8 & \multirow{2}{*}{\textit{28.57\%}} \\
& & \textit{(18.67\%)} & \textit{(5.33\%)} & \\ \cline{2-5}
& \multirow{2}{*}{\textbf{\textit{Conversation w/o BIC}}} & 41 & 9 & \multirow{2}{*}{\textit{21.95\%}} \\
& & \textit{(27.33\%)} & \textit{(6.00\%)} & \\ \cline{2-5}
& \multirow{2}{*}{\textbf{\textit{Zero-shot Prompting with BIC}}} & 34 & 12 & \multirow{2}{*}{\textit{35.29\%}} \\
& & \textit{(22.67\%)} & \textit{(8.00\%)} & \\ \cline{2-5}
& \multirow{2}{*}{\textbf{\textit{Conversation with BIC}}} & 53 & 14 & \multirow{2}{*}{\textit{26.42\%}} \\
& & \textit{(35.33\%)} & \textit{(9.33\%)} & \\ 
\Xhline{0.5pt}
\end{tabular}
\end{table*}

\begin{table*}[!htbp]
\centering
\fontsize{6.4pt}{7.2pt}\selectfont
\caption{\firstrevision{Performance Evaluation of Prompt-based APR Approaches With and Without Full Bug-Inducing Change (BIC) Information on Python Regression Bugs of the \textsc{RegressionBug4APR} benchmark.}}
\label{table:bic-results-python}
\begin{tabular}{llccc}
\Xhline{0.5pt}
\multirow{2}{*}{\textbf{Base Model}}& \multirow{2}{*}{\textbf{Method}} 
& \textbf{\#Plausible Patches} & \textbf{\#Correct Patches} & \multirow{2}{*}{\textbf{Precision (\%)}} \\
& & \textbf{(Plausible Rate \%)} & \textbf{(Correct Rate \%)} \\
\Xhline{0.5pt}
\multirow{8}{*}{\textbf{ChatGPT-4o}}
& \multirow{2}{*}{\textbf{\textit{Zero-shot Prompting w/o BIC}}} & 25 & 6 & \multirow{2}{*}{\textit{24.00\%}} \\
& & \textit{(50.00\%)} & \textit{(12.00\%)} & \\ \cline{2-5}
& \multirow{2}{*}{\textbf{\textit{Conversation w/o BIC}}} & 29 & 6 & \multirow{2}{*}{\textit{20.69\%}} \\
& & \textit{(58.00\%)} & \textit{(12.00\%)} & \\ \cline{2-5}
& \multirow{2}{*}{\textbf{\textit{Zero-shot Prompting with BIC}}} & 25 & 7 & \multirow{2}{*}{\textit{28.00\%}} \\
& & \textit{(50.00\%)} & \textit{(14.00\%)} & \\ \cline{2-5}
& \multirow{2}{*}{\textbf{\textit{Conversation with BIC}}} & 35 & 10 & \multirow{2}{*}{\textit{28.57\%}} \\
& & \textit{(70.00\%)} & \textit{(20.00\%)} & \\ 
\Xhline{0.5pt}
\multirow{8}{*}{\textbf{ChatGPT-Turbo-3.5}}
& \multirow{2}{*}{\textbf{\textit{Zero-shot Prompting w/o BIC}}} & 9 & 2 & \multirow{2}{*}{\textit{22.22\%}} \\
& & \textit{(18.00\%)} & \textit{(4.00\%)} & \\ \cline{2-5}
& \multirow{2}{*}{\textbf{\textit{Conversation w/o BIC}}} & 27 & 5 & \multirow{2}{*}{\textit{18.52\%}} \\
& & \textit{(54.00\%)} & \textit{(10.00\%)} & \\ \cline{2-5}
& \multirow{2}{*}{\textbf{\textit{Zero-shot Prompting with BIC}}} & 16 & 3 & \multirow{2}{*}{\textit{18.75\%}} \\
& & \textit{(32.00\%)} & \textit{(6.00\%)} & \\ \cline{2-5}
& \multirow{2}{*}{\textbf{\textit{Conversation with BIC}}} & 27 & 6 & \multirow{2}{*}{\textit{22.22\%}} \\
& & \textit{(54.00\%)} & \textit{(12.00\%)} & \\ 
\Xhline{0.5pt}
\end{tabular}
\end{table*}

\firstrevision{Table~\ref{table:bic-results-python} summarizes the overall performance of prompt-based APR techniques on the Python benchmark. The results are consistent with the trends observed in the Java benchmark, supporting the generalizability of our findings across programming languages.
For ChatGPT-4o, incorporating BIC information consistently improves repair effectiveness: the conversational configuration with BIC achieves 10 correct patches (28.57\% precision) compared to 6 (20.69\% precision) without BIC, and the zero-shot configuration with BIC achieves 7 correct patches (28.00\% precision) compared to 6 (24.00\%) without BIC. For ChatGPT-Turbo-3.5, the conversational configuration with BIC achieves 6 correct patches (22.22\% precision) compared to 5 (18.52\% precision) without BIC, while the zero-shot configuration with BIC achieves 3 correct patches (18.75\% precision) compared to 2 (22.22\% precision) without BIC. The limited improvement for ChatGPT-Turbo-3.5 is consistent with observations in the Java benchmark, where the benefit of BIC information is more pronounced for the stronger model. Overall, the conversational configuration with full BIC tends to achieve stronger performance, particularly for ChatGPT-4o, which is consistent with the observations in the Java benchmark. These results suggest that incorporating regression-specific context benefits repair effectiveness across programming languages, though the extent of improvement depends on the underlying model capability.}

\begin{insightbox}
\firstrevision{\textit{In total, considering both Java and Python benchmarks, the conversational configuration of ChatGPT-4o augmented with full BIC information achieves the best overall performance, correctly repairing 39 out of 200 bugs (19.50\%), compared to 24 out of 200 bugs (12.00\%) without BIC, representing a 1.6× improvement. These results demonstrate that incorporating bug-inducing change information consistently enhances repair effectiveness across both programming languages and evaluation settings.}}
\end{insightbox}

\subsection{\firstrevision{Performance Breakdown by Regression Bug Type}}
\begin{figure}[!htbp]
    \centerline{\includegraphics[width=0.9\columnwidth]{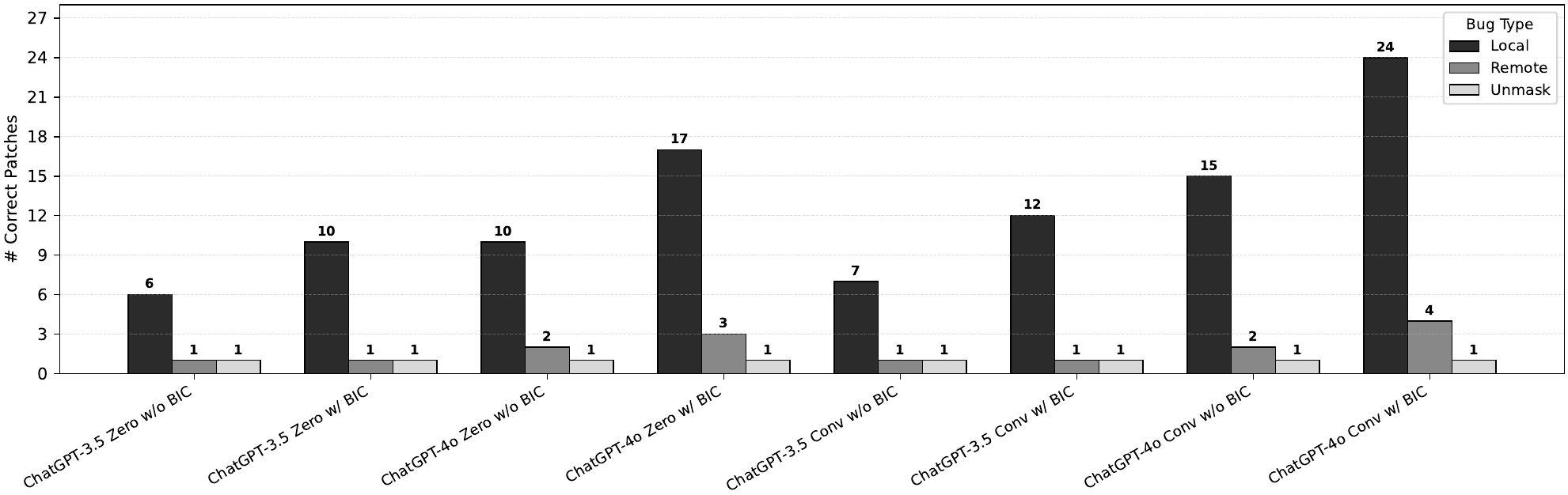}}
    \centerline{\small (a) Java}
    \centerline{\includegraphics[width=0.9\columnwidth]{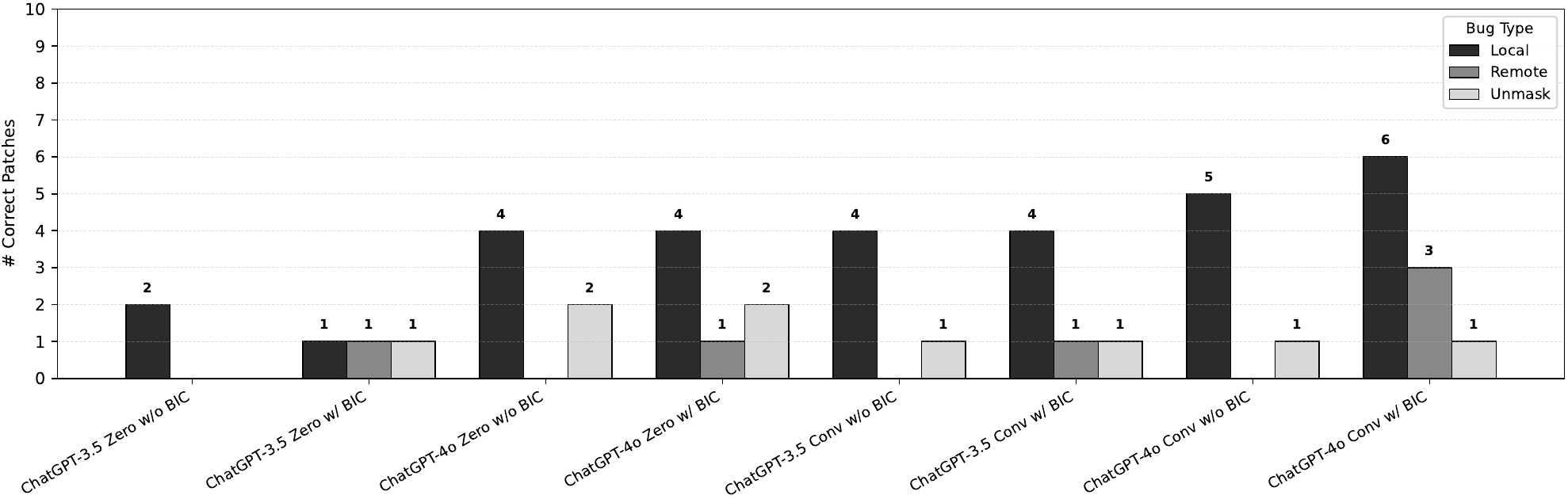}}
    \centerline{\small (b) Python}
    \vspace{-0.2cm}
    \caption{\firstrevision{Number of correct patches per regression bug type (Local, Remote, and Unmask) for each evaluated APR technique in RQ3 on the \textsc{RegressionBug4APR} benchmark. (a) Java. (b) Python.}}
    \label{fig:rq3-breakdown-bug-type}
\end{figure}

\firstrevision{To further understand how BIC information affects repair effectiveness across different regression bug types, we analyze the repair results broken down by Local, Remote, and Unmask categories on the \textsc{RegressionBug4APR} benchmark.}

\firstrevision{As shown in Figure~\ref{fig:rq3-breakdown-bug-type}, Local bugs consistently account for the majority of correct patches across all evaluated techniques and configurations in both languages. On the Java benchmark (Figure~\ref{fig:rq3-breakdown-bug-type}a), ChatGPT-4o (Conversation with BIC) achieves the highest overall repair count, generating 24, 4, and 1 correct patches for Local, Remote, and Unmask bugs, respectively. Notably, providing BIC context yields substantial gains across configurations; for instance, ChatGPT-4o (Conversation) increases its Local correct patches from 15 to 24 and its Remote correct patches from 2 to 4 when augmented with BIC information. On the Python benchmark (Figure~\ref{fig:rq3-breakdown-bug-type}b), ChatGPT-4o (Conversation with BIC) similarly achieves the best performance, generating 6, 3, and 1 correct patches for Local, Remote, and Unmask bugs, respectively. The benefit of BIC context is observed consistently across all bug types and technique configurations in both languages.}

\begin{figure}[!htbp]
    \centerline{\includegraphics[width=0.9\columnwidth]{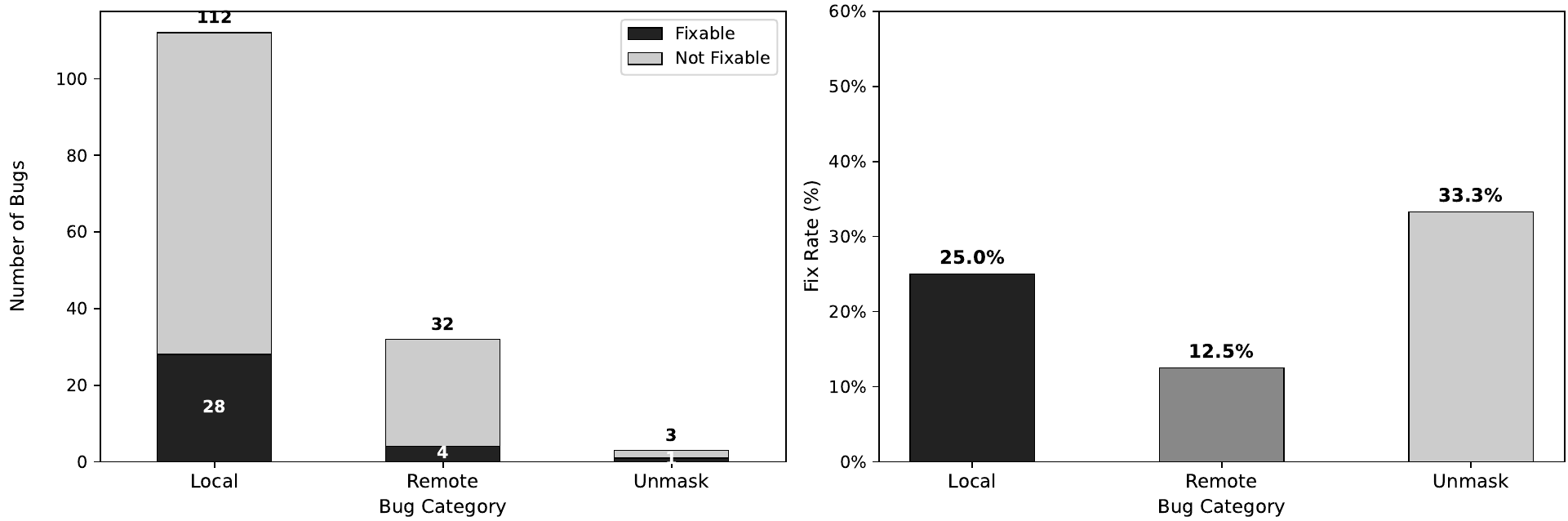}}
    \vspace{-0.1cm}
    \centerline{\small (a) Java}
    \centerline{\includegraphics[width=0.9\columnwidth]{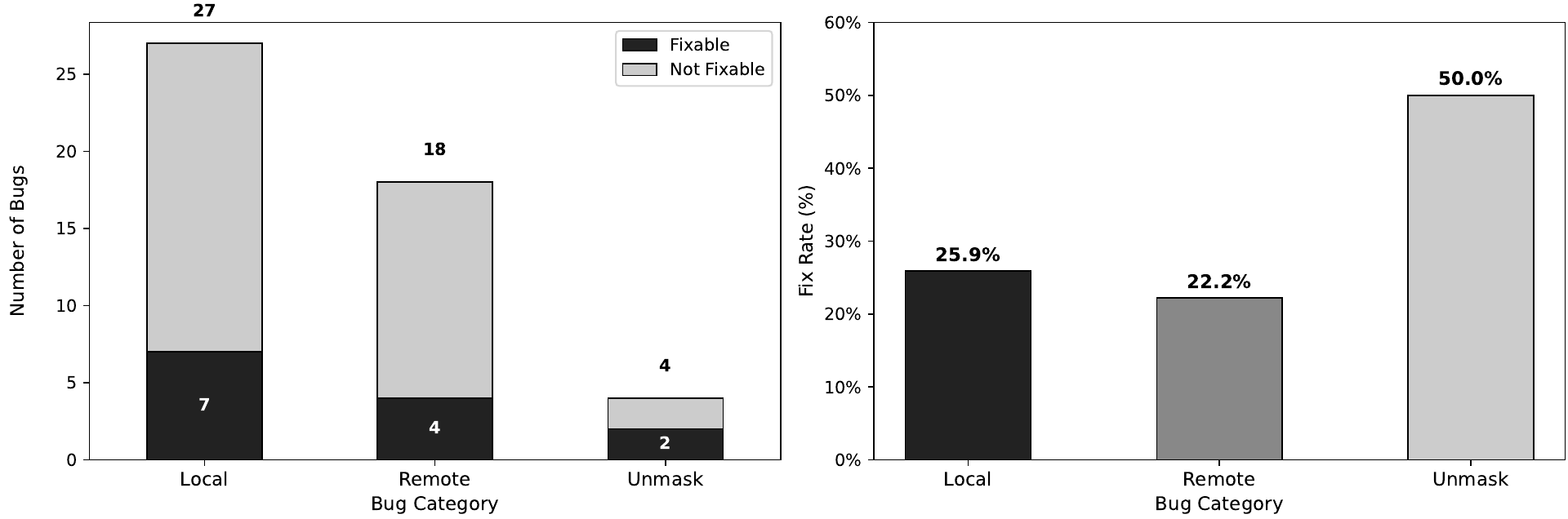}}
    \vspace{-0.1cm}
    \centerline{\small (b) Python}
    \vspace{-0.2cm}
    \caption{\firstrevision{Number of fixable bugs (left) and fix rate (right) per regression bug type (Local, Remote, and Unmask) across all evaluated prompt-based APR techniques incorporating BIC context on the \textsc{RegressionBug4APR} benchmark. (a) Java. (b) Python.}}
    \label{fig:rq3-fix-rate-bug-type}
\end{figure}

\firstrevision{Figure~\ref{fig:rq3-fix-rate-bug-type} presents the fix rates per regression bug type for both languages. On Java (Figure~\ref{fig:rq3-fix-rate-bug-type}a), a direct comparison with the RQ2 results without BIC context (Figure~\ref{fig:rq2-fix-rate-bug-type}) reveals improvements across all three bug categories when BIC information is incorporated: the fix rate for Local bugs increases from 22.3\% to 25.0\%, the fix rate for Remote bugs increases from 9.4\% to 12.5\%, and the fix rate for Unmask bugs remains at 33.3\% (1 out of 3 bugs in both settings). On the Python benchmark (Figure~\ref{fig:rq3-fix-rate-bug-type}b), Local bugs achieve a fix rate of 25.9\% (7 out of 27), Remote bugs achieve 22.2\% (4 out of 18), and Unmask bugs achieve the highest fix rate at 50.0\% (2 out of 4); however, the latter result should be interpreted with caution given the limited sample size of only 4 Unmask bugs. Across both languages, Remote bugs consistently achieve the lowest fix rate among all three bug categories, suggesting that they remain the most challenging category for automated repair, even when BIC context is provided.}

\begin{insightbox}
\firstrevision{\textit{On the Java benchmark, incorporating BIC information improves the fix rate across all regression bug types: Local bugs increase from 22.3\% to 25.0\%, and Remote bugs increase from 9.4\% to 12.5\%, while Unmask bugs remain at 33.3\%. Across the \textsc{RegressionBug4APR} benchmark (200 bugs), Local bugs achieve an overall fix rate of 25.2\% (35 out of 139 bugs), Remote bugs achieve 16.0\% (8 out of 50 bugs), and Unmask bugs achieve 42.9\% (3 out of 7 bugs). Remote bugs consistently represent the most challenging regression bug category for automated repair across both programming languages and all evaluated configurations.}}
\end{insightbox}

\subsection{Qualitative Analysis of Plausible Patches}
To better understand how bug-inducing change (BIC) information influences the behavior of LLMs in automated program repair, we conduct a qualitative analysis of plausible patches generated by the best-performing configuration, conversational ChatGPT-4o with full BIC context. Specifically, we selected cases in which regression bugs were successfully repaired only when BIC information was included, as identified in Section~\ref{subsec:results-rq3}. We organize our findings into representative case studies, each illustrating a distinct pattern in how the model leverages contextual and historical information to generate correct patches.

\subsubsection{\textbf{Observation 1:} Fully reverting to previous statements guided by bug-inducing change information}
In some cases, APR techniques successfully repair the bug by reverting all modified statements to the version that was functioning correctly before the regression was introduced. Generating such patches correctly requires the model to understand the broader context of both the bug-inducing changes and the buggy program. For example, the model must be able to distinguish between intended updates, such as bug fixes or feature additions, and unintended changes that should be reverted.

For instance, Figure~\ref{fig:regressionbug-84} presents an illustration of the \textit{Revert to Previous Statement} repair operator in \textit{RegressionBug-84} from the \textit{RegressionBug4APR} benchmark.
In this case, the developer had introduced a change that wrapped a \texttt{+} expression in parentheses (e.g., \texttt{"" + (a + b)}) to prevent premature string coercion. However, this modification was incorrect because the JavaScript \texttt{+} operator is not associative.
The fix reverted this change by removing the parentheses and directly prepending \texttt{""} to the list of expressions (i.e., \texttt{"" + a + b + c}), allowing JavaScript to perform left-to-right type coercion and thereby preserving the intended semantics.

\begin{figure}[!htbp]
\includegraphics[width=\columnwidth]{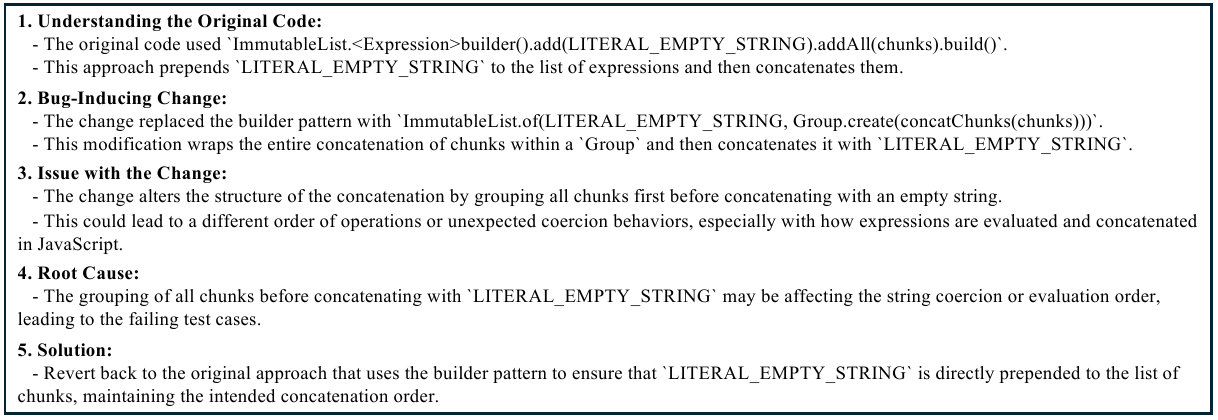}
\caption{Excerpt from ChatGPT-4o’s response under the zero-shot prompting configuration augmented with bug-inducing change information for repairing \textit{RegressionBug-84}.}
\vspace{-0.2cm}
\label{fig:response-84}
\end{figure}

Figure~\ref{fig:response-84} presents an excerpt from ChatGPT-4o’s reasoning under zero-shot prompting configuration with BIC context for \textit{RegressionBug-84}. 
Specifically, the model first analyzes the original code, then interprets the bug-inducing change, identifies the semantic issue it introduced, and ultimately concludes that the regression stems from replacing the builder-style concatenation with a grouped expression. This modification unintentionally alters the evaluation order and type coercion behavior in JavaScript, resulting in a failing test case. ChatGPT-4o correctly identifies the root cause and proposes reverting to the original builder-based expression to restore the intended semantics, aligning with the developer’s manual fix.

\begin{figure}[htbp]
    \hspace*{-0.05\columnwidth}
    \centering
    \makebox[\textwidth][c]{%
        \begin{minipage}[t]{0.58\textwidth}
            \centering
            \includegraphics[width=\linewidth]{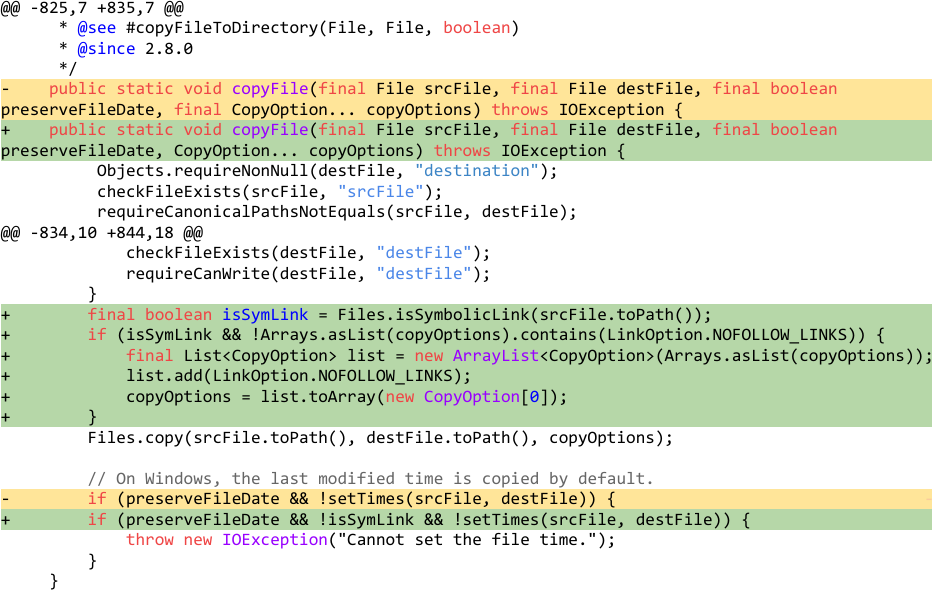}
            \vspace{2pt}
            \textit{\footnotesize\firstrevision{(a) Bug-inducing changes introduced by a developer in \textit{RegressionBug-99}.}}
            \label{fig:bic-99}
        \end{minipage}%
        \hspace{3pt}
        \begin{minipage}[t]{0.58\textwidth}
            \centering
            \vspace*{-0.45\linewidth}
            \includegraphics[width=\linewidth]{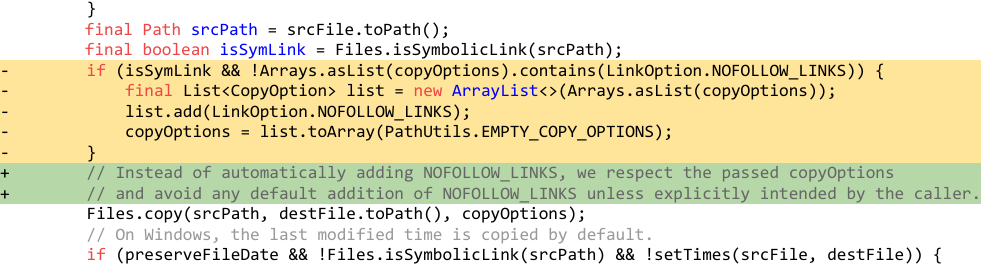}
            \vspace{2pt}
            \textit{\footnotesize\firstrevision{(b) Correct patch generated by ChatGPT-4o via conversational prompting augmented with full BIC information.}}
            \label{fig:bfc-99}
        \end{minipage}%
    }
    \caption{Illustration of the \textit{Partially Revert to Previous Statement} repair operator applied to \textit{RegressionBug-99}: (a) the bug-inducing changes and (b) the corresponding correct patch generated by ChatGPT-4o under the conversational configuration augmented with full bug-inducing change information.}
    
    \label{fig:regressionbug-99}
\end{figure}

\vspace{-0.2cm}
\subsubsection{\textbf{Observation 2:} Partially reverting edits guided by bug-inducing change information}
Beyond full reversion, partially reverting selected code fragments also proves effective in cases where only specific statements introduced in the bug-inducing commit need to be undone. This strategy allows the model to preserve valid modifications while selectively removing only the changes responsible for the regression.
For example, Figure~\ref{fig:regressionbug-99} presents an illustration of the \textit{Partially Revert to Previous Statement} repair operator in \textit{RegressionBug-99}, as generated by ChatGPT-4o under the conversation configuration enhanced with bug-inducing change information. In this case, the developers changed the behavior of \texttt{copyFile} to copy symbolic links itself (i.e., the path to the target), rather than dereferencing the link and copying the actual target file. 
This change reversed the default behavior expected by many systems and broke existing code that relied on the original semantics.
As discussed in the commit history\footnote{Available at: \url{https://github.com/apache/commons-io/commit/ec4144b01b4107d6b39f5f4d784cf05217ea4dfa}}, the developers chose to revert only the code responsible for the bug, while retaining the condition \textit{"if (preserveFileDate \&\& !isSymLink \&\& !setTimes(srcFile, destFile)) \{"}. The retained code does not change behavior, it simply avoids setting file timestamps on symbolic links to avoids errors related to modifying \textit{symlink} metadata, which is often OS-dependent and error-prone. This careful partial reversion ensures correctness without discarding unrelated, valid improvements.

\begin{figure}[!htbp]
\hspace*{-0.06\columnwidth}
\includegraphics[width=1.1\columnwidth]{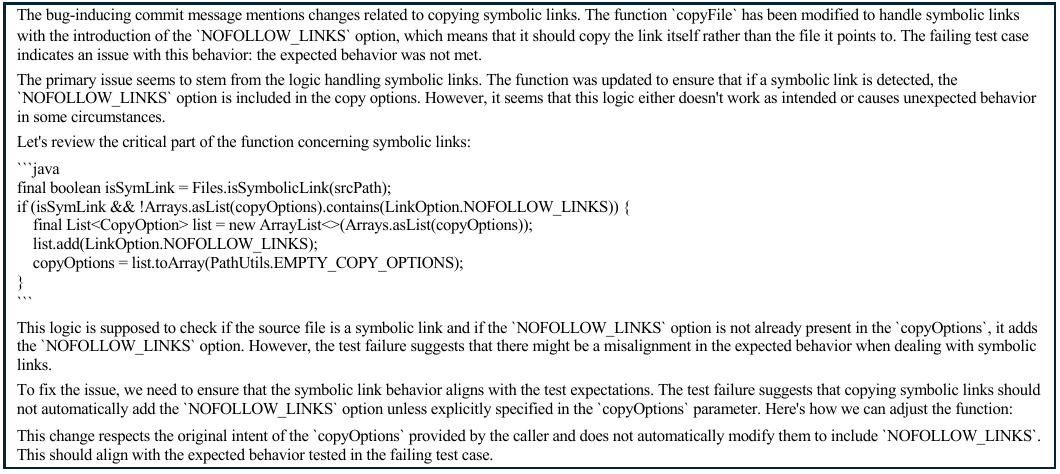}
\caption{Excerpt from ChatGPT-4o’s response under the conversation configuration augmented with full bug-inducing change information for repairing \textit{RegressionBug-99}.}
\vspace{-0.2cm}
\label{fig:response-99}
\end{figure}

Figure~\ref{fig:response-99} presents an excerpt from ChatGPT-4o’s reasoning under conversation configuration with BIC context for \textit{RegressionBug-99}. Specifically, the model explains that the regression stems from logic that forcibly adds the \texttt{NOFOLLOW\_LINKS} option when a symbolic link is detected, thereby overriding the caller’s intended behavior. It correctly identifies that this automatic injection violates expectations, particularly in cases where callers rely on the default behavior of dereferencing symbolic links. Guided by the bug-inducing change information, the model proposes reverting the conditional block responsible for modifying the copy options, while preserving the surrounding logic, including the check that skips timestamp setting for symbolic links. This patch aligns with the human-written fix, as it reverts only the faulty logic introduced in the commit while retaining valid behavior.

\subsubsection{\textbf{Observation 3:} Fully or Partially reverting edits could generate the plausible but incorrect patches}

Apart from \textit{Observation 1} and \textit{2}, our patch validation process revealed recurring patterns, notably \textit{fully or partially reverting to previously correct statements}, that lead to patches which pass the test suite but are semantically incorrect (i.e., plausible but incorrect patches). These patterns frequently appear in patches generated by techniques that leverage bug-inducing change information.
While these patterns often produce plausible patches that satisfy the test suite, they may reflect an over-reliance on syntactic rollback strategies and suggest that the model lacks a deeper understanding of the contextual semantics required to correctly fix the underlying bug.

\begin{figure}[htbp]
    \centering
    \begin{minipage}[t]{0.493\linewidth}  
        \centering
        \includegraphics[width=\linewidth]{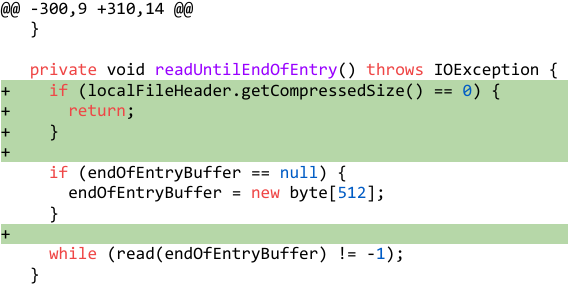}
        \vspace{2pt}
        \textit{\footnotesize\firstrevision{(a) Bug-inducing changes introduced by a developer \\ in \textit{RegressionBug-2}.}}
        \label{fig:bic-2}
    \end{minipage}
    \hfill
    \begin{minipage}[t]{0.493\linewidth}
        \centering
        \includegraphics[width=\linewidth]{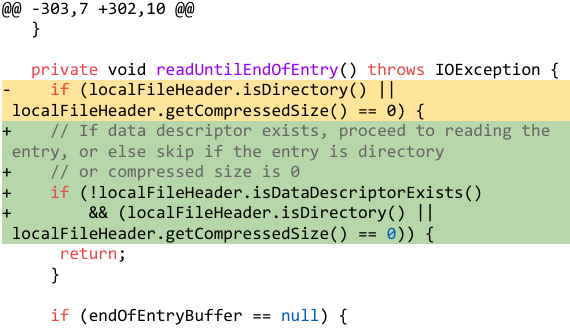}
        \vspace{2pt}
        \textit{\footnotesize\firstrevision{(b) Ground-truth bug-fixing changes manually written by a developer \\ in \textit{RegressionBug-2}.}}
        \label{fig:bfc-2}
    \end{minipage}
    \caption{The bug-inducing and bug-fixing changes in \textit{RegressionBug-2} from \textsc{RegressionBug4APR} benchmark.}
    \label{fig:regressionbug-2}
\end{figure}

\begin{figure}[!htbp]
\hspace*{-0.08\columnwidth}
\includegraphics[width=1.12\columnwidth]{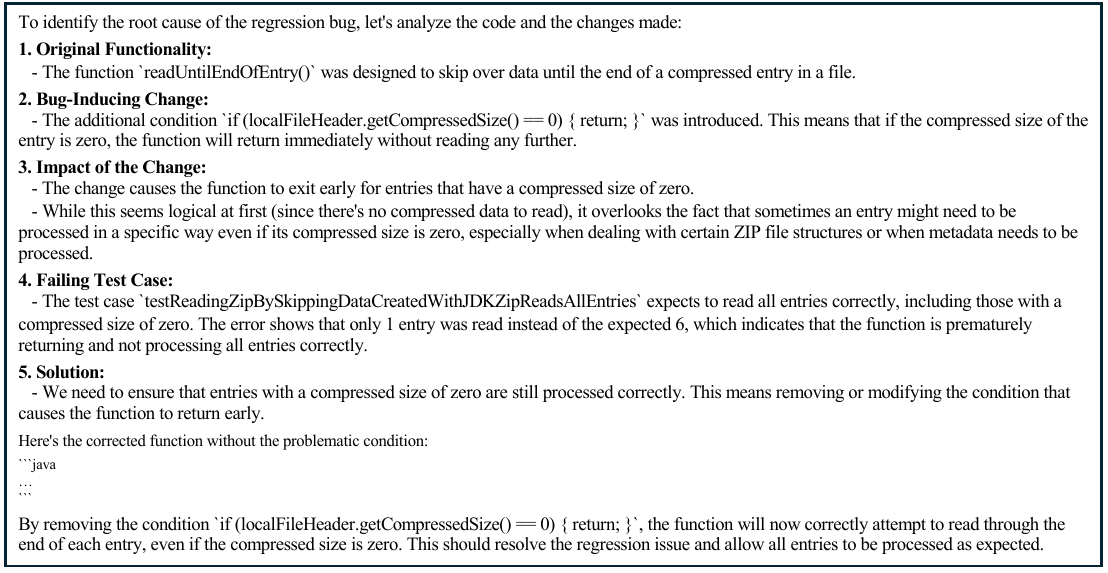}
\caption{Excerpt from ChatGPT-4o’s response under the conversation configuration augmented with full bug-inducing change information for repairing \textit{RegressionBug-2}.}
\vspace{-0.2cm}
\label{fig:response-2}
\end{figure}

To illustrate this behavior, we examine \textit{RegressionBug-2} alongside the response from ChatGPT-4o under zero-shot prompting configuration augmented with BIC context, as shown in Figures~\ref{fig:regressionbug-2} and~\ref{fig:response-2}.
In this case, the regression-inducing changes were introduced to prevent read errors when handling JAR files using ZIP4J~\footnote{Available at: \url{https://github.com/srikanth-lingala/zip4j/issues/194}}, but they unintentionally triggered a new bug.
The issue occurred when only the first entry in a ZIP archive could be read, and subsequent entries were skipped. The regression was caused by logic that returned early if an entry's compressed size was zero (Figure~\ref{fig:regressionbug-2}a). However, according to the ZIP specification, this is valid behavior when the third bit of the general purpose flag is set. This bit indicates that the actual compressed size is not available in the local file header but will instead be provided later in a separate data descriptor.
To fix this, the condition was refined to be more precise: entries are now skipped only if they do not use a data descriptor and are either directories or have a compressed size of zero (Figure~\ref{fig:regressionbug-2}b). This update preserves the original safeguard against invalid entries while allowing valid entries that rely on the data descriptor mechanism to be processed correctly.

The ChatGPT model correctly identified the root cause of the bug: the change caused valid ZIP entries to be skipped when their compressed size was zero, resulting in only one out of six entries being read. This occurred because the code returned early without properly handling ZIP structures that rely on data descriptors or contain essential metadata (as seen in parts 3 and 4 of the excerpt in Figure~\ref{fig:response-2}).
However, the model proposed removing the condition entirely, which would allow entries to be read even if the compressed size is zero. While this may avoid the regression error, it overlooks the original intent of the change.

\begin{figure}[!htbp]
    \centerline{\includegraphics[width=0.65\columnwidth]{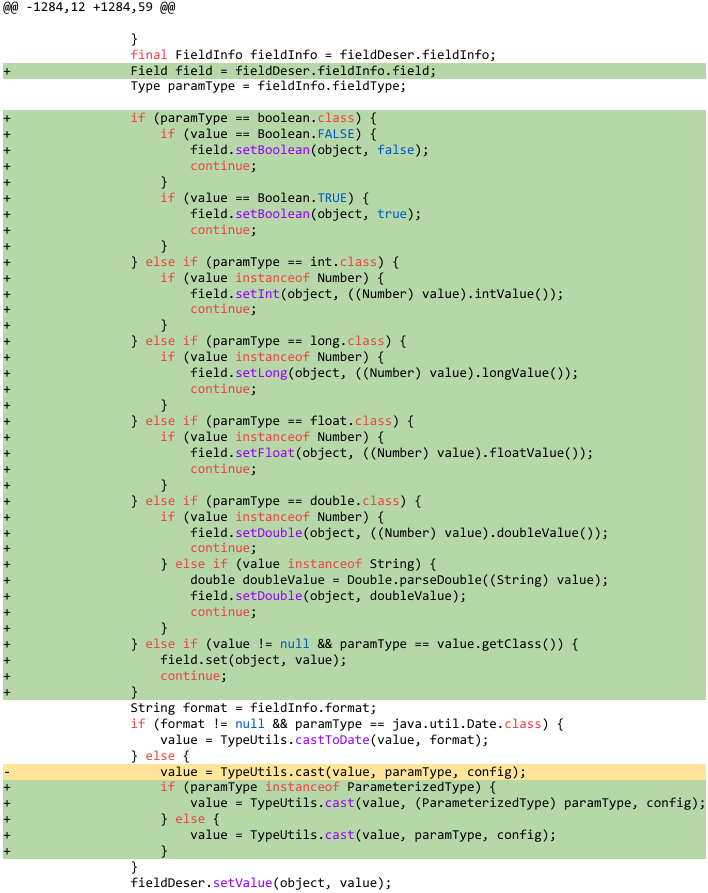}}
    \vspace{-0.1cm}
    \caption{\firstrevision{Bug-inducing changes introduced by a developer in \textit{RegressionBug-69} from \textsc{RegressionBug4APR} benchmark.}}
    \label{fig:bic-69}
\end{figure}

\subsubsection{\textbf{Observation 4:} Combining test case feedback with bug-inducing change information to narrow the fix scope}

In some cases, when the model correctly understands the context of bug-inducing changes, such as feature additions or prior bug fixes, it avoids fully or partially reverting to a previously working version. Instead, it analyzes the current buggy function while simultaneously reasoning about the root cause of the regression. This combined reasoning enables more accurate fault localization, as the bugs happens after the relevant change is introduced, making the bug-inducing changes a strong signal for identifying the issue. This, in turn, supports the generation of targeted and minimal patches.
This behavior is observed in \textit{RegressionBug-30} and \textit{RegressionBug-69}, where failing test cases provide useful information the bugs, while the bug-inducing changes provide complementary context that helps the model localize the root cause. By leveraging both sources of information, the model is able to narrow the fix scope and avoid unnecessary or overly broad modifications.

To illustrate this behavior, we present \textit{RegressionBug-69}, where the developers attempted to improve the performance of \texttt{JSONObject.toJavaObject} by introducing optimizations~\footnote{Available at: \url{https://github.com/alibaba/fastjson/commit/7b416faa1015ce04505e88d1f9b0575bcf13657f}}, as shown in Figure~\ref{fig:bic-69}. However, the modification introduced a regression due to a mismatch between the field type and the expected parameter type. Specifically, the checks used primitive types, while the actual \texttt{paramType} values were boxed class types. As a result, the new conditions failed to trigger as intended, leading to incorrect behavior during deserialization.

\begin{figure}[htbp]
    \hspace*{-0.01\columnwidth}
    \centering
    \makebox[\textwidth][c]{%
        \begin{minipage}[t]{0.65\textwidth}
            \centering
            \includegraphics[width=\linewidth]{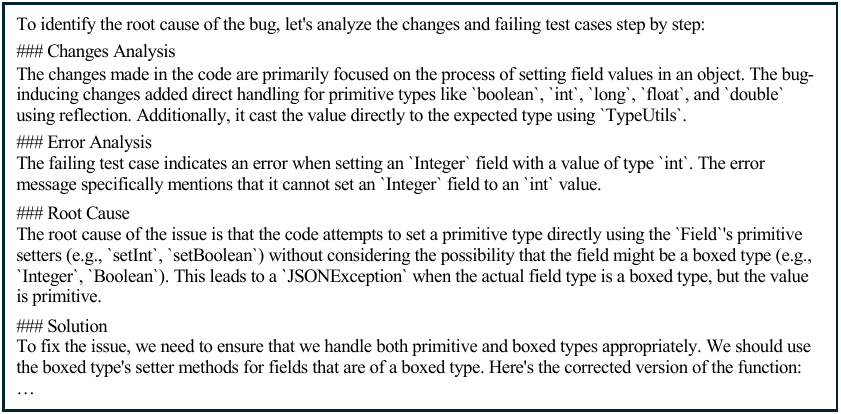}
            \textit{\footnotesize(a)}
        \end{minipage}%
        \hspace{2pt}
        \begin{minipage}[t]{0.47\textwidth}
            \centering
            \vspace*{-0.45\linewidth}
            \includegraphics[width=\linewidth]{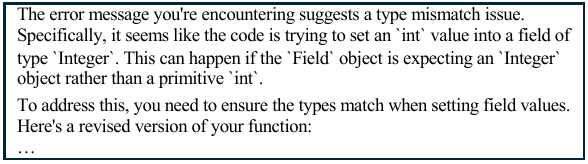}
            \textit{\footnotesize(b)}
            \label{fig:bfc-2}
        \end{minipage}%
    }
    \vspace{-0.2cm}
    \caption{Excerpts from ChatGPT-4o's response under the conversational configuration for repairing \textit{RegressionBug-69}: (a) with bug-inducing change information and (b) without BIC information.}
    \label{fig:response-69}
\end{figure}

ChatGPT-4o, under both configurations, with and without bug-inducing change (BIC) information, was able to identify the type mismatch as the root cause of the failure by analyzing the test case’s error message. However, the version augmented with BIC context demonstrated a deeper understanding of the problem and provided a clearer explanation, as shown in Figures~\ref{fig:response-69}a and~\ref{fig:response-69}b.

\begin{figure}[htbp]
    \hspace*{0.02\columnwidth}
    \centering
    \makebox[\textwidth][c]{%
        \begin{minipage}[t]{0.55\textwidth}
            \centering
            \includegraphics[width=\linewidth]{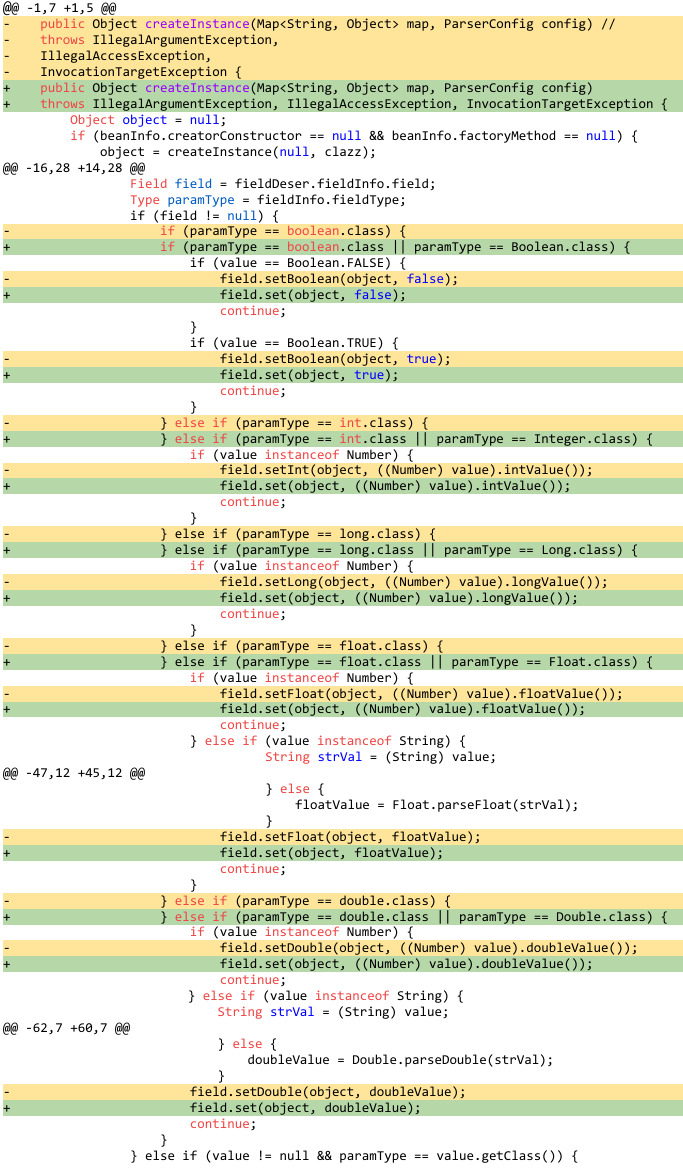}
            \textit{\footnotesize\firstrevision{(a) Patch generated by ChatGPT-4o under the conversation configuration augmented \textbf{with} bug-inducing change information.}}
        \end{minipage}%
        \hspace{10pt}%
        \begin{minipage}[t]{0.55\textwidth}
            \centering
            \vspace*{-1.2\linewidth}
            \includegraphics[width=\linewidth]{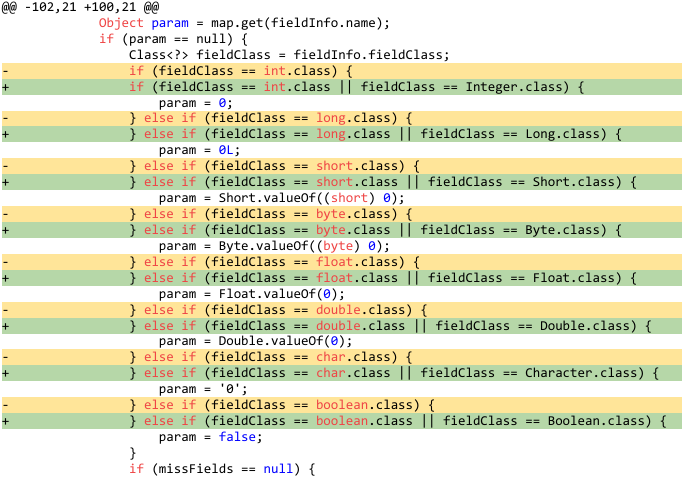}
            \textit{\footnotesize\firstrevision{(b) Patch generated by ChatGPT-4o under the conversation configuration \textbf{without} bug-inducing change information. This patch includes additional edits not present in (a), while other parts remain similar to (a).}}
            \label{fig:bfc-69}
        \end{minipage}%
    }
    \vspace{-0.1cm}
    \caption{Patches generated by ChatGPT-4o for \textit{RegressionBug-69} under different configurations.}
    \label{fig:regressionbug-69}
\end{figure}

More notably, when comparing the patches generated under both settings in Figures~\ref{fig:regressionbug-69}a and~\ref{fig:regressionbug-69}b, both configurations produced a correct fix at the appropriate location. Nevertheless, the patch generated without BIC context included unnecessary edits to structurally similar but semantically unrelated code. Although the resulting patch passed the test suite, the added changes introduced redundancy. This illustrates how the BIC context helps the model focus on the minimal fix necessary to resolve the regression without introducing superfluous modifications.

\subsubsection{\textbf{Others:}}
Apart from the specific cases discussed earlier, we also observed cases such as \textit{RegressionBug-59} and \textit{RegressionBug-65}, in which the bug-inducing change information contributed to the patch generation or model reasoning to varying degrees. In these cases, the benefit of incorporating regression-specific context was less evident.

Specifically, in \textit{RegressionBug-59}, the bug-inducing change introduced an entirely new method, \texttt{convexHull()}, intended to fix a prior issue. Unfortunately, the regression occurred within the newly added code itself. The accompanying commit message, \textit{"Stolstov/issue 172 (\#174)"}, provided little semantic guidance, making it difficult for the model to infer the intended behavior or reasoning behind the change. Under the conversational configuration, ChatGPT-Turbo-3.5 required 8 attempts and 3 rounds of interaction to produce a plausible patch. In the initial round, the model recognized the problem through the test failure message, which included a \texttt{java.lang.NullPointerException} and noted that the \texttt{convexHull()} method failed to handle certain geometry types correctly. Despite this, the bug-inducing change context offered limited additional support, and a valid patch was only produced after two additional rounds of interaction driven by test case feedback.

\begin{figure}[!htbp]
\includegraphics[width=0.9\columnwidth]{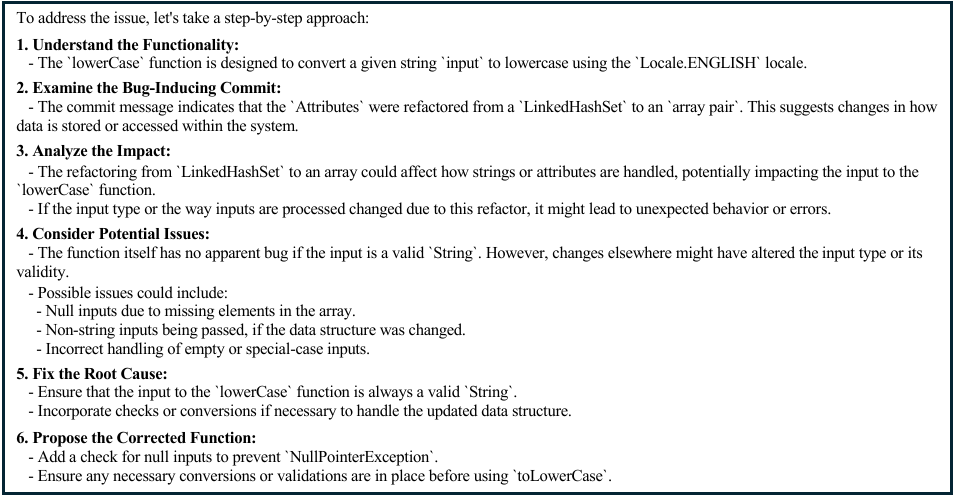}
\caption{Excerpt from ChatGPT-4o’s response under zero-shot prompting configuration augmented with bug-inducing change information for repairing \textit{RegressionBug-65}.}
\vspace{-0.2cm}
\label{fig:response-65}
\end{figure}

In \textit{RegressionBug-65}, this case represents a \textit{remote} regression bug, where code modification introduces a bug in an unchanged program element, leading to failures elsewhere in the code. The bug-inducing change information in this case includes only a vague commit message: \textit{"Refactored Attributes to be an array pair vs LinkedHashSet. Also a couple perf (cpu / garbage) tweaks"}~\footnote{Available at: \url{https://github.com/jhy/jsoup/commit/ea1fb65e9ff8eee82c4e379dc3236d09a5ab02e1}}. Based on this context, the model inferred that the \texttt{Attributes} class had been refactored from \texttt{LinkedHashSet} to an array-based structure, implying a change in how attribute data is stored or accessed (\textit{Point 2 in Figure~\ref{fig:response-65}}).
Despite the limited context, the model, guided by both the commit message and the test failure, hypothesized potential causes such as incorrect data types or null values handling (\textit{Point 4 in Figure~\ref{fig:response-65}}). It then correctly diagnosed the issue and generated an appropriate fix. This case demonstrates that even minimal bug-inducing context, when interpreted in conjunction with test case feedback, can help the model understand the underlying cause and produce a correct patch.

\begin{insightbox}
\textit{Qualitative analysis further reveals that BIC context helps the model more effectively identify the root cause of regressions and decide whether a full or partial reversion to prior code is appropriate. However, such reversion must be applied judiciously, guided by a proper understanding of the bug context. Additionally, BIC context helps narrow the fix scope, enhancing fault localization and reducing unnecessary modifications.}
\end{insightbox}

\subsection{\firstrevision{A Systematic Classification of LLM Repair Failures}}
\firstrevision{To gain a deeper understanding of the common issues and patterns found in LLM repair failures, we selected the best-performing technique in our study, specifically conversational APR with full bug-inducing change information using ChatGPT-4o model, and conducted a qualitative analysis of failure cases using open card-sorting~\cite{spencer2009card}. Open card-sorting has been widely used in prior software engineering studies~\cite{lo2015practitioners, liu2024refining, le2025towards} to generate categories or taxonomies from empirical data.}

\firstrevision{\textbf{Study Design.} We selected a representative random sample of 59 bug instances (approximately 30\% of the 200 bug instances) drawn from four groups: \textit{(i)} non-plausible patches, \textit{(ii)} plausible but incorrect patches with 1 conversation iteration, \textit{(iii)} plausible but incorrect patches with 2--3 iterations, and \textit{(iv)} plausible but incorrect patches with 4--5 iterations. To maintain representativeness, we preserved the same language ratio as the extended benchmark (1 Python : 3 Java), resulting in 16 Python and 43 Java instances. We note that group (iv) contained only 7 Java instances due to the limited number of bugs in this category. Each instance was documented as a card containing the bug ID, the number of repair attempts and conversation rounds, and the full repair log, including the failure type, the LLM response, and the generated patch.}
\firstrevision{We then applied open card-sorting to these cards through two phases. In the first phase (\textit{independent sorting}), two authors independently reviewed all 59 cards and assigned a descriptive name or theme to each based on the nature of the failure observed. In the second phase (\textit{discussion and consolidation}), the two authors compared their labellings, discussed disagreements, and converged on a final label for each card, with one author serving as a moderator. Category names were derived from patterns observed consistently across the groups.}

\firstrevision{\textbf{Results.} To validate the reliability of our classification, we measured inter-rater agreement using Cohen's Kappa~\cite{cohen1960coefficient}, obtaining a value of $\kappa=0.69$, which reflects substantial agreement. All disagreements were resolved through discussion. We observed that disagreements were most frequent in groups \textit{(i) non-plausible patches} and \textit{(iv) plausible but incorrect patches requiring 4 - 5 iterations}, where authors needed to trace through multiple conversation rounds to identify the true root cause of failure, making classification inherently more effortful and subjective. Through this analysis, we identified four categories of LLM repair failures:}

\firstrevision{\textit{(i) Wrong Diagnosis (44.07\%).} The LLM misidentifies the root cause of the bug, leading to patches that target the wrong code location or apply incorrect fix logic. This failure arises from several causes, including limited reasoning capability, misdirection from test case names or error signals, or incorrect application of reversion strategies guided by the bug-inducing information. For example, in \textit{RegressionBug-24~\footnote{Available at: \url{https://brojackvn.github.io/RegressionBug4APR-Homepage/\#!/bug/RegressionBug-24/RegressionBug-24}}}, the correct fix requires replacing \texttt{this.output.close()} with \texttt{this.output.flush()}. However, the test case is named \texttt{leftInputUnclosed()}, which misdirected the LLM toward \texttt{this.input.close()}. The LLM either removed this line or added a conditional check around it. As a result, it missed the true root cause entirely.}

\firstrevision{\textit{(ii) Weak Test Suite (23.73\%):} The patch passes all provided tests but is semantically incorrect (over-fitting), indicating the test suite is insufficient to distinguish the true fix. This can arise due to insufficient coverage of boundary conditions or alternate code paths, where the LLM exploits test gaps by simplifying logic rather than correctly addressing the underlying fault. For example, in \textit{RegressionBug-150~\footnote{Available at: \url{https://brojackvn.github.io/RegressionBug4APR-Homepage/\#!/bug/RegressionBug-150/RegressionBug-150}}}, the bug spans two locations in \texttt{getMinConcordantFragmentSize()}: the stream aggregation uses \texttt{.max()} instead of \texttt{.min()}, and inside the \texttt{if (getFile().exists())} block, \texttt{Math.max(super.} \texttt{.getMaxConcordantFragmentSize(), ...)} should be \texttt{Math.min(super.getMinConcordantFragmentSize(), ...)}. The correct fix addresses both locations with targeted one-line changes while preserving the block. However, the LLM correctly fixes the stream aggregation but removes the entire \texttt{if (getFile().exists())} block rather than correcting the logic within it. Because the test does not set up an existing assembly file, \texttt{getFile().exists()} always returns false and the deleted block is never entered, so both the correct fix and the over-removal return the same value. The critical behavioral difference, i.e., what happens when the assembly file exists, is never exercised by the test suite, allowing the structurally incorrect patch to pass all tests and appear valid.}

\firstrevision{\textit{(iii) Unreachable Fix (20.34\%):} The correct fix requires information that is neither visible to the LLM from the method body alone nor derivable from the error signal, such as domain-specific constants or internal method identifiers, making it structurally impossible to generate the correct patch from the available context. For example, in \textit{RegressionBug-122~\footnote{Available at: \url{https://brojackvn.github.io/RegressionBug4APR-Homepage/\#!/bug/RegressionBug-122/RegressionBug-122}}}, the correct fix inserts additional static field and enum constant discovery logic into the \texttt{FIELD} case of \texttt{getPropertyGet()} using a framework-internal API. The LLM correctly localizes the problem to the \texttt{FIELD} case and understands that static fields and enum constants require special handling; however, it lacks the specific framework knowledge necessary to implement the fix. It does not know which existing class handles static field introspection, what the complete \texttt{JexlPropertyGet} interface contract requires, or that the project enforces Java~1.6 source compatibility. As a result, the LLM hallucinated non-existent classes such as \texttt{EnumGetExecutor} and \texttt{StaticFieldGetExecutor}, and its attempts to use lambda expressions were rejected by the compiler due to the version constraint, producing compilation errors across nearly all 50 repair attempts and yielding no successful patch. All the information needed to generate the correct fix exists in the codebase, but none of it is visible from the provided method body alone.}

\begin{figure}[!htbp]
\centerline{\includegraphics[width=0.68\columnwidth]{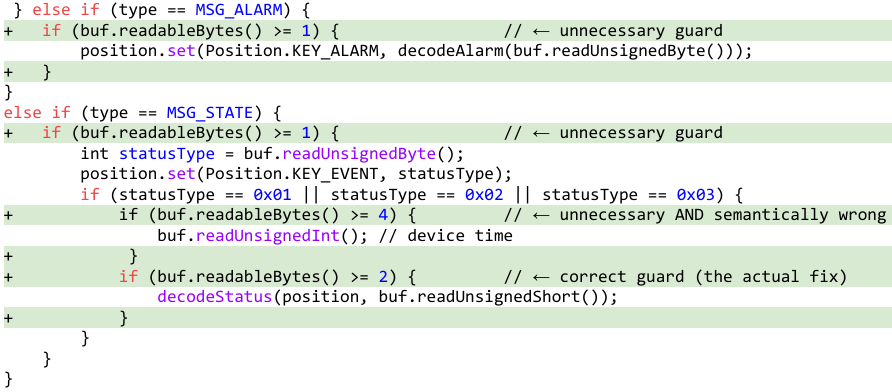}}
\vspace{-0.2cm}
\caption{\firstrevision{LLM-generated patch for \textit{RegressionBug-101}, illustrating the over-repair failure mode where six guards are added across four read operations instead of the single required guard.}}
\label{fig:rb101-llm-patch}
\end{figure}

\firstrevision{\textit{(iv) Over-repair (11.86\%)}: The LLM makes excessive modifications beyond what is needed, removing or restructuring correct code when only a minimal targeted change was required. The following example, in \textit{RegressionBug-101~\footnote{Available at: \url{https://brojackvn.github.io/RegressionBug4APR-Homepage/\#!/bug/RegressionBug-101/RegressionBug-101}}}, the ground-truth fix consists of a single readable-bytes guard inserted immediately before the \texttt{decodeStatus} call inside the \texttt{MSG\_STATE} branch, ensuring that at least 2 bytes remain in the buffer before the status short is read. Instead, as illustrated in Figure~\ref{fig:rb101-llm-patch}, the LLM added six guards across four separate read operations throughout the entire method. Three of these guards are unnecessary, as the \texttt{MSG\_ALARM} alarm byte and the \texttt{MSG\_STATE} status-type byte are always present in valid messages and do not require guarding. Worse, the LLM also wrapped the \texttt{buf.readUnsignedInt()} device-time read in a guard. This introduces a new semantic error: when 2–-3 bytes remain after the status-type byte, the guard skips the device-time read without consuming those bytes, so the status read then consumes device-time bytes as status data and produces a wrong position record. The test suite still passes because no test covers this partial-data scenario.}

\vspace{-0.5cm}
\begin{insightbox}
\firstrevision{\textit{Our analysis identifies four categories of LLM repair failures: \textit{Wrong Diagnosis} (44.07\%), \textit{Weak Test Suite} (23.73\%), \textit{Unreachable Fix} (20.34\%), and \textit{Over-repair} (11.86\%). Wrong Diagnosis is the most prevalent failure mode, arising from limited reasoning capability or misdirection from error signals. These findings suggest that future repair approaches should focus on improving root cause reasoning, strengthening test suite quality, and incorporating broader contextual knowledge beyond the method body.}}
\end{insightbox}

\begin{rqbox}{\textbf{RQ3}}{\textbf{Impact of bug-inducing change information on APR}}
\textit{Our experimental results demonstrate that incorporating bug-inducing change (BIC) information significantly enhances the effectiveness of prompt-based APR techniques in repairing regression errors. On the Java benchmark, ChatGPT-4o under the conversational configuration with BIC context achieves the highest number of correct repairs, successfully fixing 29 out of 150 bugs (19.33\%). \firstrevision{On the Python benchmark, the same configuration achieves 10 out of 50 correct repairs (20.00\%), consistent with the trends observed in Java and supporting the generalizability of our findings across programming languages. In total, across both benchmarks, this configuration correctly repairs 39 out of 200 bugs (19.50\%), representing a 1.6× improvement over the best-performing configuration without BIC.}}
\end{rqbox}
\section{\firstrevision{RQ4: Ablation Study}}
\label{sec:rq-4}

\firstrevision{Building on the findings of RQ3, we conduct an ablation study to isolate and evaluate the individual contribution of each contextual element within the bug-inducing change (BIC) information. We dissect the best-performing configuration from RQ3, specifically conversational ChatGPT-4o, as the base model. We evaluate the following four configurations:} \firstrevision{(i) \textit{w/o BIC}, where no additional context is provided beyond the buggy program and failing tests;}
\firstrevision{(ii) \textit{Commit message only}, where only the commit message of the bug-inducing commit is appended to the prompt;}
\firstrevision{(iii) \textit{Code changes only:}, where only the code changes to the buggy function and the list of changed file names are included, without the commit message; and}
\firstrevision{(iv) \textit{Full BIC:}, where the complete bug-inducing change information is provided, combining both the commit message and the code-level changes described in (ii) and (iii).}

\begin{table*}[!htbp]
\centering
\fontsize{8.5pt}{10pt}\selectfont
\caption{Ablation Study Results on Java regression bugs of the \textsc{RegressionBug4APR} benchmark}
\label{table:ablation-study-java}
\begin{tabular}{>{\raggedright\arraybackslash}p{5.2cm} c c c}
\Xhline{0.7pt}
\textbf{Configurations} 
& \textbf{\#Plausible Patches} & \textbf{\#Correct Patches} & \multirow{2}{*}{\textbf{Precision (\%)}} \\
\textbf{\footnotesize(Conversational ChatGPT-4o as a base strategy)}& \textbf{(Plausible Rate \%)} & \textbf{(Correct Rate \%)} \\
\Xhline{1pt}

\multirow{2}{*}{\makecell[l]{\textbf{w/o BIC}}} & 60 & 18 & \multirow{2}{*}{30.00\%}      \\
 & \textit{(40.00\%)} & \textit{(12.00\%)} & \\ \cline{1-4}
  
\multirow{2}{*}{\makecell[l]{\textbf{+ Commit message}}} & 67 & 21 & \multirow{2}{*}{31.34\%} \\
& \textit{(44.67\%)} & \textit{(14.00\%)} & \\ \cline{1-4}

\multirow{2}{*}{\makecell[l]{\textbf{+ Code changes}}} & 70 & 25 & \multirow{2}{*}{35.71\%} \\
& \textit{(46.67\%)} & \textit{(16.67\%)} & \\ \cline{1-4}

\multirow{2}{*}{\makecell[l]{\textbf{Full BIC}}} & 72 & 29 & \multirow{2}{*}{40.28\%} \\
& \textit{(48.00\%)} & \textit{(19.33\%)} & \\ \cline{1-4}

\Xhline{1pt}
\end{tabular}
\end{table*}

\firstrevision{Table~\ref{table:ablation-study-java} presents the ablation study results on the Java benchmark. Our results reveal several key insights. First, each contextual component contributes positively and incrementally to repair performance, confirming that the improvements observed in RQ3 are not solely attributable to a single element. Second, code-level change information consistently provides a larger benefit than the commit message alone: adding code changes yields 25 correct patches (35.71\% precision) compared to 21 (31.34\% precision) for commit message only, both representing meaningful gains over the w/o BIC baseline of 18 correct patches (30.00\% precision). This suggests that concrete code context plays a more decisive role in guiding the model toward correct patches than natural-language commit descriptions. Third, the two components are complementary: combining both yields the best outcome of 29 correct patches (40.28\% precision), an improvement of 11 correct patches over the w/o BIC baseline, demonstrating that commit messages and code changes capture different, non-redundant signals. The commit message conveys the high-level intent of the change, while the code diff provides structural and syntactic grounding that narrows the hypothesis space for the model.}

\begin{table*}[!htbp]
\centering
\fontsize{8.5pt}{10pt}\selectfont
\caption{Ablation Study Results on Python regression bugs of the \textsc{RegressionBug4APR} benchmark.}
\label{table:ablation-study-python}
\begin{tabular}{>{\raggedright\arraybackslash}p{5.2cm} c c c}
\Xhline{0.7pt}
\textbf{Configurations} 
& \textbf{\#Plausible Patches} & \textbf{\#Correct Patches} & \multirow{2}{*}{\textbf{Precision (\%)}} \\
\textbf{\footnotesize(Conversational ChatGPT-4o as a base strategy)}& \textbf{(Plausible Rate \%)} & \textbf{(Correct Rate \%)} \\
\Xhline{1pt}

\multirow{2}{*}{\makecell[l]{\textbf{w/o BIC}}} & 29 & 6 & \multirow{2}{*}{20.69\%}      \\
 & \textit{(58.00\%)} & \textit{(12.00\%)} & \\ \cline{1-4}
  
\multirow{2}{*}{\makecell[l]{\textbf{+ Commit message}}} & 33 & 7 & \multirow{2}{*}{21.21\%} \\
& \textit{(66.00\%)} & \textit{(14.00\%)} & \\ \cline{1-4}

\multirow{2}{*}{\makecell[l]{\textbf{+ Code changes}}} & 32 & 9 & \multirow{2}{*}{28.12\%} \\
& \textit{(64.00\%)} & \textit{(18.00\%)} & \\ \cline{1-4}

\multirow{2}{*}{\makecell[l]{\textbf{Full BIC}}} & 35 & 10 & \multirow{2}{*}{28.57\%} \\
& \textit{(70.00\%)} & \textit{(20.00\%)} & \\ \cline{1-4}

\Xhline{1pt}
\end{tabular}
\end{table*}

\firstrevision{Table~\ref{table:ablation-study-python} presents the ablation study results on the Python benchmark. The results are largely consistent with the trends observed in the Java benchmark, providing further evidence for the generalizability of our findings. Adding code changes yields 9 correct patches (28.12\% precision) compared to 7 (21.21\% precision) for commit message only, both representing meaningful gains over the w/o BIC baseline of 6 correct patches (20.69\% precision). The full BIC combination yields the best outcome of 10 correct patches (28.57\% precision), an improvement of 4 correct patches over the w/o BIC baseline, suggesting that the two contextual components are complementary across programming languages.}

\begin{rqbox}{\textbf{RQ4}}{\textbf{Ablation Study}}
\firstrevision{\textit{Our ablation study reveals that each contextual component within the bug-inducing change information contributes positively and incrementally to repair performance. Code-level changes consistently provide a larger gain than commit messages alone across both benchmarks. On the \textsc{RegressionBug4APR} benchmark (200 bug instances across Java and Python), the full BIC combination yields the best overall outcome of 39 correct patches (36.45\% precision), compared to 34 correct patches (33.33\% precision) for code changes only, 28 correct patches (28.00\% precision) for commit message only, and 24 correct patches (26.97\% precision) for the w/o BIC baseline. These results suggest that commit messages and code changes capture complementary, non-redundant signals that together enhance repair effectiveness.}}
\end{rqbox}

\section{Discussion}
\label{sec:discussion}

\subsection{Lesson Learned}
In this subsection, we highlight key lessons learned through our experiments and analysis that can drive future research in the field.

\textbf{Enhanced APR Workflow with Bug-inducing change Information:}  
Our empirical results highlight the significant impact of incorporating regression-inducing context on repair effectiveness. Conventionally, APR workflows operate by taking a buggy program and test suite, and attempting to generate a minimal change that passes all test cases. However, our findings suggest that enriching this workflow with additional context, specifically by detecting whether a bug is regression-related and extracting the corresponding bug-inducing commit, can substantially enhance repair outcomes. This extension is particularly beneficial in the context of continuous integration (CI) systems, where regression bugs emerge frequently and need to be addressed efficiently at scale. \firstrevision{To further illustrate this applicability, we present concrete CI use cases in Section~\ref{subsec:ci-use-case}, demonstrating how our approach can be integrated into real-world CI pipelines and showing its effectiveness on two case studies of regression bugs detected through CI systems.}

\textbf{Context-Aware Use of Reversion Operators:}
Our qualitative analysis reveals several cases where \textit{fully or partially reverting to previously correct statements} operators result in plausible but incorrect patches. These findings suggest that while reversion-based repair operators can be effective patterns, their successful application requires a proper understanding of the bug context. Blindly applying such operators, without considering the underlying cause of the regression, can lead to overfitting and semantically incorrect fixes. This also suggests that template-based APR techniques must understand the context to determine whether reversion patterns are appropriate.

\subsection{\firstrevision{Regression Repair Use Cases In CI Settings}}
\label{subsec:ci-use-case}
\firstrevision{In a typical continuous integration (CI) pipeline, e.g., Jenkins~\footnote{Available at: \url{https://www.jenkins.io}} or GitHub Actions~\footnote{Available at: \url{https://github.com/features/actions}}, every code commit triggers an automated build and regression test suite execution. When a regression is detected, i.e., a previously passing test begins to fail after a recent commit, developers must manually investigate the root cause and implement a fix, which is both time-consuming and disrupts the development workflow. In this setting, our context-aware APR workflow can be naturally integrated as an additional stage in the CI pipeline. Once the bug-inducing commit is identified, e.g., via automated tools such as SZZ, git bisect or RegMiner, the corresponding code changes, commit messages and failing regression tests can be extracted and provided to our repair tool. Our tool then generates candidate patches that explicitly account for the regression-inducing modification. These candidate patches can be automatically validated within the same CI pipeline by re-running the test suite. If a patch successfully restores the failing test while preserving previously passing behavior, it can be presented to developers as a repair suggestion (e.g., as part of CI feedback or a pull request comment). This enables a more automated and efficient regression repair process, reducing manual debugging effort and providing timely feedback during continuous development.}

\firstrevision{To further validate the practical applicability of our approach, we study two real-world regression bugs detected through CI pipelines:}

\firstrevision{\noindent\textbf{Case 1: Vectorz~\footnote{Available at: \url{https://github.com/mikera/vectorz}}} (bug-fixing commit \textit{cc07d6f6cc}, bug-inducing commit \textit{3eee45ec7a}): Vectorz is an open-source Java library for high-performance vector and matrix arithmetic. The bug-inducing commit (\textit{``extra index checks on selectClone''}) introduced a regression in the \texttt{selectClone} method of \texttt{AVector.java} by incorrectly writing \texttt{checkIndex(i)}, which validates the loop counter, instead of \texttt{checkIndex(ix)}, which should validate the caller-provided index. This caused the \texttt{testSelect} regression test to fail with an \texttt{IndexOutOfBoundsException} for valid inputs. Both conversational APR with ChatGPT-4o with and without BIC information successfully generated the correct patch, i.e., replacing \texttt{checkIndex(i)} with \texttt{checkIndex(ix)}, restoring the failing test while preserving all other existing tests. This suggests that for straightforward regression bugs where the fault is clearly localized, the test failure information alone may be sufficient to guide the repair.}

\firstrevision{\noindent\textbf{Case 2: Traccar~\footnote{Available at: \url{https://github.com/traccar/traccar}}} (bug-fixing commit \textit{30d5a99aab}, bug-inducing commit \textit{a6b8f7f7ef}). Traccar is an open-source Java-based GPS tracking platform supporting over 200 GPS protocols and more than 2,000 device models. The bug-inducing commit (\textit{``Decode course for Aquila protocol''}) extended \texttt{AquilaProtocolDecoder.java} by introducing bitwise logic to derive a composite course index from four binary bits, but guarded the bearing assignment with \texttt{if (course > 0)}, omitting an upper-bound constraint, thereby permitting out-of-range indices (values 9--15) to produce invalid bearings outside the valid compass range of 0°--359°. This caused the \texttt{testDecode} regression test to fail for protocol messages whose course bits encode an invalid value. Without BIC information, conversational APR with ChatGPT-4o failed to identify the valid range constraint and did not produce a plausible patch. Incorporating BIC information enabled our tool to correctly identify the root cause, that the feature was designed around exactly eight valid directional octants (indices 1--8), and generate the correct patch by augmenting the guard condition from \texttt{if (course > 0) \{position.setCourse((course - 1) * 45);\}} to \texttt{if (course >= 0 \&\& course <= 7) \{position.setCourse(course * 45);\}}, restoring the failing test while preserving all previously passing behavior. Notably, the generated patch is syntactically equivalent to the human-written fix.}

\firstrevision{These two case studies demonstrate complementary aspects of our approach: (i) it is directly applicable to real-world CI settings where regression bugs are naturally detected through automated pipelines; and (ii) bug-inducing change information is crucial for effective repair in non-trivial cases, as illustrated by the contrasting results between the two configurations in Case~2.}

\subsection{\firstrevision{Limitations and Future Directions}}

\firstrevision{While our empirical results demonstrate the promising potential of LLMs in repairing regression bugs, they also reveal several limitations across both fine-tuning-based and prompt-based approaches, as well as challenges that persist even when BIC context is provided. We discuss these limitations below and outline concrete directions for future work.}

\firstrevision{\textbf{Fine-tuning-based approaches} exhibit several shortcomings. First, they do not incorporate test failure information during inference, relying solely on learned bug-fix patterns, which limits their ability to reason about the specific failure context. Second, they cannot readily incorporate auxiliary information such as BIC context without retraining, limiting their adaptability to regression-specific repair scenarios. Third, as illustrated by \textit{RegressionBug-76}, these models either produce structurally malformed patches or syntactically correct but semantically meaningless ones, indicating that learned patterns are insufficient when the root cause requires library-specific knowledge. Fourth, CodeGen and InCoder achieve significantly lower plausible rates (7.33\%--9.33\%) compared to prompt-based approaches, suggesting that their generated patches frequently fail basic compilation or test execution. Fifth, smaller models (CodeGen-2B vs. CodeGen-6B, InCoder-1B vs. InCoder-6B) do not show consistent improvement with scale, suggesting the bottleneck lies in the training data rather than model capacity.}

\firstrevision{\textbf{Prompt-based approaches} also exhibit some limitations. First, conversational strategies tend to overfit to failing test cases through iterative feedback, producing a higher number of plausible patches but relatively few correct ones. Second, as illustrated by RegressionBug-76, prompt-based approaches conflate the regression fault with unrelated patterns from general domain knowledge, leading to semantically plausible but factually incorrect patches. Third, without BIC context, prompt-based techniques cannot identify the root cause of the regression when the test failure does not directly reflect the underlying fault.}

\firstrevision{\textbf{Remaining challenges with BIC context.} Even the best-performing configuration correctly repairs only 39 out of 200 bugs, indicating that there remains substantial room for improvement. Specifically, Remote bugs remain the most challenging category even when BIC information is provided, with the fix rate improving only modestly from 9.4\% to 16.0\%. Furthermore, reversion strategies must be applied judiciously, as blindly reverting to previous code can produce plausible but semantically incorrect patches when the bug-inducing commit contains a mixture of intended and unintended changes.}

\firstrevision{Based on the limitations above, we identify the following concrete directions for future LLM-based regression repair research: (i) incorporating test failure feedback and BIC context during fine-tuning training, enabling regression-aware repair beyond static bug-fix patterns; (ii) integrating domain-specific and library-level knowledge, either through retrieval-augmented generation or knowledge-enhanced prompting, to address cases where the correct fix cannot be inferred from the available context alone; (iii) developing patch diversity enforcement mechanisms to reduce overfitting to the test suite in conversational repair; and (iv) designing repair approaches that can reason across multiple program versions and distinguish intended from unintended modifications within a bug-inducing commit, to better handle Remote regression bugs and complex reversion scenarios.}

\subsection{Experimental Cost Report}
Running APR experiments using LLMs requires substantial computational resources. In our study, experiments involving RepairLLaMA, Incoder-1B/6B, and CodeGen-2B/6B were conducted using an NVIDIA A100 GPU with 80GB of VRAM. To prevent out-of-memory (OOM) crashes observed during generation, we limited the beam size to 10.

In contrast, experiments using ChatGPT models (ChatGPT-4o and ChatGPT-Turbo-3.5) did not require local computational resources, as these models were accessed via OpenAI’s API. However, using these APIs incurs direct monetary costs, as users are charged based on the number of input and output tokens processed by the model. While the exact cost depends on usage, it is worth noting that these models still rely on significant backend computing infrastructure, even if this burden is offloaded from the researcher to the service provider.
As of the time of our implementation, the cost for ChatGPT-4o was \textit{\$0.0025} per \textit{1,000} input tokens and \textit{\$0.01} per \textit{1,000} output tokens. For ChatGPT-Turbo-3.5, the cost was \textit{\$0.0005} per \textit{1,000} input tokens and \textit{\$0.0015} per \textit{1,000} output tokens.

\begin{figure}[!htbp]
\includegraphics[width=0.92\columnwidth]{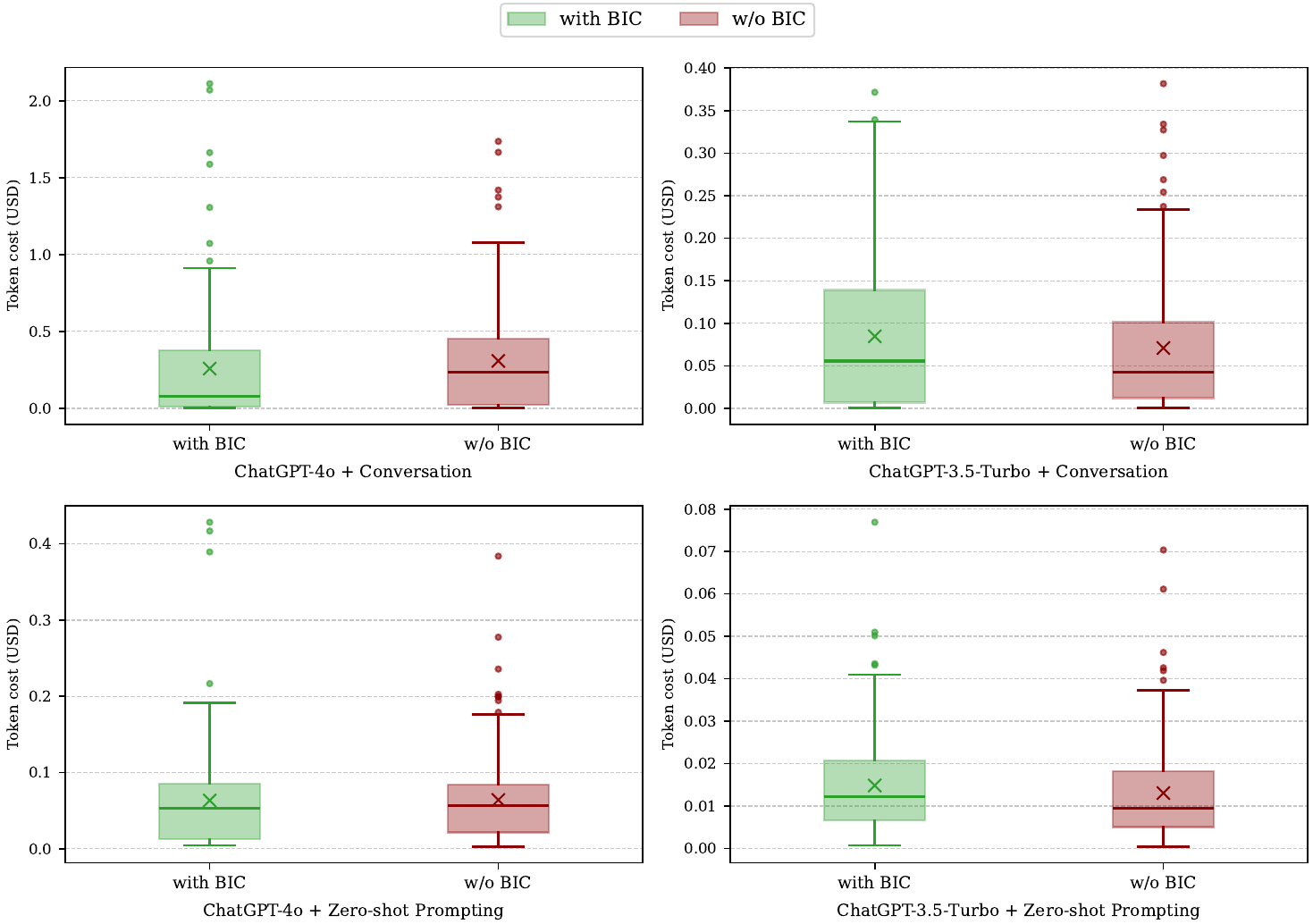}
\caption{Token cost of patch generation per bug across different methods.}
\vspace{-0.2cm}
\label{fig:cost-report}
\end{figure}

\firstrevision{Figure~\ref{fig:cost-report} presents the token cost of patch generation per bug across different configurations on the \textsc{RegressionBug4APR} benchmark}. For zero-shot prompting, the model generates up to 10 patches per bug and stops once a plausible patch is found. In the conversational strategy, the model performs up to 10 attempts per bug, with each attempt involving up to 5 rounds of interaction guided by test case feedback, terminating upon generation of a plausible patch.

Overall, the conversational strategy incurs a higher cost than zero-shot prompting due to the multiple rounds of interaction involved. \firstrevision{It also exhibits higher cost variance, with several outliers indicating that certain bugs require significantly more rounds of interaction to repair}.
Regarding the effect of BIC context, an interesting divergence is observed between the two models. For ChatGPT-Turbo-3.5, incorporating BIC context leads to slightly higher costs, as longer prompts increase the number of input tokens.
\firstrevision{In contrast, for ChatGPT-4o, incorporating BIC context reduces both the median cost and cost variance, as BIC helps the model generate plausible patches in fewer attempts, making it not only more effective but also more cost-efficient. This divergence suggests that the cost-effectiveness of BIC context depends on the underlying model capability, with stronger models being better able to leverage the additional context to reduce repair effort.}

\subsection{Threats to Validity}
We anticipate that there are the following threats to the validity of our findings.

\subsubsection{Internal Validity}
An internal threat comes from the manual validation used to determine the correctness of the plausible patches. To mitigate this threat, we conduct a thorough examination of each patch following prior works~\cite{xu2024aligning}, along with careful double-checking of the results. Still, the assessment of patch correctness remains an active area of research, and further investigation is needed to refine assessment methodologies~\cite{le2026patchguru,le2023invalidator,zhou2024leveraging}.
Another internal threat concerns the manual bug categorization, which involves analyzing, dissecting, and classifying developer-written patches according to established repair operators defined in prior works~\cite{pan2009toward,sobreira2018dissection}. To mitigate this risk, we involved an author with over three years of experience in Java and Python programming to perform this task, which was then double-checked by another author. \firstrevision{Since these repair operators represent fundamental constructs that are common across imperative programming languages, the same set of operators is applied to both Java and Python patches, and the categorization process can be considered reasonably reliable across both languages.}

\subsubsection{External Validity}  
One external threat relates to the potential exposure of our dataset to an LLM's training corpus. However, prior work~\cite{jiang2023impact} has shown that data leakage is less of a concern for APR compared to other code-related tasks. This is because LLM training corpora typically contain, at most, individual versions of a program, either buggy or fixed, but not aligned bug-fix pairs. Without such aligned data, LLMs are unlikely to learn APR tasks directly from training.
Another external threat concerns the generalizability of our findings. In this study, we evaluated several APR techniques using \firstrevision{200} regression bugs collected from widely-used Java and \firstrevision{Python} GitHub repositories. While the dataset is not large, the findings could be further strengthened by expanding to a larger and more diverse set of programs with additional real-world bugs. Nevertheless, we mitigate this threat through the analysis presented in RQ1, which demonstrates substantial diversity in both fix scope and repair operators. \firstrevision{Furthermore, our evaluation across both Java and Python regression bugs provides preliminary evidence for the generalizability of our findings across programming languages.} Moreover, the benchmark has been designed with extensibility, allowing future expansion as more regression bugs are identified and validated.

\subsubsection{Construct Validity}
One construct validity threat concerns the experimental setup used in our empirical study. To ensure a fair comparison, we adopt the original configurations (e.g., architectures and parameters) provided by the authors of the respective techniques, preserving their internal structures. The only modification made was to the sampling size, which was adjusted to control computational cost while keeping the core techniques intact.
\firstrevision{For prompt-based techniques, we use the same prompt templates and configurations across both Java and Python regression bugs, ensuring that any observed differences in repair effectiveness are attributable to the nature of the bugs rather than differences in experimental setup.}


\section{Conclusion}
\label{sec:conclusion}

In this study, we conduct an empirical study of automated program repair (APR) techniques on regression errors, including both traditional and LLM-based approaches. The evaluation is performed using our high-quality benchmark, \textsc{RegressionBug4APR}, \firstrevision{which comprises 200 real-world regression bugs collected from widely-used Java and Python GitHub repositories}. Experimental results reveal that traditional APR tools fail to repair any of the bugs, whereas LLM-based techniques demonstrate promising performance.

Motivated by the effectiveness of LLM-based approaches, we focus on prompt-based techniques due to their flexibility and efficiency compared to fine-tuning-based methods. We specifically investigate whether incorporating regression-specific context, particularly bug-inducing change (BIC) information, can enhance repair effectiveness. Our results show substantial improvements in repair performance when BIC context is included. \firstrevision{Specifically, the best-performing configuration, conversational ChatGPT-4o with full BIC information, successfully repairs 39 out of 200 regression bugs, representing a 1.6× improvement over the best-performing configuration without BIC context}. Our qualitative analysis reveals that BIC information provides crucial context that helps models better identify the root cause of regressions and decide whether to fully or partially revert to previously correct statements. It also narrows the fix scope, improving fault localization and reducing unnecessary code modifications. \firstrevision{In addition, our breakdown by regression bug type shows that Remote bugs are consistently the most challenging category to repair across both languages and all evaluated configurations, while Local bugs, being the most prevalent category, account for the majority of correct repairs.}

\firstrevision{Furthermore, our ablation study reveals that each contextual component of BIC information contributes positively and incrementally to repair performance, with code-level changes yielding a larger gain than commit messages alone, and that the two components capture complementary, non-redundant signals that together enhance repair effectiveness.}
\firstrevision{These findings are consistent across both Java and Python benchmarks, supporting the generalizability of our conclusions.}

Reflecting on the findings, we draw two key insights for future work: improving regression bug repair and leveraging additional context in APR. First, our study suggests that exposing models to how past code changes introduced a regression provides valuable cues for understanding root causes and generating correct fixes. This highlights a new opportunity to guide general-purpose LLMs in identifying regression patterns and applying suitable repairs. \firstrevision{However, our analysis also reveals that even with BIC context, certain regression bugs, such as Remote bugs and those requiring library-specific knowledge, remain difficult to repair, suggesting that BIC information alone is not sufficient and must be complemented with domain-specific or cross-component reasoning capabilities.}
Second, building on recent advances in LLMs, future APR research may benefit from systematically mining and integrating additional context, such as software evolution, into model training. \firstrevision{In this direction, future fine-tuning-based approaches should incorporate test failure feedback and BIC context directly into the training process, rather than relying solely on static bug-fix patterns, enabling models to perform regression-aware repair while retaining the efficiency benefits of fine-tuning.}

\begin{acks}
We sincerely appreciate the reviewers and editors for their constructive comments and feedback.
This material is based in part upon work supported by the Australian Research Council’s Discovery Early Career Researcher Award under project number DE220101057.
\end{acks}

\bibliographystyle{ACM-Reference-Format}
\bibliography{acm-main}


\end{document}